\documentclass[preprint]{aastex}

\usepackage{lscape}

\shorttitle{IRS spectra of disks in Cha~I} 
\shortauthors{Manoj et al.}

\begin{document}

\title{ {\it Spitzer} Infrared Spectrograph survey of young stars in the Chamaeleon I star-forming region}

\author{P. Manoj\altaffilmark{1},  K. H.  Kim\altaffilmark{1}, E. Furlan\altaffilmark{2}, M.  K. McClure\altaffilmark{3}, K. L. Luhman\altaffilmark{4,5}, Dan M. Watson\altaffilmark{1}, C. Espaillat\altaffilmark{6}, N. Calvet\altaffilmark{3},  J. R. Najita\altaffilmark{7}, P. D'Alessio\altaffilmark{8},  L. Adame\altaffilmark{8},  B. A. Sargent\altaffilmark{9}, W. J. Forrest\altaffilmark{1}, C. Bohac\altaffilmark{1},  J. D. Green\altaffilmark{10} and L.  A. Arnold\altaffilmark{1}}

\altaffiltext{1}{Dept. of Physics and Astronomy, University of Rochester, Rochester, NY 14627}

\altaffiltext{2}{{\it Spitzer} Fellow; JPL, Caltech, 4800 Oak Grove Dr., Pasadena, CA 91109}

\altaffiltext{3}{Department of Astronomy, University of Michigan, Ann Arbor, MI 48109}

\altaffiltext{4}{Department of Astronomy and Astrophysics, The Pennsylvania State University, University Park, PA 16802}

\altaffiltext{5}{Center for Exoplanets and Habitable Worlds, The Pennsylvania State University, University Park, PA 16802}

\altaffiltext{6}{NSF Astronomy \& Astrophysics Postdoctoral Fellow; Harvard-Smithsonian Center for Astrophysics, 60 Garden Street, MS-78 Cambridge, MA 02138}

\altaffiltext{7}{National Optical Astronomy Observatory, 950 N. Cherry Avenue, Tucson, AZ 85719}

\altaffiltext{8}{Instituto de Astronom\'{i}a, UNAM, Apartado Postal 70-264, Ciudad Universitaria,
04510, M\'{e}xico DF, M\'{e}xico.}

\altaffiltext{9}{Space Telescope Science Institute, 3700 San Martin Drive, Baltimore, MD 21218}


\altaffiltext{10}{The University of Texas at Austin, Dept. of Astronomy, Austin, Texas 78712-0259}

\email{manoj@pas.rochester.edu}

\begin{abstract}
We present 5 to 36~$\micron$ mid-infrared spectra of 82 young stars in
the $\sim$~2~Myr old Chamaeleon~I star-forming region, obtained with the
{\it Spitzer} Infrared Spectrograph (IRS). We have classified these
objects into various evolutionary classes based on their
spectral energy distributions and the spectral features seen in the IRS
spectra. We have analyzed the mid-IR spectra of Class II objects in
Chamaeleon I in detail, in order to study the vertical and radial
structure of the protoplanetary disks surrounding these stars. We find
evidence for substantial dust settling in most protoplanetary disks in
Chamaeleon~I.  We have identified several disks with altered radial
structures in Chamaeleon~I, among them transitional disk candidates
which have holes or gaps in their disks. Analysis of the silicate
emission features in the IRS spectra of Class~II objects in Cha~I
shows that the  dust grains in these disks have undergone significant
processing (grain growth and crystallization). However, disks with
radial holes/gaps appear to have relatively unprocessed grains. We
further find the crystalline dust content in the inner ($\la$ 1-2~AU) and
the intermediate ($\la$ 10~AU) regions of the protoplanetary disks to be
tightly correlated.  We also investigate the effects of accretion and
stellar multiplicity on the disk structure and dust
properties. Finally, we compare the observed properties of
protoplanetary disks in Cha~I with those in slightly younger Taurus
and Ophiuchus regions and discuss the effects of disk evolution in the
first 1-2 Myr.

\end{abstract}


\section{Introduction}

Planetary systems are formed out of protoplanetary disks surrounding
young stars. Understanding the processes responsible for, and the
timescales associated with, the dissipation of the disks and the
formation of planetary systems is an outstanding problem in
astronomy. Most solar-mass stars younger than $\sim$ 1 Myr appear to
harbor disks around them; by $\sim$ 3 - 5 Myr most of them shed their
disks
\citep{haisch01,hill05,hernandez08}. Therefore, detailed studies of
the structure and properties of the 1-2 Myr old protoplanetary disks
are critically important to our understanding of the key processes
governing disk evolution and planet formation.

Protoplanetary disks are natural by-products of the star formation
process, which begins when a slowly rotating cloud core collapses to
form a central protostar surrounded by an accretion disk and an
overlying, infalling envelope. The material from the envelope ``rains
down'' to the disk, which then gets accreted onto the central star
\citep[e.g.][]{shu87,hart98}. The observed spectral energy
distribution (SED) of such systems has a rising continuum in the
infrared, and they have been classified as Class I sources
\citep{lada87,wilk89}. The envelope eventually dissipates by draining
onto the disk and/or is dispersed by the outflow/wind from the star-disk
system, leaving behind a pre-main sequence star surrounded by an
accretion disk. SEDs of such objects show somewhat flat or decreasing
continuum at infrared wavelengths and they are classified as Class II
objects \citep{lada87,wilk89}. It is in these disks surrounding Class
II objects that planetary systems form.

The planet formation process begins with the sub$\micron$-sized grains
in the disks sticking together to grow to larger mm- and cm-sized
particles \citep[e.g.][]{weiden80,blum08}. As they grow, the larger
grains sink down to the disk mid-plane; sedimentation of the dust can
cause significant changes in the vertical structure of the disk
\citep{dalessio06}. Along with the grain growth, mineralization of the
initially amorphous dust grains also take place in protoplanetary
disks \citep[e.g.][]{camp89,malfait98}. The larger grains that have
settled to the disk mid-plane further grow into km-sized
planetesimals, which through collisional growth, eventually form
protoplanets \citep[e.g][]{weiden08}. Once these `planetary embryos'
have become sufficiently massive ($\sim 10$~M$_{\oplus}$), they
accrete gas from the disks to form giant planets
\citep[e.g.][]{pollack96}. A Jupiter-like gas giant formed in the disk
can gravitationally alter the disk structure by forming radial gaps
and holes in them \citep{rice03,quillen04}.  The disks eventually
dissipate either by planet formation using up the material and by
various other disk dispersal processes such as accretion onto the
central star, photoevaporation and magneto-rotational instability
induced disk clearing
\citep[e.g.][]{clarke01,alex07,chiang07}. Pre-main sequence stars
which have dissipated most of their disk material show marginal or no
excess emission above their stellar photospheres and are called Class
III objects \citep{lada87,wilk89}.

The effects of disk evolution are best studied at mid-IR
wavelengths. The mid-IR emission from a Class II object probes the
planet forming region ($\sim$ 1-10~AU) of the disk surrounding it
\citep[e.g.][]{dullemond07,dalessio06}. Dust sedimentation makes the
disks flatter, i.e. reduce the degree of flaring
\citep{dd04,dalessio06}. The shape of the mid-IR continuum, in
particular at wavelengths $\ga$ 13 $\micron$, is related to the disk
geometry and can be used to characterize the vertical distribution of
the dust in the disks \citep{kh87, dd04, dalessio06, meeus01,
  furlan06}. The changes in the radial structure of the disks also
affect the shape of the mid-IR continuum: transitional disk candidates
- objects with radial gaps or holes within their disks -
\citep{strom89, skrut90, calvet05, esp07b, najita07, brown07, esp10} and
outwardly truncated disks \citep{mcclure08} have been identified by
inspecting the mid-IR part of the SED. Additionally, the silicate
emission features centered at $\sim$ 10 and $\sim$ 20 $\micron$
generated by the optically thin dust grains in the disk surface layers
can be used to probe the size and composition of the dust in the
disks. The grain growth and crystallization, i.e. the degree of dust
processing in protoplanetary disks can be quantified using silicate
emission features \citep[e.g.][]{bouwman01, boekel04, sargent09}.

As part of a large {\it Spitzer} Infrared Spectrograph
\citep[IRS\footnote{The IRS was a collaborative venture between
    Cornell University and Ball Aerospace Corporation funded by NASA
    through the Jet Propulsion Laboratory and the Ames Research
    Center};][]{houck04} survey of star forming regions within $\sim$
500 pc, our team has obtained mid-IR spectra of several hundred YSOs
in the 1-2 Myr old, Taurus, Chamaeleon I and Ophiuchus regions. The
spectra of the young stars in Taurus and Ophiuchus have been presented
in \citet{furlan06}, \citet{furlan08} and \citet{mcclure10}. Here we
present and analyze the IRS spectra of young stars in the Chamaeleon I
star forming region; a smaller subset of the Class II disks classified and
identified in this paper along with those in Ophiuchus and Taurus has
been analyzed in \citet{furlan09}, where a comparative study of the
disk properties in these three regions is presented.

The Chamaeleon~I (henceforth Cha~I) cloud, which is one of the nearest
star forming regions to the Sun, at a distance of 160-165 pc
\citep{luhman08b}, contains a relatively isolated population of young
stars with modest levels of extinction towards them \citep{luhman04}.
More than 200 young stars are known to be members of the Cha~I star
forming region \citep[][and references therein]{luhman08b}. The young
stellar population of Cha~I has a median age of $\sim$ 2~Myr and an
age spread of 3 - 6 Myr and is thus older than that of Taurus and
Ophiuchus \citep{luhman08b}.

In this paper we present the mid-IR spectra of 82 young stars and
brown dwarfs in Cha~I, obtained with the {\it Spitzer} IRS. Our sample
was initially selected from the known members of Cha~I based on their
mid-IR flux measured by the {\it Infrared Astronomical Satellite}
(IRAS) and {\it Infrared Space Observatory} (ISO) and from the members
identified in the deep H$\alpha$ and near-IR surveys of Cha~I
\citep{whit87, whit97, persi00, comeron99, comeron00, cam98,
  ots99}. Objects with mid-IR flux F$_{14.3~\micron}$ or
F$_{12~\micron}$ $\ge$ 10~mJy were included in the sample. Several low
mass stars and brown dwarfs \citep{luhman04} were also included later
based on their IRAC and MIPS fluxes, which were available by then,
and employing a flux cutoff F$_{8~\micron}$ or F$_{24~\micron}$ $\ge$
2.5~mJy \citep[see][]{luhman08}. In Section~\ref{obs} we describe our
observations and data reduction. In Section~\ref{char}, we
characterize our sample and classify the objects into various
evolutionary classes based on their spectral energy distributions and
the IRS spectra. A detailed analysis of the IRS spectra and a
comparison of the observed disk properties of the Class~II objects in
Cha~I with those predicted by irradiated, accretion disk models are
presented in Section~\ref{disk}. The vertical and radial structure of
the disks based on the observed continuum spectral indices are
discussed in Section~\ref{index} and the properties of the silicate
emission features and dust processing in the disks are discussed in
Section~\ref{silicate}. In Section~\ref{binary}, we examine the effect
of stellar multiplicity on disk structure and in Section~\ref{cwtts}
we compare the disk properties of Classical and Weak-line T Tauri
stars.  Finally, our main results are summarized in
Section~\ref{results}.

\section{Observations and data reduction} \label{obs}

We observed 92 young stellar objects, including very low mass stars
and brown dwarfs, in the Cha~I star forming region.  All the
targets were observed in the IRS staring mode, in which a target was
successively observed in two nod positions, at $\sim$~1/3 and
$\sim$~2/3 of the slit length, along the spatial direction of the
slit. The full mid-IR spectrum from 5 to 38 $\micron$ was obtained
by using the low spectral resolution ($\lambda$/$\Delta\lambda$ $\sim$
60-120) modules of the IRS: Short-Low (SL; 5.3-14$\micron$) and
Long-Low (LL; 14-38 $\micron$).  For a few bright targets
\citep[see][]{watson09}, the high spectral resolution
($\lambda$/$\Delta\lambda$ $\sim$ 600) modules, Short-High (SH;
10-19$\micron$) and Long-High (LH; 19-37 $\micron$), were used instead
of LL. Each observation immediately followed a high-precision pointing
peak-up with the Pointing Calibration and Reference Sensor
\citep[PCRS;][]{mainzer03} yielding a single axis pointing uncertainty
of 0.28$\arcsec$, that is small compared to the pixel sizes
(1.8$\arcsec$ in SL, 4.8$\arcsec$ in LL, 2.3$\arcsec$ in SH,
5.1$\arcsec$ in LH).

To extract the spectra, we started with the stray light-corrected,
dark current-subtracted and flat-fielded basic calibrated data (BCD)
products from the Spitzer Science Center (SSC) IRS data pipeline. The
non-flat-fielded, droop products were used for some targets observed
using the combination of SL, SH, and LH during campaign 29 to enhance
the signal-to-noise. The SSC IRS pipeline versions S13.2 and S14.0
were used for objects observed in campaigns 20 to 33, and S16.1 for
campaigns 41 and 42. We first fixed the permanently bad pixels and the
so called `rogue' pixels \citep[see][]{watson09} flagged by the
pipeline by interpolation over the neighboring good pixels along the
dispersion axis (rogue pixels in SH and LH data were fixed for all
campaigns; rogue pixels in LL data for campaign 41 and 42 were fixed;
rogue pixels in SL data were not fixed). We used the SMART software
tool \citep{higdon04} to extract and calibrate the spectra. For the
low-resolution data (SL and LL), sky background was removed by
subtracting the same orders of the spectra obtained at different nod
positions before the point source spectra were extracted with a
variable-width column matched to the IRS point-spread function
\citep[see][]{sargent06,watson09}. For the high resolution data (SH
and LH), since the sky emission is negligible,
a full slit spectral extraction was performed without sky subtraction.
The extracted spectra were calibrated by multiplying them with the
Relative spectral response functions (RSRFs), constructed from the
observed spectra of calibration standards and their template spectra
\citep[see][]{watson09,sargent09}. For each object, calibrated spectra
at both the nod positions were averaged to obtain the final
spectrum. Finally, the SH spectra were truncated at 14 $\micron$ and
all the high resolution spectra (SH and LH) were rebinned to the same
resolution and sampling as SL and LL.

Special care was required for the spectral extraction of the following
objects: Cha H${\alpha}$2, Hn~10E, ISO~237 and CHXR~30A. The SL1 nod1
image of Cha~H${\alpha}$2 suffered from an artifact from a bright
object in the peak-up array, so we used only the nod 2 for the SL
spectrum. ISO~237 and CHXR~30A had problems with source contamination
in the LL slit; however, careful sky subtraction, source extraction
with a PSF, and scaling between modules generated a reliable version
of the 5-36 $\micron$ spectrum.  Hn~10E had scattered light from a nearby
bright object entering in the LL1 image and we could not completely
remove this contribution during the extraction, especially at
$\lambda$ $\ga$ 30 $\mu$m. We present the IRS spectrum of Hn~10E here,
but do not use it in our analysis.

Of the 92 sources observed, spectral extraction failed for the
following 7 objects: HM8 (2M~J11025504-7721508), ISO~13
(2M~J11025579-7724304) (probably a galaxy \citep[see][]{luhman08}), T4
(SY~Cha, 2M~J10563044-7711393), 2M~J11070768-7626326, OTS44
(2M~J11100934-7632178), 2M~J11114533-7636505 and
2M~J11122250-7714512. These sources were too faint and could not be
located in the slit either due to mispointing or due to insufficient
exposure. One of the objects, T9, turned out be a background
late M giant and not a member of Cha I \citep{luhman04}.  Two of the
sources observed, 2M~J11183572-7935548 and 2M~J11432669-7804454,
are probable members of $\epsilon$~Cha association \citep{luhman08}.
Thus in this paper, we present IRS spectra of 82 YSOs in Cha I.  The
log of our observations is given in Table~\ref{log_tbl}.

\section{Characterization of the sample} \label{char}

The basic data for the Cha~I sample are presented in
Table~\ref{basic_tbl}. We have spectral type information for 74 stars
out of 82 Cha~I members. Most of them ($\ga$ 90\%) are of spectral
type K0 or later. Of the 7 early type stars in the sample, 2 are B
type and 5 are G type. Figure~\ref{sp_hist_fig} shows the spectral
type distribution of stars in the sample which are K0 or later. The
spectral type distribution of our sample peaks at M5/M6 and is quite
similar to the distribution of the more complete sample of the known
members of Cha~I \citep{luhman08b}; our sample is representative of
the young stellar population of the Cha~I cloud. A significant
fraction of the objects in our sample with IRS spectra are very
low-mass stars or brown dwarfs: 19 of them have spectral types M5 or
later, of which 7 are likely to be brown dwarfs (M6 or later).

We classified our sample objects into classical T Tauri stars (CTTS)
and weak-line T Tauri stars (WTTS) based on their observed equivalent
width of H$\alpha$ emission (W$_{H\alpha}$) compiled from the
literature (see Table~\ref{basic_tbl}).  We
followed the empirical criteria prescribed by \citet{wb03} which is
based on W$_{H\alpha}$, but also accounts for the spectral type
dependence of the `contrast effect' which affects the observed
W$_{H\alpha}$. The W$_{H\alpha}$ values used and the CTTS/WTTS (C/W)
classification of our sample are listed in Table~\ref{basic_tbl}.

In Table~\ref{binary_tbl} we list the observed properties of the
companions of the Cha~I sources which are known to be in multiple systems.
We only list the stellar companions whose membership to Cha~I cloud
has been established.  For instance, SZ~Cha is reported to have two
candidate companions, B (2M~J10581804-7717197) and C
(2M~J10581413-7717088) separated from the primary by 5.2 $\arcsec$ and
12.5 $\arcsec$ respectively \citep{ghez97,laf08}. However, these
candidates have not been confirmed as members of Cha~I association
\citep{luhman07,luhman08b}. Similarly, T51 is reported to have a third
component T51C (2M~J11122740-7637017) separated by 12.4 $\arcsec$ from
the primary \citep{ghez97}, which is not a confirmed member of Cha~I
association \citep{luhman07,luhman08b}. We do not include them in the
table. Most of the multiple systems in our sample have companions at
separations $\la$ 3$\arcsec$. As the narrowest IRS slit width is
3.6$\arcsec$ (SL), the IRS spectra presented here for these objects
are the composite spectra of the unresolved components. We take this
into account while interpreting the spectra of these objects. However,
for most of these close binaries, one of the components is brighter
and dominates the  mid-IR spectrum.

We compiled the photometric data for our sample from the
literature. For all objects we used the J, H, and K$_s$ measurements
from the Point Source catalog of the The Two Micron All Sky Survey
\citep[2MASS;][]{skrut06,cutri03}. The I-band data were obtained from
the Third Release of the DEep Near-Infrared Survey of the southern sky
\citep[DENIS;][]{epch99} except for two for which there were no DENIS
observations. When available, we also compiled the IRAC 3.6, 4.5. 5.8,
and 8.0 $\micron$ and MIPS 24 $\micron$ flux measurements for the
objects in our sample from \citet{luhman08}.

Two mid-M type stars in Table~\ref{basic_tbl}, 2M~J11183572-7935548
and 2M~J11432669-7804454, are more likely members of the
$\epsilon$~Cha moving group as suggested by their proper motions
\citep[see][]{luhman08, luhman08b}. $\epsilon$~Cha is older ($\sim$
6~Myr) and closer to us (d $\backsimeq$ 114~pc) than Cha~I and its
members exhibit larger proper motions than those of Cha~I
\citep{luhman08,mamajek00,frink98}.

\subsection{Extinction correction}

Figure~\ref{ext_hist_fig} shows the distribution of extinction values
(A$_J$) (see Table~\ref{basic_tbl}) of our sample objects in
Cha~I. More than half of the stars have A$_J$ $> $ 0.8 mag. The
wavelength dependent extinction law towards molecular clouds is known
to diverge from that for the diffuse Interstellar medium (ISM)
\citep{mathis90} for A$_V$ $\ga$ 3 mag (A$_J$ $\ga$ 0.8 mag)
\citep{mcclure09,chapman09}. \citet{luhman04} has shown that the
extinction towards Cha~I does not follow the standard ISM extinction law
of \citet{ccm89} or \citet{mathis90}. Moreover, using the standard ISM
extinction curve to deredden the IRS spectra causes the 10 $\mu$m
silicate feature to be sharply peaked at $\sim$ 9.7 $\mu$m, especially
for objects with large extinction \citep{furlan09}. For these reasons
we adopt the composite extinction law prescribed by \cite{mcclure09}
to deredden the observed broad band flux and IRS spectra of Cha~I
objects. This method has been successfully used to deredden the IRS
spectra of young stars in the nearby young clusters
\citep{furlan09,mcclure10}. We deredden the broad band photometry and
IRS spectra as follows: objects with A$_J$ $<$ 0.8 mag (A$_{K_s}$ $<$
0.3 mag), are derddened using extinction law from \citet{mathis90} with
R$_V$ = 5 \citep[see][]{furlan09,mcclure10}. For objects with A$_J$ $\ge$ 0.8 mag (A$_{K_s}$ $\ge$ 0.3 mag), we
used the appropriate extinction curves from \citet{mcclure09} to deredden
the photometry and the IRS spectra.

\subsection{Spectral energy distributions and classification} \label{class}

We constructed the SEDs of the objects in our sample from the
photometry compiled from literature and the IRS spectra. Objects for
which the A$_J$ values are listed in Table~\ref{basic_tbl}, the SEDs
were generated from the dereddened fluxes. The stellar photospheric
flux for stars with known spectral types were derived from the colors
of main sequence stars tabulated by \citet{kenhart95}, normalized at
the dereddened J-band flux of each object. Objects for which spectral
types and extinction values are not available, we present the observed
SEDs. The SEDs of Cha~I members are shown in
Figures~\ref{class1_sed_fig} - \ref{class3_sed_fig} and those of
$\epsilon$ Cha members are shown in Figure~\ref{epscha_sed_fig}.

We first classified the Cha~I YSOs into various SED classes based on
the spectral index between 2.2 $\micron$ and 25 $\micron$
\citep{lada87,wilk89,green94}. The spectral index, $n_{2-25}$ is
computed using the relation

\begin{equation}
\label{eqn1}
n_{\lambda_1-\lambda_2}\:=\:log\left(\frac{\lambda_2F_{\lambda_2}}{\lambda_1F_{\lambda_1}}\right)\;/\;log\left(\frac{\lambda_2}{\lambda_1}\right)
\end{equation}

\noindent
from the 2MASS K$_s$-band flux ($\lambda_1$ = 2.159 $\mu$m) and the
IRS spectrum ($\lambda_2$ = 25 $\mu$m). YSOs with $n_{2-25}~\ge~0.3$
are classified as Class I sources; objects with
$0.3~\ge~n_{2-25}~\ge~-0.3$ are classified as Flat spectrum sources;
sources with $-0.3~\ge~n_{2-25}~\ge~-2.0$ as Class II objects; and
$n_{2-25}~\le~-2.0$ as Class III sources \citep[also
  see][]{luhman08}. The $n_{2-25}$ values and the SED classes based on
them are listed in Table~\ref{basic_tbl}. Objects for which A$_J$ values
are listed, the observed and the dereddened values of $n_{2-25}$ are
presented in the table; the SED classes of these objects remain
unchanged for both these values. For objects with unknown extinction,
the SED classes listed in Table~\ref{basic_tbl} are based on the
observed $n_{2-25}$. The SED classification yields 6 Class I sources,
5 Flat spectrum sources, 68 Class II sources and 3 Class III
sources. 

Our SED classification of the Cha~I objects are in good agreement with
the classification by \citet{luhman08} based on their IRAC and MIPS
photemetry of these objects. Of the 74 sources that are common between
the two samples, only 7 show disagreement between the
classifications. Four of these, T25, T29, ISO 126 and T33A, do not
have MIPS measurements available and the classification by
\citet{luhman08} is based on the 2 to 8 $\micron$ slope which differs
from our classification based on the $n_{2-25}$ index. For the other
three objects, ISO 86, Ced110 IRS6 and C1-25, our classification is
consistent with that based on the 2 to 24 $\micron$ slopes listed in
\citet{luhman08}; however, the classification of these objects based on the slope at
wavlengths $\le$ 8 $\micron$ differ from that based on  $n_{2-25}$
index, so as to make the classification ambiguous
\citep{luhman08}. Moreover, the extinction towards these objects are
uncertain and the spectral slopes are computed from the observed
SEDs. These objects are further discussed below, where we explore the
connection between the SED classes and the evolutionary status of YSOs
in Cha~I.

The empirical SED classification based on the IR-slope need not always
correspond to the physical evolutionary stages predicted by the star
formation theory \citep[e.g.][]{evans09}. As pointed out by several
authors \citep{robit07,crapsi08,mcclure10}, the spectral index
$n_{2-25}$ is affected by the viewing geometry of the source and/or
the line of sight extinction, especially in regions of high
extinction. A highly extinguished Class~II object could easily be
misclassified as a Flat spectrum or a Class~I source
\citep{robit07,crapsi08,mcclure10}. In order to disentangle the effect
of extinction, and to determine if the mid-IR emission from Class I
and Flat spectrum sources is envelope dominated, we make use of the
``extinction-free'' index between 5 and 12 $\micron$ ($n_{5-12}$)
introduced by \citet{mcclure10}. The slope of the SED measured between
these wavelengths is shown to remain unaffected by extinction
\citep{mcclure09, mcclure10}.

The $n_{5-12}$ values for our sample are listed in
Table~\ref{basic_tbl}. Figure~\ref{extinct_free_fig} shows the plot
between observed $n_{2-25}$ index used for SED classification and the
``extinction-free'' index $n_{5-12}$. The regions occupied by various
SED classes are shown in the Figure. From an analysis of the well
studied protostars in Taurus, \citet{mcclure10} have shown that for
objects with $n_{5-12}~>~-0.2$, envelope emission dominates the mid-IR
wavelengths and for objects $n_{5-12}~<~-0.2$, the mid-IR emission is
disk dominated. For $n_{5-12}~<~-2.25$, the mid-IR emission is mostly
photospheric.  Below we briefly describe the objects in various SED
classes and discuss if they are envelope or disk sources.

{\bf Class I Objects}:~~The SEDs of the 6 Class I objects in our
sample are presented in Figure~\ref{class1_sed_fig} and their IRS
spectra are presented in Figure~\ref{class1_irs}. Only 4 of them show
$n_{5-12}~>~-0.2$ indicating the presence of envelope around
them. They are Ced110~IRS4, Cha~IRN, ISO~192 and T14a. Except for T14a,
the other three show strong ice absorption features in the IRS
spectra: at 6.0 $\micron$ due to H$_2$O ice, at 6.8 $\micron$ due to
CH$_3$OH, NH$^{+}_4$, and strongly polar H$_2$O ice, and at 15.2
$\micron$ due to CO$_2$ ice. The 10 $\micron$ silicate feature is also
strongly in absorption indicating the presence of envelope material
around them. 

T14a, on the other hand, show featurless mid-IR spectrum. The
extinction towards it is relatively low ($A_J$ = 0.4 mag) compared to
that towards many of the Class II sources in our sample. The optical
spectrum of this source suggests that it is a classical T Tauri star (see
Table~\ref{basic_tbl}) with strong forbidden lines \citep{schwartz84}.
T14a is unlikely to have an envelope around it. It is a 2.35$\arcsec$
binary \citep{haisch04} and the IRS spectrum presented here is a
composite of both the components.  The contribution from the companion
is probably responsible for the rising SED and its missclassification
as a Class I source.

Two Class I sources, Ced110~IRS6 (ISO 92) and ISO~86 (CED110 IRS11),
have $n_{5-12}~<~-0.2$ indicating that their mid-IR emission is disk
dominated. Both these sources have been classified as Flat spectrum
sources by \citet{luhman08} based on their 3.6-8 $\micron$ and 3.6-24
$\micron$ SED slopes. Ced110~IRS6 is 1.95$\arcsec$ binary, but the
primary appear to dominate the mid-IR flux \citep{persi01,
  haisch06}. Both Ced110~IRS6 and ISO~86 have been reported to have
HCO+ (4-3) emission, which traces the dense envelope gas; however, the
peak intensity of the HCO+ (4-3) emission in both these sources are
significantly lower than that observed for the envelope sources
Ced110~IRS4 and Cha~IRN \citep{vankemp09,hiramatsu07}. Ced110~IRS6 is
not detected at 1.3 mm or at cm wavelengths, indicating weak envelope
emission \citep{henning93, lehtinen01, lehtinen03,
  hiramatsu07}. Likewise, ISO~86 is not detected longward of 100
$\micron$ suggesting that the source does not have dense envelope
material around it \citep{lehtinen01, lehtinen03}. However, both these
sources show ice and silicate absorption features (see
Figure~\ref{class1_irs}) indicating the presence of cold material
along the line of sight. They are likely to be disk dominated sources
with low density envelope around them, which are either viewed edge-on
and/or extinguished by foreground cloud material.

{\bf Flat spectrum sources}:~~The SEDs of the Flat spectrum sources
are presented in Figure~\ref{flat_sed_fig} and their IRS spectra are
presented in Figure~\ref{flat_irs}. All the Flat spectrum sources in
Figure~\ref{extinct_free_fig} fall close to the envelope - disk
boundary of $n_{5-12}~=~-0.2$, indicating that their envelope
emission, if any, is weak.  Their IRS spectra show weak or no ice
absorption features; the silicate features in most of them are weakly
in emission, indicating very little absorbing metrial between the
sources and the observer. They are likely disks viewed at higher
inclination angles and/or disks with remnant enevelopes.

{\bf Class II objects}:~~ All Class II objects show presence of broad
silicate emission features at 10 $\micron$ and at 20 $\micron$ in
their IRS spectra.  The SEDs of Class II objects in our sample are
presented in Figure~\ref{class2_sed_fig} and their IRS spectra are
presented in Figures~\ref{grA}~-~\ref{out}. The spectra are shown in a
sequence of decreasing strength of the 10 $\micron$ silicate emission
feature and decreasing mid-IR continuum slope. T33A does not have
$n_{2-25}$ listed in Table~\ref{basic_tbl} as the K$_s$ band flux is
an upperlimit. But the IRS spectrum shows strong silicate emission at
10 and 20 $\micron$ (Figure~\ref{grA}), and the $n_{5-12}$ index is
consistent with disk emission. We classify T33A as a Class II
source. The two B-type stars in our sample, CU~Cha and HD~97300 are
well known Herbig Be stars \citep{the94, hill92}. We classify both of
them as Class II objects \citep[also see][]{luhman08}. We present the
SEDs and IRS spectra of these Herbig Be stars, but do not include them
in our analysis.

{\bf Transitional disks}:~~ A few of the Class II disks in Cha~I have
been identified as transitional disk candidates by \citet{kim09}. The
SEDs of transitional disks are characterised by a deficit of flux at
wavelengths $<$ 8 $\micron$ compared to that of the Class II median,
and significant excess emission comparable to or higher than that of
Class II median longward of $\sim$ 13 $\micron$ \citep{calvet05,
  esp07a, esp07b, kim09}. They are protoplanetary disks with dust
depleted inner holes or gaps within them \citep{strom89, calvet05,
  najita07, esp07b, brown07}. In Section~\ref{trans}, we present a
more quantitative criterion based on the continuum indices to identify
transitional disks from a large sample of Class II disks.  We confirm
the transitional disk candidates reported by \citet{kim09}, and
identify an additional transitional disk candidate,
2M~J11241186-7630425. We reclassify one of the transitional disk
candidate reported by \citet{kim09}, T21, as a Class III object (see
below). The SEDs of 8 transitional disks in Cha I are shown in
Figure~\ref{trans_sed_fig} and their IRS spectra are shown in
Figure~\ref{trans_spec}.

{\bf Class III objects}:~~ The SEDs of the 3 Class III objects are
shown in Figure~\ref{class3_sed_fig}. They have $n_{5-12}~<~-2.25$ (see
Figure~\ref{extinct_free_fig}) consistent with their emission being
photospheric.

We assesed the completeness of the objects belonging to various SED
classes in our sample by comparing them with the Cha~I YSOs observed
and classified with Spitzer IRAC and MIPS imaging of the Cha I cloud
\citep[see][]{luhman08}. Our sample contains all the Class I and Flat
spectrum sources in \citet{luhman08} except three, two of which are
later than M3. Compared to \citet{luhman08}, our sample of Class II
objects is complete for spectral types M1 or earlier and 84\% complete
for spectral types M5 or earlier. Since our sample is flux limited,
it has only 3 Class III sources, compared to the 95 identified by
\citet{luhman08}.

{\bf $\epsilon$ Cha members}:~~  Among the $\epsilon$ Cha members, 2M
J11183572-7935548 appears to be a transitional disk candidate and 2M
J11432669-7804454 is a Class II object. The SEDs of these objects are
shown in Figure~\ref{epscha_sed_fig}.

\section{Disk structure and evolution} \label{disk}
The mid-IR excess emission observed for the Class II objects arises
from the protoplanetary disks surrounding them. The shape of the
continuum and the dust emission features in the IRS spectrum can be
used to study the evolution of disk structure and dust properties. In
the following we carry out a detailed analysis of the SEDs and the IRS
spectra of the Class II objects in Cha I and compare their observed
properties with those predicted by the irradiated, accretion disk
models \citep{dalessio06,esp09}.

\subsection{Continuum spectral indices and disk structure \label{index}}

In order to study the structure and evolution of Class II disks in
Cha~I, we quantify the shape of their mid-IR continuum by computing
various spectral indices from the dereddened 2MASS K$_s$-band flux
density and IRS spectra. The continuum spectral indices,
$n_{\lambda_1-\lambda_2}$, were computed using equation~(\ref{eqn1})
between the wavelengths 2.159 and 5.7 $\micron$ ($n_{2-6}$), 5.7 and
13.4 ($n_{6-13}$) $\micron$, and 13.4 and 31.1 $\micron$ ($n_{13-31}$)
\citep{furlan06, furlan09, watson09, mcclure10}. These wavelengths
were chosen because they are relatively free of gas and dust emission
features.  The spectral indices computed for Class II, transitional
and Class III objects in our sample are listed in
Table~\ref{tab4_index}.

In Figure~\ref{fig_indices} we present a plot of  the continuum
spectral indices $n_{13-31}$ and $n_{6-13}$ for the Class II and Class
III objects in Cha I. Transitional disk candidates
identified in Section 3.2 are labeled  separately. 
Figure~\ref{fig_indices} shows that the spectral indices of Class III
objects are very close to that of the stellar photosphere. Most transitional 
disk candidates separate out from other Class II objects in the spectral indices
plot. Two Class II disks, T51 and Hn~5, show $n_{13-31}$ $<$ -1.33, bluer
than that for an optically thick, flat disk, suggesting that the disks
around them are truncated from outside \citep[see][]{furlan09}. Gravitational 
interaction with a close companion ($\la$ 1$\arcsec$ separation at the distance 
of Cha~I) is known to cause outward truncation of the 
disks \citep [][]{artlub94, mcclure08}. T51 is a 1.9$\arcsec$ binary (physical separation $\sim$ 314~AU)
\citep{reipzinn93, ghez97, laf08}, but it is unlikely that such a  wide companion 
can cause disk truncation. Hn~5 is not known to have close 
 companions. 

\subsubsection{Dust settling in Cha~I disks} \label{settling}

Next we compare the observed continuum indices of the Class~II disks
in Cha~I with those predicted by irradiated, accretion disk
models. The models used here are based on the formalism of
\citet{dalessio06}, where the vertical disk structure is derived
self-consistently. The dust in the disk is assumed to be composed of
olivine silicate and graphite grains \citep{dorsch95,dl84} which are
assumed to follow a power law size distribution of the form
$n(a)\;\propto\;a^{-3.5}$. The disk surface layers are populated with
mostly small grains ($a_{min}$=0.005 $\micron$, $a_{max}$=0.25
$\micron$) and the disk midplane with large grains ($a_{min}$=0.005
$\micron$, $a_{max}$= 1 mm). The models incorporate disk evolutionary
effects such as dust settling by parametrizing the depletion of the
small grains from the surface layers of the disk. The dust depletion
factor, $\epsilon$, is defined as the ratio of the dust-to-gas mass
ratio in the upper layers to that of the interstellar medium.  A value
of $\epsilon$ = 1 implies no dust settling and the degree of settling
increases with decreasing value of $\epsilon$. For $\epsilon$ = 0.001,
the small grains in the upper layers of the disk are depleted by a
factor of 1000 relative to the interstellar dust-to-gas mass ratio, with a
corresponding increase of large grains close to the disk midplane
\citep [for details see][]{dalessio06, esp09}.

We computed the continuum spectral indices for a grid of disk models
for central stellar masses of 0.5 and 0.2 M$_{\odot}$ and for
different accretion rates, inclination angles and the settling
parameters $\epsilon$ \citep{esp09}. These models are the same as
those used in \citet{furlan09} and \citet{mcclure10}. The stellar
masses chosen for the models are representative of the masses of
objects in our Cha I sample \citep[][]{luhman07}. The continuum
spectral indices, n$_{6-13}$ and n$_{13-31}$, were computed from the
model SEDs following the same procedure as for the data. These
indices, computed from the models, are compared to those evaluated
from the IRS spectra of Cha I objects in
Figure~\ref{fig_index_model}. The observed values of n$_{6-13}$ and
n$_{13-31}$ indices of Class II objects in Cha I are well explained by
the disk models with settling parameter $\epsilon$ of 0.01 and 0.001
for a range of accretion rates and inclination angles, which indicates
that the surface layers of the disks in these objects are depleted of
small grains by a factor of 100 to 1000. The disks surrounding Class
II objects in Cha I seem to have undergone substantial dust
settling. Similar results have also been found for protoplanetary
disks in Taurus and Ophiuchus \citep{furlan06, watson09,
  mcclure10}. Even though the disks around Class II objects show
evidence for significant dust settling, almost all of them have values
of n$_{6-13}$ and n$_{13-31}$ indices $>\;-1.33$, appropriate for an
optically thick, flat disk, suggesting that even the most settled
disks observed are flared to some degree. Figure~\ref{fig_index_model}
shows that the disk models, clearly, cannot explain the mid-infrared
continuum indices observed for transitional disks. They have altered
disk structures and the mid-infrared continuum emission from them
originates in different regions of the disks.

\subsubsection{Comparison with disks in Taurus}

Substantial dust settling has also been observed for the
protoplanetary disks in the Taurus star forming region
\citep{furlan06,watson09}. In Figure \ref{fig_cont_indices_hist}, we
compare the continuum spectral indices of low-mass (spectral type G0
and later) Class II objects in the Cha I and the Taurus star forming
regions. The continuum indices of the Class II sources in Taurus were
computed from their IRS spectra presented in \citet{furlan06}. The
transitional disks in both regions are not included in the figure as
they have very different disk structures and the interpretation of
their continuum indices is not the same as that for full disks (see
Section~\ref{trans}).  The distribution of n$_{13-31}$ (top~panel) and
n$_{6-13}$ (middle~panel) indices are very similar for both
regions. Two-sided K-S test gives the following values for the maximum
deviation between the two cumulative distributions D and the
significance level of the K-S statistic P\%: D = 0.09 and P = 94\% for
n$_{13-31}$ and D = 0.1 and P = 83\% for n$_{6-13}$, which indicate
that the distribution of these two indices are statistically the same
for objects in Cha I and Taurus. The vertical dust distribution in the
disks in these two regions appear to be similar. Thus protoplanetary
disks in Taurus region, which has a median age of $\sim$ 1 Myr
\citep{kenhart95,hart01}, show similar degree of dust settling as that
of those in Cha I, which is slightly older, at a median age of 2 Myr
\citep{luhman08b}. Recent analysis of the IRS spectra of $\la$ 1 Myr
Ophiuchus star forming region has shown that the distribution of
mid-IR continuum indices of Class II objects in Ophiuchus is
statistically not different from that for objects in Taurus
\citep{mcclure10}. This indicates that most protoplanetary disks have
undergone significant dust settling by $\la$ 1 Myr. This is not
surprising as theoretical models predict timescales as short as $\ll$
1 Myr for grain growth and dust sedimentation in planet forming disks
\citep{dd05,dominik07}. However, protoplanetary disks in the older
Cha~I region show no evidence for higher degree of dust settling
compared to the disks in younger Taurus and Ophiuchus regions,
indicating that the dust settling is not just a monotonic function of
age. Even within a region, protoplanetary disks show a large range in
the observed degree of dust settling. This suggests that other
processes such as fragmentation of large grains in the midplane and
turbulent mixing must be at work \citep[e.g.][]{oliv10} along with
grain growth and sedimentation so as to produce the observed
statistical distribution of dust settling for $\la 1 - 2$~Myr old
protoplanetary disks .

The distribution of n$_{2-6}$ index in
{Figure~\ref{fig_cont_indices_hist} (bottom panel), however,  is quite different
  for the disks in Cha~I and Taurus. A K-S test gives D = 0.24 and P =
  5\% indicating that the distribution of n$_{2-6}$ of Class II (full)
  disks in Cha~I is statistically  different from that of Taurus. A significant
  fraction of Taurus objects have redder (more positive) n$_{2-6}$
  values than Cha I objects. To illustrate this further, in Figure
  \ref{fig_tau_cha_cont_indices} we plot the n$_{13-31}$ index against
  n$_{2-6}$ index for Class II objects in Taurus and Cha~I \citep[also
    see][]{mcclure10}. On average, the observed values of n$_{2-6}$
  index of Cha I objects are bluer (more negative) than that for
  objects in Taurus. The median n$_{2-6}$ for Cha I (n$_{2-6_{med}}$ =
  -1.61) is lower than the Taurus median (n$_{2-6_{med}}$ = -1.41).
  This is partly because of the difference in the underlying spectral
  type distribution of the Taurus and the Cha I sample: there are many
  more objects in the M3-M8 spectral type range in the Cha~I sample
  than in the Taurus sample resulting in the spectral type
  distribution peaking around M5 for Cha I and M0 for Taurus.  For
  Cha~I the median $n_{2-6}$ index for M3-M8 objects (n$_{2-6_{med}}$
  = -1.69) is significantly bluer than that for the K5-M2 objects
  (n$_{2-6_{med}}$ = -1.48) (also see Section \ref{median}) and this
  shifts the overall median to the bluer side.

Figure \ref{fig_tau_cha_cont_indices} also shows that Cha I lacks
objects with extreme red $n_{2-6}$: as many as 7 Class II objects in
Taurus have $n_{2-6}$ $\ga$ -0.5; no objects in Cha I show $n_{2-6}$
$\ga$ -0.5. Most of the excess emission in the 2 - 6 $\micron$ range
in Class II objects comes from the innermost parts of the disks - from
the inner rim at the dust sublimation radius heated directly by the
stellar radiation. When the accretion rate is high, the added
contribution from the accretion shock on the stellar surface increases
the net emission from the inner rim \citep{muz03} which tends to make
$n_{2-6}$ index flatter (more positive) and when the accretion
luminosity is comparable to or higher than the intrinsic stellar
luminosity, the $n_{2-6}$ $\thickapprox$ 0 \citep{dalessio09}. As can
be seen from Figure~\ref{fig_tau_cha_cont_indices} {\it(bottom
  panel)}, the upper most octile of the n$_{2-6}$ distribution (12.5\%
of the points lie above this value) for the Taurus sample is
significantly redder than that for Cha I. Some of the objects in
Taurus which have n$_{2-6}$ $\ga$ n$_{2-6_{upper\;oct}}$, viz.,
DP~Tau, DR~Tau and DG~Tau are the highest accretors in that
region. High accretors with large n$_{2-6}$ values have also been
found among the Class II objects in the Ophiuchus region
\citep{mcclure10}. However, there is clearly a paucity of such high
accretors in Cha I as indicated by the absence of objects with large
n$_{2-6}$ indices.  This is unlikely to be a selection effect. The
high accretors in Taurus with $n_{2-6}$ $\ga$ -0.5 have spectral types
earlier than M1. As discussed in Section 3.2, the Class II disks in
Cha~I sample is complete for spectral types M1 or earlier.  Moreover,
T Tauri stars with high accretion rates tend to have higher mid-IR
excess emission \citep{dalessio06}. Since our sample is mid-IR flux
limited, it is unlikely that we have preferentially missed out high
accretors in Cha I with spectral types M1 or earlier. The lack of high
accretors in Cha~I in this spectral range is possibly a result of
temporal evolution such that Cha I which is older, on average, than
both Ophiuchus and Taurus appears to have fewer high accretors. This
is consistent with the general finding that accretion rate on average
decreases with system age \citep{hart98b}.

\subsubsection{Transitional and pre-transitional disks} \label{trans}

The structural diversity among the transitional disks has been
demonstrated from the detailed modeling of these objects, particularly
in the Taurus region \citep{dalessio05,calvet05,esp07b,esp08,esp10}. Objects
like CoKu~Tau4, DM~Tau and GM~Aur show no or at most small continuum
excess emission at wavelengths $\la$ 8 $\micron$ indicating that their
inner optically thick disks have been almost entirely dissipated
\citep{dalessio05, nagel10, calvet05}. The term `transitional disks' (hereafter
TDs) has been specifically used to designate such disks with inner
holes. Some of these disks may contain small amounts of optically thin
dust within their inner holes as in the case of GM~Aur. The $n_{2-6}$
index of TDs typically are smaller than the  lowest octile
($n_{2-6_{lower\;octile}}$) as can be seen from
Figure~\ref{fig_tau_cha_cont_indices} (top panel).

In contrast, objects like LkCa~15 and UX~TauA, which show a flux
deficit at wavelengths $\la$ 8 $\micron$ relative to the the Class II
median, have near-IR ($\la$ 4 $\micron$) excess emission consistent
with that from an optically thick disk.  These objects have an inner
optically thick disk which is separated from the outer optically thick
disks by an optically thin gap. Such disks with radial gaps in them
have been called `pre-transitional' disks (PTDs) \citep{esp07b,esp08,esp10},
with the name implying an evolutionary stage earlier to that of
TDs. PTDs show higher excess at shorter wavelengths than the TDs and
their $n_{2-6}$ values fall between the lowest octile and the median
($n_{2-6_{lower\;octile}}\;\ga\;\;n_{2-6_{PTD}}\;\ga\;n_{2-6_{median}}$)
(see Figure~\ref{fig_tau_cha_cont_indices}). Both TDs and PTDs in
Taurus show $n_{13-31}$ $\ga$ the upper octile
($n_{13-31_{upper\;octile}}$) \citep{mcclure10}.

Based on the locations of well modeled TDs and PTDs in Taurus in the
$n_{13-31}$ -- $n_{2-6}$ plot (Figure~\ref{fig_tau_cha_cont_indices};
bottom panel), we classify CS~Cha, T25, T54, CHXR~22E and
2M~J11241186-7630425 in Cha I as TDs with inner holes \citep[also
  see][]{kim09,furlan09}. All these objects have $n_{2-6}$ $\la$
$n_{2-6_{lower\;octile}}$ and $n_{13-31}$ $\ga$ $n_{13-31_{upper
    \;octile}}$.  The disk around CS~Cha has already been modeled and
is found to have an inner hole of $\sim$~43~AU in radius
\citep{esp07a}. The locations of SZ~Cha, T35 and T56 in
Figure~\ref{fig_tau_cha_cont_indices} (bottom panel), would qualify
them as PTDs\citep[also see][]{kim09,furlan09}. Their $n_{2-6}$ values
fall between the median and the lower octile and their $n_{13-31}$
values are above the upper octile. Modeling of the individual SEDs and
high resolution submm/mm imaging of the TDs and PTDs are required to
constrain the detailed structure of their disks.

\subsection{Analysis of the silicate emission features} \label{silicate}

\subsubsection{Silicate emission and disk structure}

In this section we explore the connection between the disk structure
as revealed by the mid-IR continuum spectral indices and the strength
of the silicate emission features in the observed spectra of Class~II
disks in Cha~I \citep[see also][]{watson09,furlan09,mcclure10}. To
quantify the strength of the silicate emission we computed the
equivalent width of the dust emission features centered at 10
$\micron$ and 20 $\micron$. The equivalent width is defined as

\begin{equation}
\label{eqn2}
W\;=\; \int_{\lambda_1 }^{\lambda_2} \frac{F_{\lambda} - F_{\lambda,cont}}{F_{\lambda,cont}} \;d\lambda
\end{equation}

\noindent
and we integrate between $\lambda_1$ = 8 $\micron$ and $\lambda_2$ =
13 $\micron$ to obtain the equivalent width of the 10 $\micron$
feature,W$_{10}$ and between $\lambda_1$ = 16 $\micron$ and
$\lambda_2$ = 28 $\micron$ to obtain the equivalent width of the 20
$\micron$ feature,W$_{20}$. We also computed the integrated flux of
the 10 and 20 $\micron$ features as

\begin{equation}
\label{eqn3}
 F\;=\; \int_{\lambda_1 }^{\lambda_2} (F_{\lambda} - F_{\lambda,cont}) \;  d\lambda 
\end{equation}

\noindent
and integrating between the same limits as above to obtain F$_{10}$
and F$_{20}$. In order to compute the equivalent widths and the
integrated flux of the silicate emission features, we first fit a
continuum to the observed spectra as a polynomial of degree 3, 4 or 5
depending on the shape of the continuum. The points used for the
continuum fit were chosen so that the dust emission features are small
within those wavelength ranges: 5.61 - 7.94 $\micron$, 13.02 - 13.50
$\micron$, 14.32 - 14.80 $\micron$ , 30.16 - 32.19 $\micron$ and 35.07
- 35.92 $\micron$ \citep{watson09}. Objects for which the continuum
could not be fit by a single polynomial, we used a two polynomial fit,
one between wavelength interval 5 - 14 $\micron$ and the other between
14 - 36 $\micron$, making sure that the overall continuum is smoothly
varying. The values of integrated flux and equivalent widths of the
silicate emission features thus obtained are tabulated in
Table~\ref{tab5_index}.  The dominant source of errors in the
quantities estimated from the observed silicate features is the
uncertainty in the determination of the underlying continuum. In
particular, the equivalent width of the 20 $\micron$ feature, which is
broader and generally flatter than the 10 $\micron$ feature, is much
more sensitive to the choice of the continuum. The errors listed in
Table~\ref{tab5_index} assumes an uncertainty of 20\% in the estimated
continuum underlying the 20 $\micron$ feature and a 10\% uncertainty
in the continuum for other quantities.

Figure~\ref{gr_ew10_index} shows a plot between equivalent width of
the 10 $\micron$ silicate feature, W$_{10}$ against the continuum
index $n_{13-31}$ for the Class II objects in Cha I. Also shown in the
Figure, marked as the black polygon, is the region defined by the
models in the W$_{10}$ - n$_{13-31}$ plane, for irradiated, accretion
disk models for a range of values of accretion rates, stellar masses,
inclination angles and settling parameter $\epsilon$
\citep{esp09,furlan09}. It can be seen from the figure that most of
the Class II objects fall within the region predicted by the
models. Not surprisingly, objects identified as having altered radial
disk structures - the outwardly truncated disks and TDs and PTDs -
fall outside the polygon.  Only 4 out of the 8 TDs and PTDs, SZ~Cha,
CS~Cha, T56 and 2M~J11241186-7630425, show the 10 $\micron$ silicate
features in emission and only they are shown here (see
Figure~\ref{trans_spec}). The silicate emission in these disks are
generated either by the optically thin dust grains within the inner
holes/gaps or at the atmosphere of the inner `wall' of the outer
optically thick disk \citep{calvet05,esp08,kim09}. TDs and PTDs in
general show relatively high values of W$_{10}$ and n$_{13-31}$, i.e.,
they are outliers in W$_{10}$ and/or n$_{13-31}$ as can be seen from
Figure~\ref{gr_ew10_index} \citep[see also][]{furlan09}. Six objects,
TW~Cha, ISO~91, CR~Cha, T33A, B43 and CV~Cha, (gray solid circles in
Figure~\ref{gr_ew10_index}) show enhanced 10 $\micron$ silicate
emission.  They have W$_{10}$ values greater than the upper octile of
the W$_{10}$ distribution (dashed line in Figure~\ref{gr_ew10_index})
of all the Class~II objects. Their W$_{10}$ values are significantly
higher than those predicted by the models for typical disks; however,
they have $n_{13-31}$ values within the range allowed by the models.

 Figure~\ref{flx10_ew10} compares the W$_{10}$ values of Class~II
 objects in Cha~I with the integrated flux of the 10 $\micron$ feature
 F$_{10}$. The equivalent width of the silicate feature is a measure
 of the amount of the optically thin dust per unit projected area of
 the optically thick disk whereas the integrated flux of the silicate
 feature is a measure of the amount of optically thin dust. Both
 W$_{10}$ and F$_{10}$ are measures of the strength of the 10
 $\micron$ feature, but W$_{10}$ expresses it in the units of the
 underlying continuum. For radially continuous `full' disks both the
 equivalent width W$_{10}$ and the integrated flux F$_{10}$ track each
 other well as shown in Figure~\ref{flx10_ew10}.  TDs and PTDs show
 moderate to low values of F$_{10}$ indicating that the amount of
 optically thin dust contributing to the 10 $\micron$ emission is
 relatively small; and yet they show large W$_{10}$ values. TDs and
 PTDs have holes/gaps in their disks which lowers the continuum
 emission thereby increasing their W$_{10}$. The objects identified as
 having enhanced 10 $\micron$ emission (solid gray circles in
 Figure~\ref{flx10_ew10}) have F$_{10}$ values similar to other
 Class~II objects in the sample, but show significantly high W$_{10}$
 values. The large W$_{10}$ values of these objects are likely caused
 by the reduced continuum emission in these objects. Most of the
 continuum emission underlying the 10 $\micron$ feature comes from
 $\la$ 1-2~AU of the disks \citep{dalessio06}. One possible
 explanation is that these disks are opening up gaps in the inner
 regions ($\la$ 1-2 AU) of their optically thick disks, which will
 lower their continuum emission. If the gaps are sufficiently small (a
 few AU), one will not be able to distinguish these objects from the
 `full' disks purely based on the mid-IR continuum; however, they will
 stand out as outliers in W$_{10}$ \citep{esp09,furlan09}. Such
 objects which are possibly opening gaps within them have also been
 identified in the Taurus and Ophiuchus star forming regions as
 outliers in W$_{10}$ \citep{furlan09}. However, detailed modeling of
 the SEDs and high resolution imaging are required before the disk
 structure of these objects can be confirmed.

In Figure~\ref{gr_ew20_index}, we plot the equivalent width of the 20
$\micron$ feature, W$_{20}$ against the continuum index
$n_{13-31}$. Most TDs and PTDs show large W$_{20}$ values. The
holes/gaps in them are sufficiently large to lower the continuum
emission underlying both the 10 and 20 $\micron$ silicate emission
features. The objects with W$_{10}$ $\ga$ W$_{10,upper\:octile}$
(gray solid circles), do not have such high W$_{20}$; their W$_{20}$
values, though higher than the median, are similar to those of many
other Class~II disks in the sample.  This is consistent with the
supposition that gaps in the these disks are small and are at smaller
disk radii so that the continuum emission underneath the 20 $\micron$
feature in these disks is not significantly lowered so as to enhance
W$_{20}$.

Leaving aside the TDs/PTDs and the objects with enhanced W$_{10}$,
Figure~\ref{gr_ew10_index} shows that, in general, the equivalent
width of the 10 $\micron$ feature, W$_{10}$, increases with increasing
value of n$_{13-31}$ index for `full' disks. A similar positive
correlation is also found between W$_{20}$ and the n$_{13-31}$ index
as shown in Figure~\ref{gr_ew20_index}. The linear (Pearson)
correlation coefficient for these relations is r = 0.3 and the
probability that a randomly drawn sample of the same size could show
such a correlation, is P $\la$ 5\%, indicating that the correlations
are marginally significant.  Such correlations between the strengths
of the silicate features and the disk geometry are expected for
radially continuous `full disks' from the disk models: when
n$_{13-31}$ increases (less dust settling in the disk) degree of
flaring of the disk increases and the location of the silicate
emission moves farther out in the disk resulting in a larger emitting
area; thus the amount optically thin emission from the warm grains in
the disk surface layers will increase. However, when the degree of
flaring increases, the height of the irradiation surface also increases,
which increases the amount of heating of the disk interior and as a
result the continuum level goes up. The net effect appears to be an
overall increase in the equivalent width of the silicate feature and
the model predictions agree with such an increase
\citep[see][]{dalessio06, furlan09}.  We note here that the spread in
stellar masses, accretion rates and inclination angles of the objects
in our sample would act to weaken the trend between equivalent widths
of silicate features and n$_{13-31}$.  Despite this, marginally
significant correlation could be found between the strength of the
silicate emission and the underlying disk structure; this makes the
case for it even stronger.

Next we look at how the changes in the disk structure affects the strength
of the 20 $\micron$ silicate feature relative to that of the 10 $\micron$
feature. In Figure \ref{gr_f20f10_index} we plot the ratio of the
integrated flux of the 20 $\micron$ silicate emission to that of 10
$\micron$ emission as a function of the n$_{13-31}$ index. We find a
tight correlation between the F$_{20}$/F$_{10}$ ratio and the
n$_{13-31}$ index for the `full disks'. The Pearson correlation
coefficient is r = 0.6 and the corresponding probability is P $\ll$
0.05\% indicating that the correlation is highly significant. The
correlation between the integrated flux ratio F$_{20}$/F$_{10}$ and
the continuum index n$_{13-31}$ is much tighter than the correlations
that we find between W$_{10}$ and W$_{20}$ with n$_{13-31}$. This
indicates that even though both the 10 and the 20 $\micron$ feature
strength decreases with increasing degree of dust sedimentation in the
disk, the strength of the 20 $\micron$ silicate emission feature drops
faster relative to the 10 $\micron$ feature. This is because the 20
$\micron$ emission arises from larger disk radii compared to 10
$\micron$ emitting region. As the disks become less flared because of
sedimentation, the optically thin emitting area decreases; however,
the reduction in the emitting area at larger disk radii is more than
the reduction in the emitting area closer to the star and therefore
the the 20 $\micron$ emission is affected more than the 10 $\micron$
emission \citep[see][]{dalessio06,esp09}.

TDs and PTDs separate out from the `full' disks in F$_{20}$/F$_{10}$ -
n$_{13-31}$ plane. This is not surprising as these disks have different
radial structures and both the mid-IR continuum and the silicate
emission arise in different parts of the disks as compared to `full'
disks. The objects identified as having high W$_{10}$ (gray solid
circles) do not follow the trend seen between F$_{20}$/F$_{10}$ and
n$_{13-31}$ for `'full' disks. They show smaller F$_{20}$/F$_{10}$
ratio than that for objects with similar n$_{13-31}$ values. This is
suggestive of their disk structure being different from that of
radially continuous `full' disks.

\subsubsection{Silicate emission and dust processing in disks}

We further analyzed the silicate features in the mid-IR spectra of
Class~II objects in Cha~I to study the amount of dust processing in
protoplanetary disks. Figure~\ref{gr_shape_ew10} shows a plot of the
flux ratio F$_{11.3}$/F$_{9.8}$ versus the equivalent width of the 10
$\micron$ silicate feature, W$_{10}$. The flux ratio
F$_{11.3}$/F$_{9.8}$ has long been considered as a measure of the
degree of dust processing (grain growth and crystallization) in the
inner 1-2~AU of the disk \citep{boekel05,silacci06,honda06,olof09}. The
larger values of the F$_{11.3}$/F$_{9.8}$ ratio indicates the presence
of highly processed grains and the smaller values imply relatively
unprocessed grains. The flux densities F$_{9.8}$ and F$_{11.3}$ are
computed from the IRS spectra by integrating the flux densities within
an interval of 0.4~$\micron$ centered at 11.3 and 9.8~$\micron$
respectively. The F$_{11.3}$/F$_{9.8}$ values are listed in
Table~\ref{tab5_index}.  One of the TD/PTD objects, SZ~Cha is not
shown in the Figure~\ref{gr_shape_ew10} because it has strong PAH
emission at 11.3 $\micron$ which makes reliable estimate of the
F$_{11.3}$/F$_{9.8}$ difficult. We also show TDs and PTDs in the
Taurus region (open star symbols) in Figure~\ref{gr_shape_ew10} for
comparison.  The `full' disks (open circles) show a strong negative
correlation between the dust processing indicator F$_{11.3}$/F$_{9.8}$
and W$_{10}$. The Pearson coefficient for the correlation is r = -0.5
with a corresponding probability P $\la$ 0.05\%, indicating a highly
significant correlation. Objects with low W$_{10}$ values show
evidence for highly processed dust grains and objects with high
W$_{10}$ values show relatively unprocessed grains.

In the last Section we showed that in `full' disks W$_{10}$ is a
measure of the amount of the optically thin dust responsible for the
10 $\micron$ silicate emission, and less settled disks (higher degree
of flaring), in general, have high W$_{10}$. The correlation that we
find between F$_{11.3}$/F$_{9.8}$ and W$_{10}$ also suggests that more
settled disks show higher degree of dust processing (grain growth and
crystallization).

The TDs and PTDs in both Cha I and Taurus have relatively low values
of F$_{11.3}$/F$_{9.8}$, indicating that the dust grains contributing
to the 10 $\micron$ emission in them are relatively unprocessed.
Interestingly, the objects identified as having enhanced 10 $\micron$
emission (solid gray symbols) also show low F$_{11.3}$/F$_{9.8}$
values, indicating that the optically thin dust in them are relatively
unprocessed. The high W$_{10}$ values and the pristine dust
composition that these objects show is quite similar to that exhibited
by PTDs in Taurus and Cha~I \citep{watson09,kim09,sargent09} and very
different from that shown by `full' disks.  This suggests that as in
PTDs, the 10 $\micron$ silicate emission in these disks is likely
generated by optically thin dust within the gaps in the disks. A more
detailed discussion on the pristine dust composition of these disks
will be presented in a forthcoming paper (Manoj et al. 2010, in
preparation).

As pointed out earlier, both grain growth and crystallization in the
inner disk ($\la$ 1-2~AU) affect the F$_{11.3}$/F$_{9.8}$ ratio. In
order to disentangle the effect of crystallization from that of the
grain growth, we make use of the isolated forsterite feature at
33.6~$\micron$, which is an independent measure of the degree of
crystallinity and, in addition, probes the crystalline content of the
cooler dust farther out ($\la$ 10~AU) in the disk \citep{olof09,
  sargent09}.  We compute the equivalent width of the 33.6~$\micron$
feature, W$_{33}$, from equation~(\ref{eqn2}) by integrating the flux
between $\lambda_1$ = 32~$\micron$ and $\lambda_2$ = 35~$\micron$
\citep{watson09,mcclure10}; the W$_{33}$ values are listed in
Table~\ref{tab5_index}.  Figure~\ref{gr_shape_ew33} shows W$_{33}$
plotted against F$_{11.3}$/F$_{9.8}$ for Class~II objects in
Cha~I. W$_{33}$ and F$_{11.3}$/F$_{9.8}$ appear to be tightly
correlated; a linear (Pearson) coefficient of r = 0.4 and an
associated probability of $\ll$ 1\%, indicates that the correlation is
highly significant. The higher the degree of dust processing (grain
growth and crystallization) in the inner disk, the higher is the
crystalline content of the outer disk.  F$_{11.3}$/F$_{9.8}$ and
W$_{33}$ have also been found to be correlated for Class~II disks in
other star forming regions.  In Figure~\ref{tau_shape_ew33} we show
W$_{33}$ plotted against F$_{11.3}$/F$_{9.8}$ for low mass Class~II
objects in Taurus. These quantities were computed from the IRS spectra
of Class II objects in Taurus presented in \citet{furlan06} and
\citet{watson09}.  Again we find a tight correlation between
F$_{11.3}$/F$_{9.8}$ and W$_{33}$ with a (Pearson) correlation
coefficient r = 0.5 and an associated probability of $\ll$
1\%. Similar result has also been found for Class~II disks in
Ophiuchus \citep{mcclure10}.

Although both grain growth and crystallization in the inner disk can
affect the F$_{11.3}$/F$_{9.8}$ ratio, the correlation that we find is
more likely due to the crystalline content in the outer and inner
regions of the disks tracking each other
\citep[see][]{watson09}. Since the crystalline grains are generally
thought to be formed in the inner ($\la$ 1~AU) disks, where the
temperature required for crystallization can be reached ($\ga$ 1000~K)
\citep{grossman72,gail01,wooden05}, higher crystalline content of the
outer disk should indicate that the crystalline grains are also
present in the inner disk. The above results suggest that an increase
in the inner disk crystallinity is accompanied by a similar increase
in the outer disk crystallinity in protoplanetary disks surrounding
low mass stars \citep{watson09,mcclure10}. If the crystalline grains
are produced only close to the central star, then the correlation that
we find between the inner and the outer disk crystallinity should
suggest efficient radial mixing of the dust in the disk at least up to
a radius of $\la$ 10~AU. If, on the other hand, the crystalline grains
are produced {\it insitu} in the outer parts ($\la$ 10~AU) of the
disks \citep{hark02, wooden05}, then both the inner and the outer disk
crystallization mechanisms must operate on similar timescales.

The changes in the F$_{11.3}$/F$_{9.8}$ ratio have at times been
interpreted as almost entirely due to grain growth in the inner disk
\citep[e.g.][]{przy03,honda06, silacci06}. However, the correlation
that we find between F$_{11.3}$/F$_{9.8}$ and W$_{33}$ would suggest
that F$_{11.3}$/F$_{9.8}$ is as good or a better indicator for the
crystalline content of the inner disk as it is for grain growth
\citep{watson09,sargent09}. Since changes in the grain size affect the
F$_{11.3}$/F$_{9.8}$ ratio, our results also imply that the grain
growth and crystallization in the inner disk must occur concurrently,
on similar timescales \citep[e.g.][]{apai05}. Detailed dust
composition analysis of the IRS spectra of T Tauri stars in various
star forming regions has shown that both large ($\ga$ 1~$\micron$) and
crystalline grains are present in most protoplanetary disks
\citep[e.g.][]{sargent09}.

\subsection{Median SED of Class II objects} \label{median}

We constructed median SEDs for the Class II objects in Cha~I from 0.9
$\micron$ to 36 $\micron$ using $I,\;J,\;H\;$ and $K_s$ photometry and
IRS spectra in order to study the average disk properties. Two
separate medians were computed for objects belonging to the spectral
type bins of K5-M2 and M3-M8 to compare the emission from
protoplanetary disks with different central star properties. The
median was computed for 25 objects with spectral types between K5 to
M2 and for 29 objects between M3 to M8. Objects which have only upper
limits in $I,\;J,\;H\;$ and $K_s$ measurements were excluded from the
median.  The medians for each of these bins were obtained by first
normalizing the dereddened flux densities at $I,\;H\;$ and $K_s$ and
IRS spectrum of each object to the dereddened J-band flux of that
object, and then computing the median at $I,\;J,\;H\;$ and $K_s$ and
the IRS wavelengths. We also obtained the quartiles for the medians:
the upper and lower quartiles define the range above and below the
median within which 50\% of the flux values lie. The median SEDs for
the Class II objects in the spectral bins of K5-M2 and M3-M8, along
with the quartiles, are presented in Figure~\ref{fig_medsed}. The gray
shaded regions in the figure are bound by the lower and upper
quartiles of the medians. The photospheres plotted in the figure are
that for a K7 star for the K5-M2 median and and that for a M5 star for
the M3-M8 median, both computed from the photospheric colors listed in
\citet{kenhart95} and normalized to the J-band flux.

The median SEDs computed for the T Tauri stars (K5-M2) and very
low-mass stars and brown dwarfs (M3-M8) clearly show mid-IR excess
emission and 10 and 20 $\micron$ silicate emission features. The
mid-IR continuum spectral indices, n$_{6-13}$ and n$_{13-31}$, of both
the medians are very similar: n$_{6-13}$ = $-0.9$ for both K5-M2 and
M3-M8 medians and n$_{13-31}$ = $-0.26$ for K5-M2 and $-0.28$ for M3-M8
median. However, the nature of excess emission below 8 $\micron$ is
very different for these two medians; the excess emission in K5-M2
median begins at $\sim$ 2 $\micron$ whereas for the M3-M8 median 
significant excess starts to show up only at 4-5 $\micron$. This is
because the ratio of the disk emission to the stellar photospheric
emission scales as T$^3_{\star}$, and for cooler stars (M3 or later)
it is difficult to see the disk excess above photospheric emission at
$\lambda$ $\la$ 6 $\micron$ \citep{ercolano09}. In addition, we also
find the continuum slope at wavelengths $\la$ 8 $\micron$ for M3-M8
median is much steeper (bluer) than that for the K5-M2 median: K5-M2
median has the continuum index n$_{2-6}$ = $-1.42$ while M3-M8 median
has n$_{2-6}$ = $-1.74$.

It is important to recognize that very low-mass stars and brown dwarfs
on average have fainter and steeper SEDs at wavelengths $\la$ 8
$\micron$, especially when identifying TD and PTD candidates. They are
identified by virtue of the flux deficit and steeper slope that they
show at wavelengths $\la$ 8 $\micron$ compared to the median Class II
SED. A very low-mass object or a brown dwarf with normal disk could
easily be misidentified as TD or PTD candidate if compared to a T
Tauri (K5-M2) median \citep[see][]{ercolano09}.

The effect of dust sedimentation in protoplanetary disks is best seen
in the SED at wavelengths $\ga$ 13 $\micron$ \citep{dalessio06,
  furlan06,watson09}. To demonstrate this, in Figure~\ref{fig_medsed}
where the median SED of K5 - M2 Class II objects in Cha I is presented
({\it top panel}), we also show synthetic SEDs generated from the
accretion disk models for a 0.5 M$_{\odot}$ star with accretion rates of
10$^{-7}$, 10$^{-8}$ and 10$^{-9}$ M$_{\odot}$yr$^{-1}$ and the settling
parameter $\epsilon$ of 0.001, 0.01 and 0.1.  It can be seen that the
median SED longward of 13 $\micron$ is best described by disk models
with settling parameter $\epsilon$ = 0.01 and 0.001 for accretion
rates of 10$^{-7}$ and 10$^{-8}$ M$_{\odot}$yr$^{-1}$; even for an
accretion rate of 10$^{-9}$ M$_{\odot}$yr$^{-1}$, most disks have
$\epsilon$ $\la$ 0.1.  The median and the quartiles suggest
significant dust settling in the Class II disks in Cha I: small grains
in the upper layers of more than 75\% of the disks have been depleted
by a factor of at least 10; about 50\% of the disks have depletion
factor $\ga$ 100.

In Figure~\ref{fig_medsed} {\it (bottom panel)}, we show the
synthetic SEDs generated from the accretion disk models for a 0.2
M$_{\odot}$ star with accretion rates of 10$^{-8}$, 10$^{-9}$ and
10$^{-10}$ M$_{\odot}$yr$^{-1}$ and the settling parameter $\epsilon$
of 0.001, 0.01 and 0.1. The observed median and the quartiles at
wavelengths $\ga$ 13 $\micron$, where the dust sedimentation effects
are seen, are best explained by disk models with settling parameter
values of $\epsilon$ = 0.01 and 0.001 for the above accretion rates
indicating significant dust depletion in the upper layers of these
disks. Both the K5-M2 stars and the low-mass objects with spectral
types M3-M8 seem to have dust depletion factors of 100-1000 in their
upper layers, similar to what was found for Taurus \citep{furlan06}.
 
However, the match between the model predictions and the median SEDs
is not satisfactory at shorter wavelengths. The accretion disk models
for the 0.5 M$_{\odot}$ significantly under predicts the observed flux of
the K5-M2 median at wavelengths $\la$ 8 $\micron$. Also, the upper
quartile of the M3-M8 median at wavelengths $\la$ 13 $\micron$ cannot
be explained by the 0.2 M$_{\odot}$ models even for an accretion rate
10$^{-8}$ M$_{\odot}$yr$^{-1}$. These discrepancies between the models and
the observed medians and the possible reasons for them are discussed in
detail by \citet{esp09}. Here we only discuss behavior of the median
at wavelengths $\ga$ 13 $\micron$ which is relevant for the dust
sedimentation effects.

\section{Multiplicity and disk structure} \label{binary}

In this Section we discuss the effect of multiplicity on the structure
and dust properties of the Class~II disks in Cha~I.  Close companions
can gravitationally perturb and truncate the disks around individual
components \citep{lubarty00}. For orbital separations of $\sim$
200~AU, companions truncate the disk from outside (for typical disk
sizes of $\sim$ 100 AU); for even smaller orbital separations ( $\la$
a few tens of AU) the disk could be truncated from inside as in the
case of CoKu Tau/4 \citep{ik08}. An outwardly truncated disk is likely
to show a steeper mid-IR continuum slope, i.e.  n$_{13-31}$ less than
that for a flat disk \citep{mcclure08}. Similarly, disks truncated
from inside will have significantly steeper n$_{6-13}$ continuum
index. There have also been suggestions that close companions could
affect the dust properties in protoplanetary disks
\citep{meeus03,sterzik04}.

We compared the observed disk properties of the objects in multiple
systems to those in single systems in order to study the effect of
companions on the disk structure. Figure~\ref{bin_ind} compares the
distribution of continuum indices of close ($<$ 1$\arcsec$ $\sim$
165~AU) binaries and wide ($\ge$ 1$\arcsec$) binaries among Class~II
objects with those with no known companions. A two-sided K-S test
shows that the distributions of the $n_{2-6}$ and $n_{13-31}$ indices
for the sub-arcsecond binaries are similar to those for the wide
binaries and single stars. However, the $n_{6-13}$ distribution for
the sub-arcsecond binaries appears to be different from that for the
wide binaries and single stars at a significance level of 97\%. The
$n_{6-13}$ distribution for the sub-arcsecond binaries peaks at bluer
value than that for other objects.  We also compared the silicate
feature properties, W$_{10}$ and W$_{20}$, and dust processing
indicators, F$_{11.3}$/F$_{9.8}$ and W$_{33}$, of the sub-arcsecond
binary systems with those of other Class~II objects. The silicate
feature properties and the dust processing indicators of objects with
close ($<$ 1$\arcsec$) companions are statistically similar to those
of wide binaries and single stars. These results suggest that the disk
properties are only weakly affected by the presence of companions.
The multiple systems in our sample have companions with projected
separations $\ga$ 20~AU (see Table~\ref{binary_tbl}), and the above
results hold for such systems. However, only 12 out of the 68 Class~II
objects in our sample are known to have sub-arcsecond companions. The
small number makes the statistical comparison less certain. In
addition, the multiplicity census of Cha~I sample is likely to be
incomplete, as it is harder to rule out extremely close ($\ll$ 1
$\arcsec$) companions. Therefore, we are not able to draw any definite
conclusions on the effect of multiplicity on the disk structure from
the comparison based on our sample.  Below we discuss the disk
properties of Class~II objects which show clear evidence for disk
truncation and examine if close companions are responsible for it.

We first examine the multiplicity of TDs and PTDs which have
holes/gaps within their disks. The SED of a close (separation $\sim$ a
few AU) binary system with a circumbinary disk can appear TD-like as
has been shown in the case of CoKu Tau/4 \citep{ik08}. However, only
two of the TDs/PTDs in Cha~I, viz., CS~Cha and T54 are known to have
close stellar companions (see Table~\ref{binary_tbl}). T54 has a
0.26$\arcsec$ companion ($\sim$ 43~AU at 165~pc) and the disk around
it could be truncated from inside by the gravitational influence of
the close companion; the mid-IR excess that we observe could be from
the circumbinary disks surrounding these systems. On the other hand,
CS~Cha, which has an estimated inner hole of 43~AU in radius
\citep{esp07a,kim09}, is a spectroscopic binary \citep{guenther07}. It
is not clear, though, if such a close companion can carve out the hole
of this size in the disk \citep[see][]{kim09, furlan09}. CHXR~22E,
T25, T35, and T56 do not have known companions around them: companions
up to 5,12,12 and 15 times as faint as the primary in the K$_s$ band
have been ruled out for separations $\ga$ 0.1$\arcsec$ ($\sim$17~AU)
for CHXR~22E, T25, T35, and T56 respectively \citep{laf08}. Further,
the companions reported in the literature for SZ~Cha are not confirmed
members of the Cha I association \citep{luhman07}.  Comparative
studies of various mechanisms that could possibly create holes or gaps
in disks have shown that most of the observed characteristics of a
large sample of TDs/PTDs from various star forming regions are best
explained by Jovian mass planets opening up gaps and holes in the
disks \citep{kim09, najita07, rice03, calvet05, esp07b}.

Next, we discuss objects which show clear evidence for disk
truncation, i.e, continuum indices bluer than that for an optically
thick, flat disk, to see if known companions are responsible for disk
truncation.  Among the Class~II disks in our sample, 9 show $n_{6-13}$
$<$ -1.33, steeper than that for an optically thick, flat disk. 5 of
them are TDs, only two of which , CS~Cha and T54, are sub-arcsecond
binaries as discussed above. Among the remaining 4 objects,
2M~J11062942-7724586, CHXR~30A, C7-1 and ISO 282, only CHXR~30A is
known to have a close ($\sim$ 0.46$\arcsec$) companion which could be
responsible for the disk truncation. Two class~II objects, T51 and
Hn~5, show $n_{13-31}$ $<$ -1.33, indicating that they have outwardly
truncated disks; no sub-arcsecond companions have been reported for
these two objects (see Section~\ref{index})
\citep[see][]{furlan09}. Thus, there is evidence that disk truncation
is caused by a close companion in a few objects; however, no
sub-arcsecond companions are known for most objects with truncated
disks. While undetected companions at small separations ($\ll$
1$\arcsec$) could still be responsible for disk truncation in these
objects, available multiplicity data for Class~II objects in Cha~I do
not clearly demonstrate that close (projected separation $\ga$ 20~AU)
companions affect disk structure. Similar results have also been found
for Class~II disks in Taurus, where no statistically significant
differences were found in the structure and dust properties of disks
in single and multiple systems \citep{pascucci08}.

\section{Accretion and disk structure} \label{cwtts}
In Section 3, we classified the Class~II objects in Cha~I into WTTS
and CTTS using the H$\alpha$ emission strength as an accretion
indicator (see Table~\ref{basic_tbl}). WTTS are young T Tauri stars
that have little if any accretion and probably have
dissipated their inner disk, unlike CTTS that  are actively accreting
from the disk.  Here we compare the mid-IR emission of the WTTS and
CTTS in our sample as revealed by their IRS spectra. There are 48 CTTS
and 15 WTTs in our Cha I sample: Two of the WTTS are Class III
objects and two are TDs.

All of the WTTS in our sample except the Class III objects show mid-IR
($\ga$ 6 $\micron$) excess emission as can be seen from their SEDs
(Figure~\ref{class2_sed_fig}). All of them except the 2 TDs, CHXR~22E
and T25, and CHSM~10862, also show both 10 and 20 $\micron$ silicate
feature in emission. The mid-IR excess and the silicate features seen
in the IRS spectra of the WTTs indicate that they still have dusty
disks around them. However, most WTTS show low or no near-IR ($\la$ 6
$\micron$) excess emission (Figure~\ref{class2_sed_fig}).  In
Figure~\ref{fig_cwtts}, we compare the distribution of spectral
indices $n_{2-6}$, $n_{6-13}$ and $n_{13-31}$ for WTTS and CTTS.  The
distribution of all three continuum indices for the WTTS is shifted to
more negative values compared to those for the CTTS, indicating that the
WTTS have steeper continuum slope of the SED in the infrared
wavelengths.  As can be seen from Figure~\ref{fig_cwtts}, this effect
is more conspicuous at shorter wavelengths, in the $n_{2-6}$ index
than in the other two indices. A two-sided K-S test shows that the
distribution of the $n_{2-6}$ index of CTTS and WTTS are statistically
different at 99.99\% significance level.

Most TDs and PTDs in Cha I, except CHXR~22E and T25, could be
classified as CTTS based on the strength of their H$\alpha$ emission
(see Table~\ref{basic_tbl}). T54, for which H$\alpha$ measurement is
not available in the literature, has been reported to be accreting at
a relatively high rate based on its U band flux \citep{kim09}. Thus
most TDs/PTDs is Cha~I are active accretors. All the objects
identified as having enhanced 10 $\micron$ emission (W$_{10}$ $\ga$
W$_{10,upper\:octile}$; see Figure~\ref{gr_ew10_index}), except T33A, are
CTTS based on their H$\alpha$ equivalent widths indicating that they
are actively accreting from their disks. T33A is a close binary and
the IRS spectra has contribution from both the components. The
interpretation of W$_{10}$ and the continuum index in this system is
inconclusive.  Thus most objects in Cha~I which show evidence for
holes or gaps in their disks appear to be actively accreting.

Apart from the TDs and PTDs discussed earlier, there are a few other
objects in our sample which show significant flux deficits at
wavelengths $\la$ 8 $\micron$, but show silicate emission features and
continuum excesses longward of 8 $\micron$ with n$_{13-31}$ index
comparable to that of Class II objects. They are CHXR~20, CHXR~30A,
C7-1, and CHXR~47. The slope of their continuum between 2 and 13
$\micron$ is steeper (more negative) than that for most ($\sim$ 80\%)
Class II disks and they show little excess emission below 8
$\micron$. The dust grains in their inner disks appear to be greatly
depleted creating an opacity deficit.  They differ from the TDs and
PTDs in that their excess flux, even at wavelengths longer than 13
$\micron$, is considerably lower than that of the Class II
median. Also, they do not have a rising continuum longward of 13
$\micron$ indicating that the outer disk probably is not as optically
thick as that of Class II disks and the `wall' emission is not
significant. This class of disks has been identified by various names
in the literature: `evolved disks', `anemic disks', `homologously
depleted disks' \citep{lada06, hernandez07, currie09a,
  luhman10}. Among these objects, CHXR~20,CHXR~30A and CHXR~47 are
WTTS (see Table~\ref{basic_tbl}); they have stopped accreting or are
accreting at very low rates. C7-1, however, has an H$\alpha$
equivalent width of 60 $\AA$ \citep{luhman04} and is accreting. It
probably has gas in the inner most parts of the disk. It is not clear
if evolved disks fit into an evolutionary sequence in which the
optically thick disk develops inner gaps and holes and eventually
dissipates inside out. It is possible that they represent a different
evolutionary path for the primordial optically thick disks where the
disk dissipation does not proceed inside out. In this sense, evolved
disks could also be considered to be in a transitional phase between
the primordial optically thick disks and the optically thin debris
disks \citep{lada06, currie09a, muzerolle10}. A more detailed
discussion of these objects and the question of disk dissipation in
Cha~I will be addressed in a forthcoming paper (Furlan et al. in
preparation).

\section{Summary and conclusions} \label{results}

We have presented the mid-IR spectra of 82 young sources (most of them
later than K5, including very low mass stars and brown dwarfs) in the
Chamaeleon I star forming region, obtained with the IRS onboard the
{\it Spitzer Space Telescope}. The IRS spectra and the photometry
compiled from the literature have been used to construct the SEDs of
these objects. Based on the observed SEDs, we classified the young
stars in Cha~I into various evolutionary classes. Our sample consists
of 6 Class I objects, 5 Flat spectrum sources, 68 Class II objects and
3 Class III stars. Eight of the Class II sources appear to be
transitional disk candidates.

We analyzed the IRS spectra of the Class II objects in order to study
the structure and evolution of their disks. To do this, we compared
the shape of the observed mid-IR continuum to that predicted
by the irradiated, accretion disk models which incorporates effects of
disk evolution, like grain growth and sedimentation. We analyzed the
silicate emission features in the IRS spectra to study the structure
and dust properties of protoplanetary disks in Cha~I. We computed
median SEDs representative of T Tauri stars (K5-M2) and very low mass
objects (M3-M8) in our sample and compared them to the model SEDs and
to each other to investigate the differences in disk properties as a
function of stellar mass. We also examined the effects of multiplicity
and accretion on the observed disk properties. Finally, we compared
the properties of protoplanetary disks in Cha~I (age $\sim$ 2~Myr)
with those in slightly younger (ages $\la$ 1~Myr) Taurus and Ophiuchus
star forming regions. The main results of this study are summarized below.

\begin{itemize}
\item 
The observed continuum indices $n_{6-13}$ and $n_{13-31}$ for most
Class II objects are best explained by the disk models where the
surface layers of the disk are depleted of small grains by a factor of
100 to 1000 compared to that of the standard ISM dust-to-gas mass
ratio. Substantial dust settling has occurred in most protoplanetary
disks surrounding Class II objects in Cha I.  The distribution of
$n_{6-13}$ and $n_{13-31}$ continuum indices of Class II objects in
the $\sim$ 2 Myr old Cha I, $\sim$ 1 Myr Taurus and $\la$ 1 Myr
Ophiuchus are statistically indistinguishable indicating that the
degree of dust settling in the disks and the frequency of settled
disks are quite similar in these three regions. Most protoplanetary
disks undergo significant dust settling by $\la$~1~Myr. No evidence
for temporal evolution is seen in the observed degree of settling in
$\la 1-2$~Myr old disks.  Grain growth and sedimentation in
protoplanetary disks appear to be balanced by counteracting processes
such as fragmentation of large grains in the midplane and turbulent
mixing.

\item
The absence of objects showing high $n_{2-6}$ index in Cha I indicates
a paucity of high accretors among the Class II objects in Cha I as
compared to Taurus and Ophiuchus. The apparent lack of high accretors
in Cha I is consistent with the accretion rates in Class~II disks
declining with age.

\item
We confirmed the TDs and PTDs in Cha~I reported by \citet{kim09} and
identified and additional TD candidate of spectral type M5,
2M~J11241186-7630425. TDs and PTDs show presence of holes or gaps in
their inner disks, but have optically thick outer disks similar to
that in Class II disks. Most TDs and PTDs are active accretors. Some
of them have close stellar companions which are likely responsible for
their altered disk structure; but in most TDs and PTDs, the holes and
gaps are probably caused by very low mass, hitherto undetected
substellar or planetary mass objects.

We also identified a few objects which show significantly high 10
$\micron$ silicate equivalent widths (W$_{10}$ $\ga$ W$_{10,upper\:octile}$)
than what the models for `full' disks would predict, but have
$n_{13-31}$ index within the allowable range for the 'full' disks.
These objects are possibly opening up gaps in their disk
\citep{esp09,furlan09}. However, detailed modeling of their disk
structure is required to confirm this. Most of these objects are
actively accreting.

We further confirmed 4 objects which could be classified as `evolved'
disks based on their observed SEDs. Their continuum emission is
considerably lower than the `full ' disks in the entire infrared
wavelength range (2 - 36 $\micron$) indicating that their outer disks are relatively
optically thin. Most of these objects have stopped accreting or are
accreting at a low rate.

\item
Almost all Class II disks show mid-IR continuum excess and silicate
emission features centered at 10 and 20~$\micron$. The strength of
both the 10 and 20 $\micron$ silicate features, on average,  decreases as the degree
of dust settling increases.  Flared disks with less sedimentation show
stronger silicate features whereas flatter disk with high degree of
dust settling show weaker silicate features. We further find that the
relative strength of the 20~$\micron$ feature over the 10 $\micron$
feature drops with increasing degree of dust sedimentation in the
disks.

\item
The 10 and 20 $\micron$ silicate features in Class II disks display a
wide variety of feature strengths and shapes. The flux ratio
F$_{11.3}$/F$_{9.8}$ and the equivalent width of the 33.6 $\micron$
forsterite feature indicate that the dust grains in the disk in Cha~I
have undergone significant processing (grain growth and
crystallization). Additionally, we find a tight correlation between the
equivalent width of 10 $\micron$ silicate feature and the flux ratio
F$_{11.3}$/F$_{9.8}$, which is a measure of the degree of dust
processing in the inner ($\la$ 1-2~AU) disks. Less settled disks, with
stronger silicate emission show low degree of dust processing. The
more settled disks in our sample, in general, show higher degree of
dust processing.

However, disks with radial holes/gaps in general, show larger silicate
emission equivalent widths and relatively pristine dust composition
compared to that found for radially continuous `full' disks.

We find a tight correlation between the flux ratio
F$_{11.3}$/F$_{9.8}$ and the equivalent width of the forsterite
feature at 33.6 $\micron$ which suggests that the inner ($\la$ 1-2~AU)
and outer ($\la$ 10~AU) disk crystallinity track each other well for
Class~II disks in Cha~I. The correlation between crystalline content
of the inner and outer regions of the disks has also been found for
Class~II objects in Taurus and Ophiuchus region.

\item
Median SEDs computed for Class II objects in the spectral type range
of K5-M2 and M3-M8 show crucial difference in the shape and strength
of the continuum emission at wavelengths $\la$ 8 $\micron$: mid-M type
stars and brown dwarfs show excess emission only at wavelengths $\ga$
4-5 $\micron$ while in T Tauri stars (K5-M2) the excess emission
begins at $\sim$ 2 $\micron$; the observed continuum at wavelengths
$\la$ 8 $\micron$ in mid-M type stars and brown dwarfs is considerably
steeper than that observed for T Tauri stars. However, both the
medians look similar at longer wavelengths: they show 10 and 20
$\micron$ silicate emission features and their continuum slopes at
wavelengths $\ga$ 8 $\micron$ are not very different.

\item
Most WTTS in our sample show mid-IR ($\ga$ 8 $\micron$) excess
emission and the silicate features at 10 and 20 $\micron$, indicating
the presence of disks around them. However, most of them do not show
measurable excess at near-IR ($\la$ 6 $\micron$) wavelengths. Their
mid-IR continuum indices are steeper (bluer) than those for CTTS. The
observed mid-IR excess of WTTS is considerably weaker than that of
CTTS at all wavelengths.

\end{itemize}

\acknowledgments

This work is based on observations made with the {\it Spitzer Space
  Telescope}, which is operated by the Jet Propulsion Laboratory
(JPL), California Institute of Technology (Caltech), under NASA
contract 1407. Support for this work was provided by NASA through
contract number 1257184 issued by JPL/Caltech, JPL contract 960803 to
Cornell University, and Cornell subcontracts 31419-5714 to the
University of Rochester. E.F. was partly supported by a NASA
Postdoctoral Program Fellowship, administered by Oak Ridge Associated
Universities through a contract with NASA, and partly supported by
NASA through the Spitzer Space Telescope Fellowship Program, through a
contract issued by JPL/Caltech under a contract with NASA. K. L. was
supported by grant AST-0544588 from the National Science Foundation.
The Center for Exoplanets and Habitable Worlds is supported by the
Pennsylvania State University, the Eberly College of Science, and the
Pennsylvania Space Grant Consortium. C.~E. was supported by the National
Science Foundation under Award No. 0901947. P.D. acknowledges a grant from
PAPIIT-UNAM, M\'{e}xico.  This publication makes use of the SIMBAD and
VizieR databases, operated at CDS (Strasbourg, France) and NASA's
Astrophysics Data System Abstract Service.

{\it Facilities:} \facility{Spitzer (IRS)}


\clearpage

\begin{deluxetable}{lllllcll}
\tabletypesize{\scriptsize}
\rotate
\tablewidth{490pt}
\tablecaption{Log of IRS observations of the Cha I sample \label{log_tbl}}
\tablehead{
\colhead{Object Name} & \colhead{2MASS ID} & \colhead{RA} & \colhead{Dec} &\colhead{Date of}
&\colhead{Campaign} &\colhead{AOR} &\colhead{Module}\\
\colhead{} & \colhead{} & \colhead{(2000)} & \colhead{(2000)} &\colhead{observation}
&\colhead{No.} &\colhead{ID} &\colhead{}\\
\colhead{(1)} & \colhead{(2)} & \colhead{(3)} & \colhead{(4)} &\colhead{(5)}
&\colhead{(6)} &\colhead{(7)} &\colhead{(8)}\\
}

\startdata

              SX Cha&    J10555973-7724399 &   10 55 59.73&  -77 24 39.90&   2005 May 26&  21&   12697345&    SLSHLH\\
                  T5&    J10574219-7659356 &   10 57 42.20&  -76 56 35.70&   2005 Jul 13&  22&   12696320&      SLLL\\
             \nodata&    J10580597-7711501 &   10 58  5.98&  -77 11 50.10&   2007 Jun 26&  41&   21914368&      SLLL\\
              SZ Cha&    J10581677-7717170 &   10 58 16.77&  -77 17 17.10&   2005 Jul 12&  22&   12696832&      SLLL\\
              TW Cha&    J10590108-7722407 &   10 59  1.09&  -77 22 40.70&   2005 Jul 10&  22&   12696576&      SLLL\\
              CR Cha&    J10590699-7701404 &   10 59  6.99&   -77 1 40.40&   2005 May 26&  21&   12697345&    SLSHLH\\
             \nodata&    J11011926-7732383 &    11 1 19.22&  -77 32 38.60&   2007 Jun 18&  41&   21920512&      SLLL\\
              CS Cha&    J11022491-7733357 &    11 2 24.91&  -77 33 35.70&   2005 Jul 11&  22&   12695808&      SLLL\\
              CT Cha&    J11040909-7627193 &    11 4  9.09&  -76 27 19.40&   2005 May 26&  21&   12697345&    SLSHLH\\
                T14a&    J11042275-7718080 &    11 4 22.76&  -77 18  8.10&   2005 Jul 11&  22&   12695808&      SLLL\\
              ISO 52&    J11044258-7741571 &    11 4 42.58&  -77 41 57.10&   2005 Jul 12&  22&   12691200&      SLLL\\
                 T21&    J11061540-7721567 &    11 6 15.41&  -77 21 56.80&   2005 Jul 13&  22&   12696320&      SLLL\\
             \nodata&    J11062942-7724586 &    11 6 29.43&  -77 24 58.60&    2006 Aug 2&  33&   18360832&      SLLL\\
           CHSM 7869&    J11063276-7625210 &    11 6 32.77&  -76 25 21.10&   2007 Jun 26&  41&   21913600&      SLLL\\
              ISO 79&    J11063945-7736052 &    11 6 39.45&  -77 36  5.20&   2005 Jul 11&  22&   12690176&      SLLL\\
                Hn 5&    J11064180-7635489 &    11 6 41.81&  -76 35 49.00&   2005 Jul 13&  22&   12696320&      SLLL\\
              UX Cha&    J11064346-7726343 &    11 6 43.47&  -77 26 34.40&   2005 Jul 13&  22&   12696064&      SLLL\\
             CHXR 20&    J11064510-7727023 &    11 6 45.10&  -77 27  2.30&   2005 Jul 12&  22&   12695296&      SLLL\\
         Ced110 IRS4&    J11064658-7722325 &    11 6 46.58&  -77 22 32.60&   2005 Apr 17&  20&   12692224&      SLLL\\
              ISO 86&    J11065803-7722488 &    11 6 58.03&  -77 22 48.90&   2005 Apr 17&  20&   12692224&      SLLL\\
              UY Cha&    J11065906-7718535 &    11 6 59.07&  -77 18 53.60&   2005 Apr 17&  20&   12692224&      SLLL\\
             \nodata&    J11065939-7530559 &    11 6 59.40&  -75 30 56.00&   2007 Jun 15&  41&   21915136&      SLLL\\
             \nodata&    J11070369-7724307 &    11 7  3.69&  -77 24 30.70&    2006 Aug 2&  33&   18360576&      SLLL\\
         Ced110 IRS6&    J11070919-7723049 &    11 7  9.20&  -77 23  5.00&   2005 Jul 12&  22&   12696832&      SLLL\\
              ISO 91&    J11070925-7718471 &    11 7  9.25&  -77 18 47.10&   2005 Apr 24&  20&   12691712&      SLLL\\
              UZ Cha&    J11071206-7632232 &    11 7 12.07&  -76 32 23.20&   2005 Jul 13&  22&   12696320&      SLLL\\
            CHXR 22E&    J11071330-7743498 &    11 7 13.33&  -77 43 49.80&    2006 Jul 3&  32&   18361344&      SLLL\\
    Cha H${\alpha}$1&    J11071668-7735532 &    11 7 16.69&  -77 35 53.30&   2005 Jul 13&  22&   12689920&      SLLL\\
                 T25&    J11071915-7603048 &    11 7 19.15&   -76 3  4.80&   2005 Apr 24&  20&   12695552&      SLLL\\
              DI Cha&    J11072074-7738073 &    11 7 20.74&  -77 38  7.40&   2005 May 26&  21&   12697345&    SLSHLH\\
                 B35&    J11072142-7722117 &    11 7 21.43&  -77 22 11.80&   2005 Jul 13&  22&   12696320&      SLLL\\
              VV Cha&    J11072825-7652118 &    11 7 28.26&  -76 52 11.90&   2005 Jul 13&  22&   12696064&      SLLL\\
    Cha H${\alpha}$2&    J11074245-7733593 &    11 7 42.45&  -77 33 59.40&   2005 Jul 12&  22&   12690688&      SLLL\\
                 T28&    J11074366-7739411 &    11 7 43.66&  -77 39 41.10&   2005 Jul 12&  22&   12696832&      SLLL\\
          CHSM 10862&    J11074656-7615174 &    11 7 46.56&  -76 15 17.50&   2007 Jun 15&  41&   21913856&      SLLL\\
            CHXR 30B&    J11075730-7717262 &    11 7 57.31&  -77 17 26.20&   2005 Jul 10&  22&   12696576&      SLLL\\
                 T29&    J11075792-7738449 &    11 7 57.93&  -77 38 44.90&   2005 May 26&  21&   12697345&    SLSHLH\\
            CHXR 30A&    J11080002-7717304 &    11 8  0.02&  -77 17 30.50&   2005 May 30&  21&   12692481&      SLLL\\
              VW Cha&    J11080148-7742288 &    11 8  1.49&  -77 42 28.80&   2005 Jul 12&  22&   12696832&      SLLL\\
             ISO 126&    J11080297-7738425 &    11 8  2.98&  -77 38 42.60&   2005 May 26&  21&   12697345&    SLSHLH\\
              CU Cha&    J11080329-7739174 &    11 8  3.30&  -77 39 17.40&    2006 Mar 7&  29&   12697088&    SLSHLH\\
                T33A&    J11081509-7733531 &    11 8 15.10&  -77 33 53.20&    2006 Mar 7&  29&   12697088&    SLSHLH\\
             ISO 138&    J11081850-7730408 &    11 8 18.50&  -77 30 40.80&   2005 Jul 12&  22&   12690432&      SLLL\\
             ISO 143&    J11082238-7730277 &    11 8 22.38&  -77 30 27.70&   2005 Jul 12&  22&   12690432&      SLLL\\
             Cha IRN&    J11083896-7743513 &    11 8 38.20&  -77 43 51.80&   2005 May 26&  21&   12697345&    SLSHLH\\
                 T35&    J11083905-7716042 &    11 8 39.05&  -77 16  4.20&   2005 Jul 13&  22&   12696320&      SLLL\\
              VY Cha&    J11085464-7702129 &    11 8 54.64&   -77 2 13.00&   2005 May 26&  21&   12691969&      SLLL\\
                C1-6&    J11092266-7634320 &    11 9 22.67&  -76 34 32.00&   2005 May 26&  21&   12697345&    SLSHLH\\
              VZ Cha&    J11092379-7623207 &    11 9 23.79&  -76 23 20.80&   2005 Jul 12&  22&   12696832&      SLLL\\
             ISO 192&    J11092855-7633281 &    11 9 28.55&  -76 33 28.10&    2006 Mar 7&  29&   12697088&    SLSHLH\\
               C1-25&    J11094192-7634584 &    11 9 41.93&  -76 34 58.40&   2005 Jul 12&  22&   12696832&      SLLL\\
                C7-1&    J11094260-7725578 &    11 9 42.60&  -77 25 57.90&   2005 Jul 12&  22&   12686336&      SLLL\\
              Hn 10E&    J11094621-7634463 &    11 9 46.21&  -76 34 46.40&   2005 Jul 13&  22&   12696320&      SLLL\\
                 B43&    J11094742-7726290 &    11 9 47.42&  -77 26 29.10&   2005 Apr 17&  20&   12691456&      SLLL\\
            HD 97300&    J11095003-7636476 &    11 9 50.03&  -76 36 47.70&    2006 Mar 7&  29&   12697088&    SLSHLH\\
             ISO 220&    J11095336-7728365 &    11 9 53.37&  -77 28 36.60&    2006 Mar 7&  29&   12686080&      SLLL\\
                 T42&    J11095340-7634255 &    11 9 53.41&  -76 34 25.50&    2006 Mar 7&  29&   12697088&    SLSHLH\\
                 T43&    J11095407-7629253 &    11 9 54.08&  -76 29 25.30&   2005 Apr 17&  20&   12692224&      SLLL\\
             ISO 225&    J11095437-7631113 &    11 9 54.38&  -76 31 11.40&   2005 Jul 12&  22&   12690432&      SLLL\\
                C1-2&    J11095505-7632409 &    11 9 55.06&  -76 32 41.00&   2005 Jul 12&  22&   12696832&      SLLL\\
              WX Cha&    J11095873-7737088 &    11 9 58.74&  -77 37  8.90&   2005 Jul 10&  22&   12696576&      SLLL\\
              WW Cha&    J11100010-7634578 &   11 10  0.11&  -76 34 57.90&    2006 Mar 7&  29&   12697088&    SLSHLH\\
               Hn 11&    J11100369-7633291 &   11 10  3.69&  -76 33 29.20&   2005 Jul 13&  22&   12696064&      SLLL\\
                T45a&    J11100469-7635452 &   11 10  4.69&  -76 35 45.30&   2005 May 26&  21&   12691969&      SLLL\\
              WY Cha&    J11100704-7629376 &   11 10  7.04&  -76 29 37.70&   2005 Jul 13&  22&   12696064&      SLLL\\
             ISO 235&    J11100785-7727480 &   11 10  7.85&  -77 27 48.10&   2005 Jul 12&  22&   12690432&      SLLL\\
             ISO 237&    J11101141-7635292 &   11 10 11.42&  -76 35 29.30&   2005 Jul 10&  22&   12696576&      SLLL\\
             CHXR 47&    J11103801-7732399 &   11 10 38.02&  -77 32 39.90&   2005 Jul 13&  22&   12696064&      SLLL\\
             ISO 252&    J11104141-7720480 &   11 10 41.41&  -77 20 48.00&    2005 Jul 9&  22&   12685825&      SLLL\\
                 T47&    J11104959-7717517 &   11 10 49.60&  -77 17 51.70&   2005 Jul 10&  22&   12696576&      SLLL\\
              WZ Cha&    J11105333-7634319 &   11 10 53.34&  -76 34 32.00&   2005 Apr 17&  20&   12692224&      SLLL\\
             ISO 256&    J11105359-7725004 &   11 10 53.59&  -77 25  0.50&   2005 May 26&  21&   12691969&      SLLL\\
               Hn 13&    J11105597-7645325 &   11 10 55.97&  -76 45 32.60&   2005 Jul 12&  22&   12690688&      SLLL\\
              XX Cha&    J11113965-7620152 &   11 11 39.66&  -76 20 15.20&   2005 Jul 10&  22&   12696576&      SLLL\\
             ISO 282&    J11120351-7726009 &   11 12  3.51&  -77 26  1.00&   2005 Jul 13&  22&   12686848&      SLLL\\
                 T50&    J11120984-7634366 &   11 12  9.85&  -76 34 36.50&   2005 Jul 13&  22&   12696064&      SLLL\\
                 T51&    J11122441-7637064 &   11 12 24.41&  -76 37  6.40&   2005 Jul 10&  22&   12696576&      SLLL\\
              CV Cha&    J11122772-7644223 &   11 12 27.72&  -76 44 22.30&    2006 Mar 7&  29&   12697088&    SLSHLH\\
                 T54&    J11124268-7722230 &   11 12 42.69&  -77 22 23.00&   2005 Apr 24&  20&   12695552&      SLLL\\
              Hn 21W&    J11142454-7733062 &   11 14 24.54&  -77 33  6.20&   2007 Jun 16&  41&   21914112&      SLLL\\
                 T56&    J11173700-7704381 &   11 17 37.01&   -77 4 38.10&   2005 Jul 13&  21&   12696064&      SLLL\\
             \nodata&    J11183572-7935548 &   11 18 35.73&  -79 35 54.90&   2007 Jun 10&  41&   21915392&      SLLL\\
             \nodata&    J11241186-7630425 &   11 24 11.87&  -76 30 42.60&    2007 Aug 2&  42&   21914624&      SLLL\\
             \nodata&    J11432669-7804454 &   11 43 26.70&   -78 4 45.50&   2007 Jun 17&  41&   21914880&      SLLL\\

\enddata
\tablecomments{Columns (1) \& (2) list the
object names and their 2MASS identifiers. Observed coordinates are
listed in Columns (3) and (4) and the date of observations in column
(5). The {\it Spitzer} IRS observing campaigns during which the objects
were observed are listed in Column (6) and the corresponding
Astronomical Observation Requests (AORs) in Column (7). Column (8) lists
the IRS modules used in the observations.}
\end{deluxetable}

\clearpage

\begin{deluxetable}{llccclccc}
\tabletypesize{\scriptsize}
\tablewidth{490pt}
\tablecaption{Basic data for the Cha I sample \label{basic_tbl}}
\tablehead{
\colhead{Object Name} & \colhead{Sp. Type} & \colhead{A$_J$} & \colhead{$n_{2-25}$} & \colhead{$n_{2-25}$} &\colhead{SED Class}
& \colhead{$n_{5-12}$} &\colhead{W(H$_{\alpha}$)} &\colhead{C/W}\\
\colhead{} & \colhead{} & \colhead{(mag)} & \colhead{(observed)} &\colhead{(dereddened)} & \colhead{}
& \colhead{}&\colhead{(\AA)} &\colhead{}\\
\colhead{(1)} & \colhead{(2)} & \colhead{(3)} & \colhead{(4)} &\colhead{(5)}
&\colhead{(6)} &\colhead{(7)} &\colhead{(8)} &\colhead{(9)}\\
}

\startdata

              SX Cha&        M0&      0.7&     -0.61&     -0.59&             II&     -0.71&     31.7&        C\\
                  T5&     M3.25&      0.3&     -1.02&     -1.02&             II&     -0.87&     22.0&        C\\
2M J10580597-7711501&     M5.25&      0.4&     -0.94&     -0.93&             II&     -1.18&  \nodata&  \nodata\\
              SZ Cha&        K0&      0.4&     -0.59&     -0.58&         II$/$T&     -1.20&      7.0&        C\\
              TW Cha&        K8&      0.3&     -1.00&     -1.00&             II&     -1.35&     34.5&        C\\
              CR Cha&        K2&      0.0&     -0.82&     -0.82&             II&     -0.62&     38.5&        C\\
2M J11011926-7732383&     M7.25&      0.4&     -2.38&     -2.37&            III&     -2.70&     25.0&        W\\
              CS Cha&        K6&      0.0&     -0.64&     -0.63&         II$/$T&     -1.52&     20.0&        C\\
              CT Cha&        K5&      0.4&     -0.64&     -0.63&             II&     -0.50&     34.5&        C\\
                T14a&        K7&      0.4&      0.50&      0.51&              I&      0.52&    159.0&        C\\
              ISO 52&        M4&      0.3&     -0.81&     -0.81&             II&     -0.47&      6.4&        W\\
                 T21&        G5&      0.9&     -2.53&     -2.50&            III&     -2.65&  \nodata&  \nodata\\
2M J11062942-7724586&        M6&      5.6&     -1.36&     -1.06&             II&     -1.29&  \nodata&  \nodata\\
           CHSM 7869&        M6&      0.4&     -1.09&     -1.08&             II&     -1.13&    120.0&        C\\
              ISO 79&     M5.25&      0.7&     -0.70&     -0.69&             II&     -0.76&      8.0&        W\\
                Hn 5&      M4.5&      0.3&     -0.83&     -0.82&             II&     -0.72&     48.0&        C\\
              UX Cha&        M3&      0.9&     -2.62&     -2.54&            III&     -2.56&      3.0&        W\\
             CHXR 20&        K6&      1.0&     -1.30&     -1.27&             II&     -0.81&      1.8&        W\\
         Ced110 IRS4&   \nodata&  \nodata&      1.61&   \nodata&              I&      0.31&  \nodata&  \nodata\\
              ISO 86&   \nodata&  \nodata&      0.91&   \nodata&              I&     -0.95&  \nodata&  \nodata\\
              UY Cha&     M4.25&      0.0&     -1.00&     -1.00&             II&     -0.52&      8.5&        W\\
2M J11065939-7530559&     M5.25&      0.1&     -1.00&     -0.99&             II&     -1.23&     51.0&        C\\
2M J11070369-7724307&      M7.5&      4.6&     -0.94&     -0.70&             II&     -0.75&  \nodata&  \nodata\\
         Ced110 IRS6&   \nodata&  \nodata&      0.70&   \nodata&              I&     -0.81&  \nodata&  \nodata\\
              ISO 91&        M3&      4.5&     -1.15&     -0.91&             II&     -0.53&  \nodata&  \nodata\\
              UZ Cha&      M0.5&      0.6&     -1.04&     -1.03&             II&     -1.02&      9.7&        W\\
            CHXR 22E&      M3.5&      1.3&     -1.79&     -1.74&         II$/$T&     -2.62&      4.5&        W\\
    Cha H${\alpha}$1&     M7.75&      0.0&     -0.52&     -0.52&             II&     -0.18&    112.0&        C\\
                 T25&      M2.5&      0.4&     -0.81&     -0.80&         II$/$T&     -1.97&      5.0&        W\\
              DI Cha&        G2&      0.7&     -0.99&     -0.98&             II&     -1.12&     18.3&        C\\
                 B35&   \nodata&  \nodata&     -0.02&   \nodata&             FS&     -0.02&  \nodata&  \nodata\\
              VV Cha&        M3&      0.1&     -0.94&     -0.94&             II&     -0.40&     72.0&        C\\
    Cha H${\alpha}$2&     M5.25&      1.0&     -1.28&     -1.24&             II&     -1.07&     60.0&        C\\
                 T28&        M0&      1.3&     -1.11&     -1.07&             II&     -1.01&     78.0&        C\\
          CHSM 10862&     M5.75&      0.4&     -0.84&     -0.83&             II&     -0.90&     18.0&        W\\
            CHXR 30B&     M1.25&      3.1&     -0.96&     -0.79&             II&     -1.42&     80.0&        C\\
                 T29&        K6&      1.2&     -0.50&     -0.47&             II&     -0.36&     61.5&        C\\
            CHXR 30A&        K8&      2.9&     -1.75&     -1.59&             II&     -1.58&      1.0&        W\\
              VW Cha&        K8&      0.7&     -0.89&     -0.87&             II&     -1.02&     64.0&        C\\
             ISO 126&     M1.25&      1.3&     -0.13&     -0.09&             FS&     -0.49&    120.0&        C\\
              CU Cha&      B9.5&      0.7&     -0.14&     -0.13&             II&     -0.03&  \nodata&  \nodata\\
                T33A&        G7&      0.8&   \nodata&   \nodata&             II&     -0.47&      0.9&        W\\
             ISO 138&      M6.5&      0.0&     -1.00&     -1.00&             II&     -0.75&     19.0&        W\\
             ISO 143&        M5&      1.1&     -1.16&     -1.12&             II&     -0.85&    118.0&        C\\
             Cha IRN&   \nodata&  \nodata&      1.24&   \nodata&              I&      1.46&  \nodata&  \nodata\\
                 T35&        K8&      1.3&     -1.23&     -1.19&         II$/$T&     -2.48&    100.0&        C\\
              VY Cha&      M0.5&      0.9&     -0.91&     -0.88&             II&     -1.18&     55.0&        C\\
                C1-6&     M1.25&      3.2&     -0.64&     -0.46&             II&     -0.50&    110.0&        C\\
              VZ Cha&        K6&      0.5&     -1.07&     -1.06&             II&     -0.94&     41.3&        C\\
             ISO 192&   \nodata&  \nodata&      1.24&   \nodata&              I&      1.49&  \nodata&  \nodata\\
               C1-25&   \nodata&  \nodata&     -0.26&   \nodata&             FS&     -0.11&  \nodata&  \nodata\\
                C7-1&        M5&      1.9&     -1.59&     -1.53&             II&     -1.78&     60.0&        C\\
              Hn 10E&     M3.25&      1.0&     -0.44&     -0.41&             II&     -0.14&     68.0&        C\\
                 B43&     M3.25&      2.2&     -0.97&     -0.90&             II&     -1.08&    200.0&        C\\
            HD 97300&        B9&      0.7&     -1.32&     -1.31&             II&      0.17&  \nodata&  \nodata\\
             ISO 220&     M5.75&      1.7&     -1.23&     -1.18&             II&     -0.94&    150.0&        C\\
                 T42&        K5&      1.4&     -0.39&     -0.34&             II&     -0.38&    205.0&        C\\
                 T43&        M2&      1.4&     -0.94&     -0.89&             II&     -0.92&     43.0&        C\\
             ISO 225&     M1.75&      1.2&      0.10&      0.14&             FS&      0.00&     12.7&        C\\
                C1-2&   \nodata&  \nodata&      0.10&   \nodata&             FS&     -0.43&     45.0&  \nodata\\
              WX Cha&     M1.25&      0.5&     -1.06&     -1.05&             II&     -1.16&     63.0&        C\\
              WW Cha&        K5&      1.3&     -0.46&     -0.42&             II&     -0.78&     61.0&        C\\
               Hn 11&        K8&      2.1&     -0.84&     -0.77&             II&     -0.58&     19.0&        C\\
                T45a&        M1&      0.5&     -1.00&     -1.00&             II&     -0.93&      7.5&        W\\
              WY Cha&        M0&      1.1&     -1.18&     -1.15&             II&     -0.83&     51.0&        C\\
             ISO 235&      M5.5&      2.4&     -1.31&     -1.24&             II&     -0.71&     18.0&        W\\
             ISO 237&      K5.5&      1.9&     -0.78&     -0.72&             II&     -0.88&      3.8&        C\\
             CHXR 47&        K3&      1.4&     -1.24&     -1.20&             II&     -1.05&      1.0&        W\\
             ISO 252&        M6&      0.9&     -1.01&     -0.98&             II&     -0.71&    320.0&        C\\
                 T47&        M2&      1.1&     -0.63&     -0.60&             II&     -0.19&     89.5&        C\\
              WZ Cha&     M3.75&      0.1&     -0.79&     -0.79&             II&     -0.53&    279.0&        C\\
             ISO 256&      M4.5&      2.5&     -0.75&     -0.67&             II&     -0.57&     55.0&        C\\
               Hn 13&     M5.75&      0.2&     -1.02&     -1.02&             II&     -1.08&     22.5&        C\\
              XX Cha&        M2&      0.3&     -0.90&     -0.89&             II&     -0.90&     90.0&        C\\
             ISO 282&     M4.75&      1.0&     -1.13&     -1.10&             II&     -1.51&     50.0&        C\\
                 T50&        M5&      0.1&     -1.08&     -1.08&             II&     -1.01&     19.5&        W\\
                 T51&      K3.5&      0.0&     -1.08&     -1.08&             II&     -1.04&      2.5&        W\\
              CV Cha&        G9&      0.4&     -0.78&     -0.77&             II&     -0.79&     62.5&        C\\
                 T54&        G8&      0.6&     -1.68&     -1.67&         II$/$T&     -2.32&  \nodata&  \nodata\\
              Hn 21W&        M4&      0.7&     -1.34&     -1.33&             II&     -1.26&     68.0&        C\\
                 T56&      M0.5&      0.1&     -0.93&     -0.93&         II$/$T&     -1.22&     56.0&        C\\
2M J11183572-7935548&     M4.75&      0.0&     -0.66&     -0.66&         II$/$T&     -0.67&     11.0&        W\\
2M J11241186-7630425&        M5&      0.7&     -0.91&     -0.89&         II$/$T&     -1.13&     30.0&        C\\
2M J11432669-7804454&        M5&      0.2&     -1.48&     -1.47&             II&     -1.77&     75.0&        C\\

\enddata \tablecomments{ Spectral types and extinction values (A$_J$)
  listed in Columns~(2)~\&~(3) are from \citep{luhman07,luhman08b}. `T' in Column~(5) denotes transitional disk candidates (see Section 3.2). Column~(6) lists H$\alpha$ equivalent widths compiled from the literature \citep{luhman04,herbigbell88,comeron99,walter92,covino97,feigkris89,guenther07}. Positive values indicate H$\alpha$ in emission.
}
\end{deluxetable}

\clearpage
\begin{deluxetable}{lcllll}
\tabletypesize{\tiny}
\rotate
\tablewidth{550pt}
\tablecaption{Objects in multiple systems \label{binary_tbl}}
\tablehead{
\colhead{} &\colhead{} &\colhead{Separation} &\colhead{} &\colhead{Brightness} &\colhead{}\\
\colhead{Name} &\colhead{N} &\colhead{($\arcsec$)} &\colhead{Spectral Type} &\colhead{Comparison} &\colhead{References}\\
\colhead{(1)} &\colhead{(2)} &\colhead{(3)} &\colhead{(4)} &\colhead{(5)} &\colhead{(6)}\\
}

\startdata

                            SX Cha (A,B)&  2&                                                 2.2 &                        M0+M3.5 &                          A$>$B &             1,2 \\
                                T5 (A,B)&  2&                                                0.16 &                          M3.25 &                        A$\sim$B&                2\\
              2M J11011926-7732383 (A,B)&  2&                                                  1.4&                     M7.25+M8.25&                          A$>$B &                3\\
                            CS Cha (A,B)&  2&                                                   SB&                              K6&                           A$>$B&                4\\
                              T14a (A,B)&  2&                                                2.35 &                              K7&                          A$>$B &                5\\
                               T21 (A,B)&  2&                                                0.14 &                              G5&                         A$>>$B &                2\\
                        CHXR 20 + UX Cha&  2&                                               28.46 &                           K6+M3&                CHXR20$>$UX Cha &               6 \\
                       Ced110 IRS6 (A,B)&  2&                                                1.95 &                         ClassI &                          A$>$B &             5,7 \\
                        DI Cha (A,Ba,Bb)&  3&                           4.68(A,Ba+Bb),0.07(Ba,Bb) &              G2(A),M4.25(Ba+Bb)&          A$>>$Ba+Bb,Ba$\sim$Bb &            1,8,2\\
                            VV Cha (A,B)&  2&                                                 0.8 &                             M3 &                          A$>$B &              1,2\\
                           Cha Ha2 (A,B)&  2&                                                0.17 &                          M5.25 &                        A$\sim$B&              9,2\\
                               T28 (A,B)&  2&                                               28.87 &                        M0+M5.75&                         A$>>$B &                6\\
                     T29 + ISO 126 (A,B)&  3&                             17.6, 0.29(ISO 126 A,B) &                        K6+M1.25&       T29$>$ISO 126A$>$ISO 126B&                8\\
                       CHXR 30 (Aa,Ab,B)&  3&                                0.46(Aa,Ab),9.93(A,B)&              K8(Aa+Ab)+M1.25(B)&                Aa$>$B,Aa$>>$Ab &             6,2 \\
                    VW Cha (A,Ba,Bb,C,D)&  4&   0.66(A,Ba),0.7(A,Bb),0.1(Ba,Bb),2.7(A,C),16.9(A,D)&       K8(A+B),K7(Ba+Bb),M2.5(D)&     A$>$B,Ba$>$Bb,A$>$C,A$>>$D &   1,8,10,11,12,2\\
                               T33 (A,B)&  2&                                                2.48 &                          G7+K6 &                          A$>$B &        10,11,5,2\\
                       ISO 143 + ISO 138&  2&                                               18.16 &                        M5+M6.5 &               ISO 143$>$ISO 138&                6\\
                          Hn 10E + C1-25&  2&                                               19.17 &                            M3.5&               Hn 10E$\sim$C1-25&            6,13 \\
                          HD 97300 (A,B)&  2&                                                 0.82&                              B9&                         A$>>$B &              8,2\\
                               T43 (A,B)&  2&                                                 0.8 &                             M2 &                          A$>$B &                2\\
                            WX Cha (A,B)&  2&                                                0.78 &                           M1.25&                         A$>>$B &              8,2\\
                            WY Cha (A,B)&  2&                                                0.12 &                             M0 &                          A$>$B &               2 \\
                          ISO 237 + T45a&  2&                                               28.32 &                        K5.5+M1 &                 ISO 237$>$T45a &               6 \\
          CHXR 47 (A,B)\tablenotemark{a}&  2&                                     1.2(2),0.18(13) &                              K3&          A$\sim$B(2),A$>$B(13) &              8,2\\
                               T47 (A,B)&  2&                                               12.09 &                        M2+M4.5 &                          A$>$B &                6\\
                             Hn 13 (A,B)&  2&                                                0.13 &                           M5.75&                        A$\sim$B&              9,2\\
                            XX Cha (B,A)&  2&                                               24.38 &                           M2+K6&                          B$<$A &               6 \\
                               T51 (A,B)&  3&                                                 1.85&                            K3.5&                         A$>>$B &            1,8,2\\
                         CV Cha + CW Cha&  2&                                               11.55 &                          G9+M1 &               CV Cha$>>$CW Cha &              1,8\\
                               T54 (A,B)&  2&                                                0.26 &                              G8&                         A$>>$B &             8,2 \\
                             Hn 21 (W,E)&  2&                                                5.52 &                       M4+M5.75 &                           W$>$E&          10,6,2 \\

\enddata

\tablecomments{When the companion(s) have common names, we list those
  names in Column(1). When there are more than one measurement for the
  separations between components, and if they are consistent, we have
  listed the mean value in Column(3) and cited all the references. The
  spectral types of the primary and of the other components listed in
  Column (4) are from \citet{luhman07}, except for VW Cha and
  2M~J11011926-7732383, for which they are from the corresponding
  references listed. The relative brightness of the components listed
  in Column (6) is based on the K-band flux.}

\tablenotetext{a}{~The separation and the position angle of the CHXR 47 binary system reported by \citet{ghez97} and \citet{laf08} are very different. We have separately listed the binary separation and flux ratio from both the authors with the references in parenthesis.}

\tablerefs{
(1) \citealt{reipzinn93}; (2) \citealt{laf08}; (3) \citealt{luhman04b}; (4) \citealt{guenther07}; (5) \citealt{haisch04}; (6) \citealt{kraushill07}; (7) \citealt{persi01}; (8) \citealt{ghez97}; (9) \citealt{ahmic07}; (10) \citealt{brand96}; (11) \citealt{brandeker01}; (12) \citealt{brandzinn97}; (13) \citealt{hn93}
}

\end{deluxetable}

\clearpage

\begin{deluxetable}{lrrrrrr}
\tablewidth{380pt}
\tablecaption{Continuum spectral indices of Class II and Class III objects \label{tab4_index}}
\tablehead{
\colhead{Object name} & \colhead{$n_{2-6}$} & \colhead {$\sigma_{n_{2-6}}$} & \colhead{$n_{6-13}$} & \colhead {$\sigma_{n_{6-13}}$} & \colhead{$n_{13-31}$} & \colhead {$\sigma_{n_{13-31}}$} \\

}
\startdata

                             SX Cha&     -0.81&      0.05&   -0.65&    0.05&   -0.49&    0.07\\
                                 T5&     -1.73&      0.06&   -0.87&    0.08&   -0.32&    0.07\\
               2M J10580597-7711501&     -2.08&      0.08&   -0.93&    0.11&    0.26&    0.09\\
                             SZ Cha&     -2.07&      0.08&   -1.04&    0.09&    1.72&    0.03\\
                             TW Cha&     -1.56&      0.08&   -1.27&    0.10&   -0.17&    0.10\\
                             CR Cha&     -1.82&      0.06&   -0.48&    0.06&   -0.12&    0.04\\
               2M J11011926-7732383&     -2.11&      0.08&   -2.79&    0.29&   -1.50&    0.81\\
                             CS Cha&     -2.62&      0.07&   -1.46&    0.13&    2.90&    0.11\\
                             CT Cha&     -1.15&      0.03&   -0.46&    0.02&   -0.32&    0.07\\
                             ISO 52&     -1.98&      0.07&   -0.42&    0.08&    0.15&    0.05\\
                                T21&     -2.70&      0.09&   -2.71&    0.14&   -1.07&    0.14\\
               2M J11062942-7724586&     -1.61&      0.11&   -1.40&    0.27&   -0.16&    0.27\\
                          CHSM 7869&     -1.81&      0.07&   -1.01&    0.13&   -0.56&    0.22\\
                             ISO 79&     -1.21&      0.08&   -0.71&    0.11&   -0.04&    0.06\\
                               Hn 5&     -0.67&      0.05&   -0.77&    0.08&   -1.44&    0.12\\
                             UX Cha&     -2.64&      0.10&   -2.66&    0.44&   -2.01&    0.94\\
                            CHXR 20&     -2.15&      0.07&   -0.78&    0.08&   -0.96&    0.06\\
                             UY Cha&     -1.70&      0.07&   -0.48&    0.09&   -0.92&    0.07\\
               2M J11065939-7530559&     -1.75&      0.09&   -1.27&    0.16&    0.37&    0.19\\
               2M J11070369-7724307&     -1.55&      0.12&   -0.68&    0.12&    0.15&    0.07\\
                             ISO 91&     -2.38&      0.09&   -0.22&    0.12&   -0.40&    0.11\\
                             UZ Cha&     -1.76&      0.08&   -0.95&    0.10&   -0.22&    0.07\\
                           CHXR 22E&     -2.51&      0.09&   -2.70&    0.17&    0.58&    0.14\\
                            Cha Ha1&     -1.50&      0.08&    0.07&    0.12&    0.01&    0.09\\
                                T25&     -2.75&      0.12&   -1.75&    0.26&    2.80&    0.23\\
                             DI Cha&     -1.35&      0.06&   -1.01&    0.06&   -0.67&    0.05\\
                             VV Cha&     -1.74&      0.06&   -0.38&    0.07&   -0.78&    0.07\\
                            Cha Ha2&     -1.66&      0.08&   -1.04&    0.10&   -0.99&    0.07\\
                                T28&     -1.61&      0.06&   -0.99&    0.08&   -0.60&    0.07\\
                         CHSM 10862&     -1.41&      0.05&   -0.87&    0.11&   -0.05&    0.11\\
                           CHXR 30B&     -0.81&      0.06&   -1.33&    0.08&   -0.39&    0.09\\
                                T29&     -1.06&      0.03&   -0.07&    0.03&   -0.32&    0.03\\
                           CHXR 30A&     -2.49&      0.08&   -1.51&    0.11&   -0.59&    0.09\\
                             VW Cha&     -1.48&      0.06&   -0.93&    0.06&   -0.17&    0.06\\
                             CU Cha&     -1.26&      0.05&    0.02&    0.06&    1.03&    0.05\\
              T33A\tablenotemark{a}&   \nodata&   \nodata&   -0.40&    0.06&   -0.55&    0.03\\
                            ISO 138&     -1.69&      0.58&   -0.69&    0.76&   -0.27&    0.42\\
                            ISO 143&     -1.50&      0.06&   -0.85&    0.08&   -1.04&    0.10\\
                                T35&     -2.01&      0.08&   -2.56&    0.20&    1.47&    0.19\\
                             VY Cha&     -1.35&      0.06&   -1.13&    0.07&   -0.19&    0.05\\
                               C1-6&     -0.59&      0.04&   -0.45&    0.04&   -0.61&    0.05\\
                             VZ Cha&     -1.41&      0.06&   -0.85&    0.05&   -0.91&    0.07\\
                               C7-1&     -2.02&      0.08&   -1.81&    0.10&   -0.56&    0.11\\
                             Hn 10E&     -1.41&      0.07&   -0.05&    0.08&    0.37&    0.05\\
                                B43&     -1.84&      0.06&   -1.11&    0.09&    0.43&    0.10\\
                           HD 97300&     -2.50&      0.09&   -0.15&    0.22&   -0.85&    0.23\\
                            ISO 220&     -1.67&      0.07&   -0.98&    0.09&   -0.82&    0.10\\
                                T42&     -0.73&      0.04&   -0.35&    0.03&    0.00&    0.02\\
                                T43&     -1.36&      0.05&   -0.89&    0.06&   -0.37&    0.05\\
                             WX Cha&     -1.15&      0.06&   -1.14&    0.07&   -1.02&    0.07\\
                             WW Cha&     -0.90&      0.07&   -0.69&    0.05&    0.42&    0.02\\
                              Hn 11&     -1.05&      0.04&   -0.65&    0.05&   -0.73&    0.05\\
                               T45a&     -2.06&      0.05&   -0.89&    0.07&    0.14&    0.06\\
                             WY Cha&     -1.53&      0.04&   -0.88&    0.06&   -1.16&    0.08\\
                            ISO 235&     -1.69&      0.06&   -0.75&    0.09&   -1.26&    0.09\\
                            ISO 237&     -1.65&      0.06&   -0.81&    0.08&    0.52&    0.06\\
                            CHXR 47&     -1.87&      0.07&   -1.08&    0.08&   -0.66&    0.06\\
                            ISO 252&     -1.40&      0.05&   -0.72&    0.07&   -0.87&    0.09\\
                                T47&     -1.46&      0.04&   -0.04&    0.04&   -0.27&    0.05\\
                             WZ Cha&     -1.29&      0.05&   -0.39&    0.08&   -0.75&    0.08\\
                            ISO 256&     -0.92&      0.07&   -0.65&    0.08&   -0.62&    0.07\\
                              Hn 13&     -1.58&      0.05&   -1.04&    0.06&   -0.40&    0.05\\
                             XX Cha&     -1.48&      0.04&   -0.81&    0.06&   -0.38&    0.07\\
                            ISO 282&     -1.16&      0.08&   -1.48&    0.10&   -0.54&    0.07\\
                                T50&     -1.85&      0.10&   -0.98&    0.14&   -0.33&    0.11\\
                                T51&     -1.09&      0.04&   -1.08&    0.07&   -1.52&    0.11\\
                             CV Cha&     -1.41&      0.04&   -0.69&    0.04&   -0.27&    0.07\\
                                T54&     -2.88&      0.07&   -2.38&    0.17&    1.10&    0.16\\
                             Hn 21W&     -1.98&      0.08&   -1.17&    0.09&   -0.77&    0.10\\
                                T56&     -2.13&      0.12&   -1.14&    0.15&    0.90&    0.07\\
               2M J11241186-7630425&     -2.33&      0.08&   -0.81&    0.13&    0.68&    0.11\\

\enddata

\tablenotetext{a}{2MASS K$_s$ flux is an upper limit, therefore $n_{2-6}$ value is not listed.}

\end{deluxetable}

\clearpage

\begin{deluxetable}{lcccccccccccc}
\tabletypesize{\tiny}
\rotate
\tablewidth{680pt}
\tablecaption{Properties of the Silicate features of Class II objects \label{tab5_index}}
\tablehead{
\colhead{Object name} & \colhead{$W_{10}$} & \colhead {$\sigma_{W_{10}}$} & \colhead{$F_{10}$} & \colhead {$\sigma_{F_{10}}$} & \colhead{$F_{11.3}\:/\:F_{9.8}$} & \colhead {$\sigma_{F_{11.3}\:/\:F_{9.8}}$}  &  \colhead{$W_{20}$} & \colhead {$\sigma_{W_{20}}$} & \colhead{$F_{20}$} & \colhead {$\sigma_{F_{20}}$} & \colhead{$W_{33}$} & \colhead {$\sigma_{W_{33}}$}\\
    & \colhead{($\micron)$} & \colhead{($\micron$)} & \colhead{($erg\:s^{-1}\:cm^{-2})$} & \colhead{$(erg\:s^{-1}\:cm^{-2})$} & & &\colhead{($\micron$)} & \colhead{($\micron$)} & \colhead{($erg\:s^{-1}\:cm^{-2}$)} & \colhead{($erg\:s^{-1}\:cm^{-2}$)} &\colhead{($\micron$)} & \colhead{($\micron$)}\\
}
\startdata

              SX Cha&      2.46&      0.08&   3.0E-11&   6.6E-13&      0.49&      0.03&      1.69&      0.23&   8.4E-12&   6.6E-13&      0.09&      0.06\\
                  T5&      1.25&      0.07&   2.7E-12&   1.3E-13&      0.74&      0.07&      2.28&      0.31&   1.7E-12&   1.3E-13&      0.12&      0.08\\
2M J10580597-7711501&      2.86&      0.15&   2.4E-13&   1.1E-14&      0.79&      0.05&      4.37&      0.37&   2.1E-13&   1.1E-14&      0.70&      0.11\\
              SZ Cha&      3.28&      0.09&   1.4E-11&   2.5E-13&      0.61&      0.03&      4.79&      0.35&   2.8E-11&   2.5E-13&     -0.01&      0.07\\
              TW Cha&      5.47&      0.12&   1.7E-11&   1.8E-13&      0.47&      0.02&      3.76&      0.34&   5.6E-12&   1.8E-13&     -0.08&      0.08\\
              CR Cha&      4.49&      0.11&   5.0E-11&   6.4E-13&      0.44&      0.02&      2.77&      0.25&   1.8E-11&   6.4E-13&      0.04&      0.06\\
              CS Cha&      3.19&      0.09&   3.1E-12&   6.4E-14&      0.61&      0.03&      3.07&      0.32&   8.1E-12&   6.4E-14&     -0.01&      0.07\\
              CT Cha&      2.00&      0.08&   2.0E-11&   5.1E-13&      0.42&      0.03&      2.95&      0.26&   1.1E-11&   5.1E-13&     -0.00&      0.07\\
              ISO 52&      2.16&      0.08&   1.1E-12&   3.0E-14&      0.55&      0.04&      2.51&      0.32&   8.0E-13&   3.0E-14&      0.27&      0.08\\
2M J11062942-7724586&   \nodata&   \nodata&   \nodata&   \nodata&   \nodata&   \nodata&   \nodata&   \nodata&   \nodata&   \nodata&   \nodata&   \nodata\\
           CHSM 7869&      1.39&      0.08&   8.4E-14&   4.6E-15&      0.76&      0.06&      2.31&      0.38&   4.3E-14&   4.6E-15&     -0.14&      0.36\\
              ISO 79&      1.25&      0.07&   3.3E-13&   1.5E-14&      0.61&      0.06&      2.52&      0.32&   2.9E-13&   1.5E-14&     -0.13&      0.08\\
                Hn 5&      2.09&      0.08&   6.6E-12&   1.8E-13&      0.66&      0.04&      2.71&      0.32&   1.7E-12&   1.8E-13&      0.09&      0.10\\
             CHXR 20&      3.63&      0.10&   1.1E-11&   1.8E-13&      0.53&      0.03&      3.62&      0.34&   3.4E-12&   1.8E-13&      0.04&      0.08\\
              UY Cha&      2.95&      0.09&   4.0E-12&   7.8E-14&      0.65&      0.04&      2.17&      0.30&   9.0E-13&   7.8E-14&      0.40&      0.10\\
2M J11065939-7530559&      1.55&      0.07&   1.0E-13&   4.0E-15&      0.63&      0.06&      3.09&      0.33&   8.0E-14&   4.0E-15&      0.19&      0.08\\
2M J11070369-7724307&   \nodata&   \nodata&   \nodata&   \nodata&   \nodata&   \nodata&   \nodata&   \nodata&   \nodata&   \nodata&   \nodata&   \nodata\\
              ISO 91&      6.93&      0.14&   5.1E-12&   4.4E-14&      0.35&      0.01&      4.76&      0.35&   3.1E-12&   4.4E-14&      0.03&      0.07\\
              UZ Cha&      2.15&      0.08&   4.3E-12&   1.2E-13&      0.61&      0.04&      3.82&      0.35&   2.6E-12&   1.2E-13&     -0.14&      0.07\\
            CHXR 22E&   \nodata&   \nodata&   \nodata&   \nodata&   \nodata&   \nodata&   \nodata&   \nodata&   \nodata&   \nodata&   \nodata&   \nodata\\
             Cha Ha1&      2.01&      0.08&   5.3E-13&   1.6E-14&      0.58&      0.04&      1.95&      0.31&   3.2E-13&   1.6E-14&     -0.03&      0.07\\
                 T25&   \nodata&   \nodata&   \nodata&   \nodata&   \nodata&   \nodata&      3.39&      0.33&   1.6E-12&   2.2E-14&      0.15&      0.07\\
              DI Cha&      1.87&      0.08&   9.2E-11&   2.7E-12&      0.60&      0.04&      1.98&      0.24&   3.6E-11&   2.7E-12&      0.01&      0.06\\
              VV Cha&      2.29&      0.08&   5.0E-12&   1.2E-13&      0.45&      0.03&      2.05&      0.31&   1.6E-12&   1.2E-13&      0.06&      0.08\\
             Cha Ha2&      1.08&      0.25&   8.4E-13&   1.9E-13&      0.72&      0.31&      1.83&      0.51&   3.5E-13&   1.9E-13&      0.16&      0.26\\
                 T28&      2.07&      0.08&   1.6E-11&   4.2E-13&      0.51&      0.04&      1.56&      0.29&   4.2E-12&   4.2E-13&      0.25&      0.08\\
          CHSM 10862&   \nodata&   \nodata&   \nodata&   \nodata&   \nodata&   \nodata&      1.68&      0.30&   1.1E-13&   1.1E-14&      0.14&      0.10\\
            CHXR 30B&      2.88&      0.09&   1.4E-11&   2.9E-13&      0.61&      0.03&      3.29&      0.32&   6.5E-12&   2.9E-13&      0.23&      0.08\\
                 T29&      1.74&      0.07&   1.6E-10&   5.0E-12&      0.49&      0.04&      2.41&      0.24&   9.7E-11&   5.0E-12&     -0.01&      0.06\\
            CHXR 30A&      2.25&      0.08&   4.7E-12&   1.2E-13&      0.40&      0.03&      2.19&      0.30&   1.5E-12&   1.2E-13&      0.04&      0.07\\
              VW Cha&      2.42&      0.08&   5.6E-11&   1.3E-12&      0.60&      0.04&      2.45&      0.31&   2.5E-11&   1.3E-12&      0.16&      0.07\\
              CU Cha&   \nodata&   \nodata&   \nodata&   \nodata&   \nodata&   \nodata&   \nodata&   \nodata&   \nodata&   \nodata&   \nodata&   \nodata\\
                T33A&      5.25&      0.12&   6.7E-10&   7.1E-12&      0.44&      0.02&      3.79&      0.27&   2.3E-10&   7.1E-12&     -0.00&      0.06\\
             ISO 138&   \nodata&   \nodata&   \nodata&   \nodata&   \nodata&   \nodata&   \nodata&   \nodata&   \nodata&   \nodata&   \nodata&   \nodata\\
             ISO 143&      1.69&      0.07&   1.2E-12&   3.9E-14&      0.63&      0.05&      1.02&      0.28&   2.1E-13&   3.9E-14&      0.08&      0.09\\
                 T35&   \nodata&   \nodata&   \nodata&   \nodata&   \nodata&   \nodata&      5.87&      0.39&   2.5E-12&   6.8E-14&      0.74&      0.09\\
              VY Cha&      2.61&      0.09&   6.7E-12&   1.4E-13&      0.53&      0.03&      3.49&      0.33&   3.7E-12&   1.4E-13&      0.20&      0.08\\
                C1-6&      1.83&      0.08&   7.3E-11&   2.2E-12&      0.63&      0.05&      1.87&      0.24&   3.2E-11&   2.2E-12&      0.02&      0.06\\
              VZ Cha&      1.26&      0.07&   1.1E-11&   4.9E-13&      0.57&      0.05&      1.23&      0.28&   3.0E-12&   4.9E-13&     -0.03&      0.07\\
                C7-1&      0.95&      0.07&   5.0E-13&   2.8E-14&      0.53&      0.08&      4.97&      0.37&   4.7E-13&   2.8E-14&      0.01&      0.07\\
              Hn 10E&      3.20&      0.09&   8.3E-12&   1.5E-13&      0.39&      0.02&      1.43&      0.28&   3.0E-12&   1.5E-13&     -0.13&      0.07\\
                 B43&      7.19&      0.14&   7.6E-12&   6.2E-14&      0.48&      0.02&      3.18&      0.32&   2.4E-12&   6.2E-14&      0.11&      0.07\\
            HD 97300&   \nodata&   \nodata&   \nodata&   \nodata&   \nodata&   \nodata&   \nodata&   \nodata&   \nodata&   \nodata&   \nodata&   \nodata\\
             ISO 220&      1.35&      0.07&   2.9E-13&   1.3E-14&      0.62&      0.06&      1.28&      0.31&   9.0E-14&   1.3E-14&      0.21&      0.14\\
                 T42&      1.22&      0.07&   1.8E-10&   8.0E-12&      0.81&      0.08&      2.38&      0.24&   1.9E-10&   8.0E-12&     -0.02&      0.06\\
                 T43&      2.00&      0.08&   9.0E-12&   2.5E-13&      0.45&      0.03&      1.78&      0.29&   3.1E-12&   2.5E-13&      0.08&      0.07\\
              WX Cha&      2.49&      0.08&   2.9E-11&   6.5E-13&      0.60&      0.04&      1.89&      0.29&   5.9E-12&   6.5E-13&     -0.07&      0.07\\
              WW Cha&      4.03&      0.10&   4.9E-10&   6.7E-12&      0.45&      0.02&      3.63&      0.26&   2.8E-10&   6.7E-12&     -0.04&      0.06\\
               Hn 11&      2.60&      0.08&   2.2E-11&   4.5E-13&      0.54&      0.03&      3.39&      0.33&   7.8E-12&   4.5E-13&      0.20&      0.08\\
                T45a&      1.35&      0.07&   2.3E-12&   9.4E-14&      0.69&      0.07&      3.63&      0.33&   2.7E-12&   9.4E-14&      0.10&      0.08\\
              WY Cha&      1.61&      0.07&   1.3E-11&   4.4E-13&      0.62&      0.05&      2.47&      0.31&   4.4E-12&   4.4E-13&      0.14&      0.07\\
             ISO 235&      1.52&      0.07&   1.3E-12&   4.5E-14&      0.45&      0.04&      2.02&      0.31&   3.2E-13&   4.5E-14&     -0.00&      0.08\\
             ISO 237&      3.48&      0.10&   2.2E-11&   3.6E-13&      0.44&      0.02&      2.01&      0.30&   9.7E-12&   3.6E-13&     -0.06&      0.07\\
             CHXR 47&      2.18&      0.08&   1.4E-11&   3.5E-13&      0.63&      0.04&      2.82&      0.32&   5.5E-12&   3.5E-13&      0.16&      0.08\\
             ISO 252&      1.38&      0.07&   3.8E-13&   1.5E-14&      0.59&      0.06&      1.78&      0.31&   1.3E-13&   1.5E-14&      0.06&      0.10\\
                 T47&      2.16&      0.08&   1.5E-11&   3.8E-13&      0.40&      0.03&      1.51&      0.29&   5.2E-12&   3.8E-13&     -0.00&      0.07\\
              WZ Cha&      1.88&      0.08&   3.9E-12&   1.2E-13&      0.49&      0.04&      1.57&      0.29&   1.1E-12&   1.2E-13&      0.00&      0.08\\
             ISO 256&      3.30&      0.09&   6.1E-12&   1.0E-13&      0.40&      0.02&      3.31&      0.33&   2.0E-12&   1.0E-13&      0.03&      0.07\\
               Hn 13&      1.25&      0.07&   1.5E-12&   6.9E-14&      0.79&      0.08&      3.84&      0.35&   1.4E-12&   6.9E-14&      0.25&      0.08\\
              XX Cha&      1.68&      0.07&   6.8E-12&   2.2E-13&      0.59&      0.05&      2.57&      0.32&   3.9E-12&   2.2E-13&      0.13&      0.07\\
             ISO 282&      1.75&      0.08&   5.3E-13&   1.9E-14&      0.69&      0.05&      2.92&      0.34&   2.1E-13&   1.9E-14&      0.14&      0.09\\
                 T50&      1.95&      0.08&   1.8E-12&   6.1E-14&      0.68&      0.06&      3.73&      0.34&   1.2E-12&   6.1E-14&      0.55&      0.11\\
                 T51&      3.95&      0.10&   4.5E-11&   6.1E-13&      0.65&      0.03&      3.21&      0.32&   7.9E-12&   6.1E-13&      0.21&      0.08\\
              CV Cha&      5.57&      0.12&   1.6E-10&   1.6E-12&      0.49&      0.02&      3.38&      0.26&   4.7E-11&   1.6E-12&      0.20&      0.07\\
                 T54&   \nodata&   \nodata&   \nodata&   \nodata&   \nodata&   \nodata&   \nodata&   \nodata&   \nodata&   \nodata&   \nodata&   \nodata\\
              Hn 21W&      1.15&      0.07&   5.6E-13&   2.7E-14&      0.77&      0.08&      1.96&      0.30&   2.6E-13&   2.7E-14&      0.35&      0.10\\
                 T56&      7.10&      0.14&   7.9E-12&   7.3E-14&      0.50&      0.02&      6.80&      0.40&   4.0E-12&   7.3E-14&      0.24&      0.08\\
2M J11241186-7630425&      4.84&      0.11&   3.6E-13&   5.1E-15&      0.48&      0.02&      4.58&      0.35&   3.5E-13&   5.1E-15&      0.13&      0.07\\

\enddata
\tablecomments{Values are not listed for objects for which a satisfactory continuum fit could not be obtained}
\end{deluxetable}


\clearpage

\begin{figure}
\epsscale{0.8}
\plotone{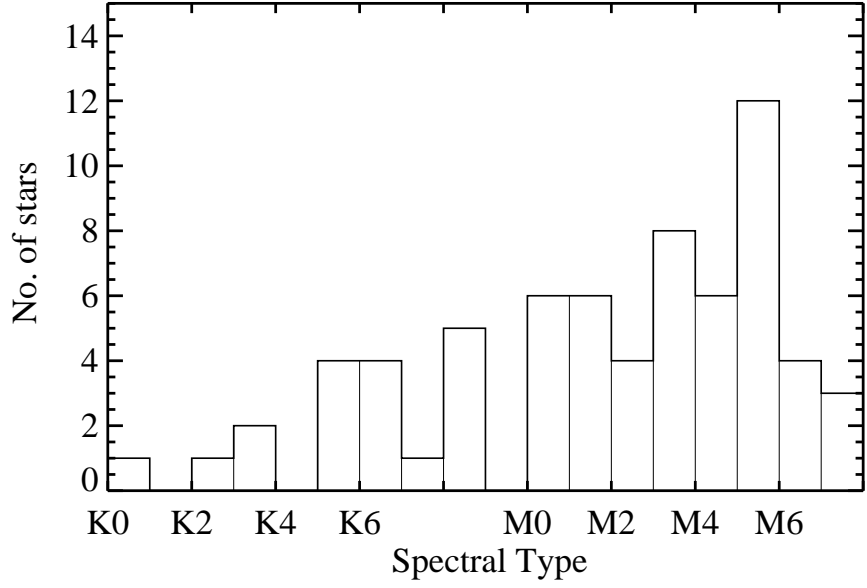}
\caption{Spectral type distribution of the Cha~I sample for stars with spectral types later than K0. \label{sp_hist_fig}}
\end{figure}

\begin{figure}
\epsscale{0.8}
\plotone{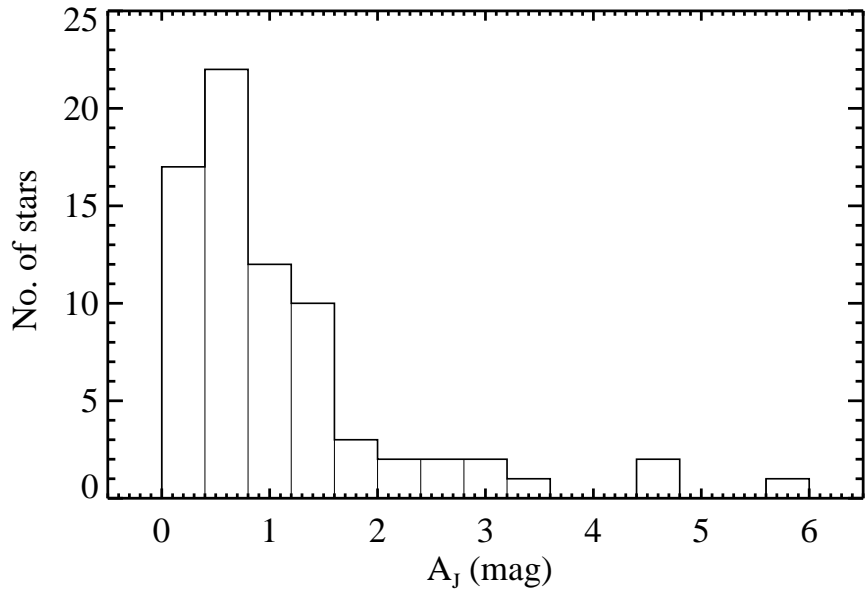}
\caption{Distribution of extinction (A$_J$) towards the young stars in
  Cha~I \label{ext_hist_fig}}
\end{figure}

\clearpage

\begin{figure}
\centering
\epsscale{1.0}
\plotone{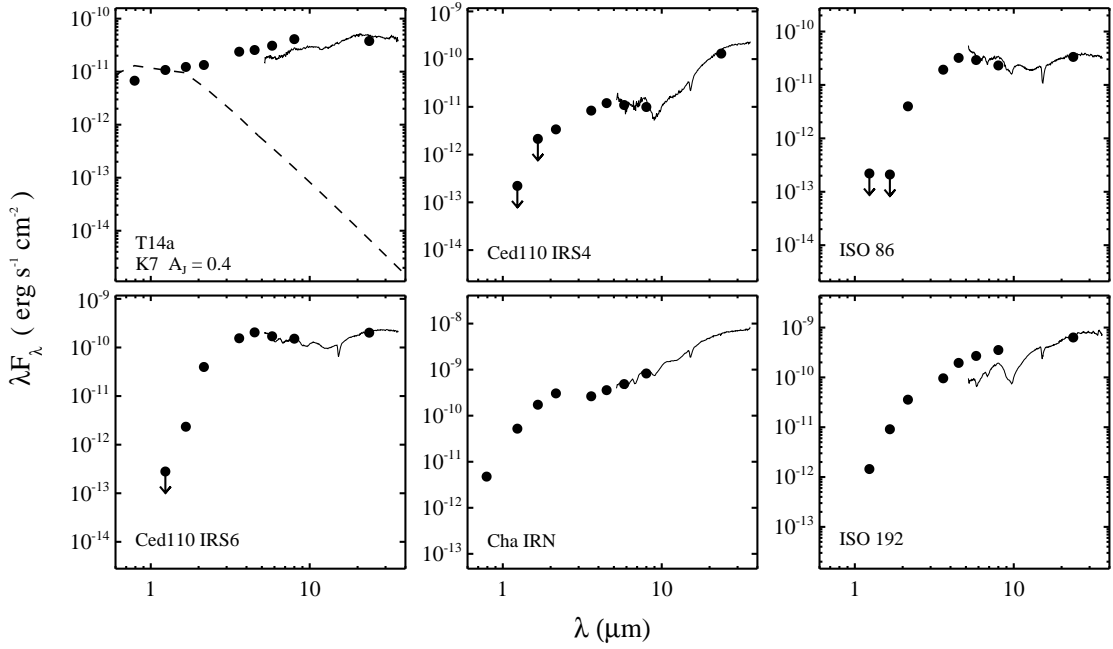}

\caption{SEDs of Class I objects in the Chamaeleon~I
  sample. Photometric (DENIS, 2MASS, IRAC and MIPS) measurements are
  shown as solid cricles and the IRS spectrum as solid line. The
  dashed line represents the stellar photosphere.}
\label{class1_sed_fig}
\end{figure}

\begin{figure}
\centering

\plotone{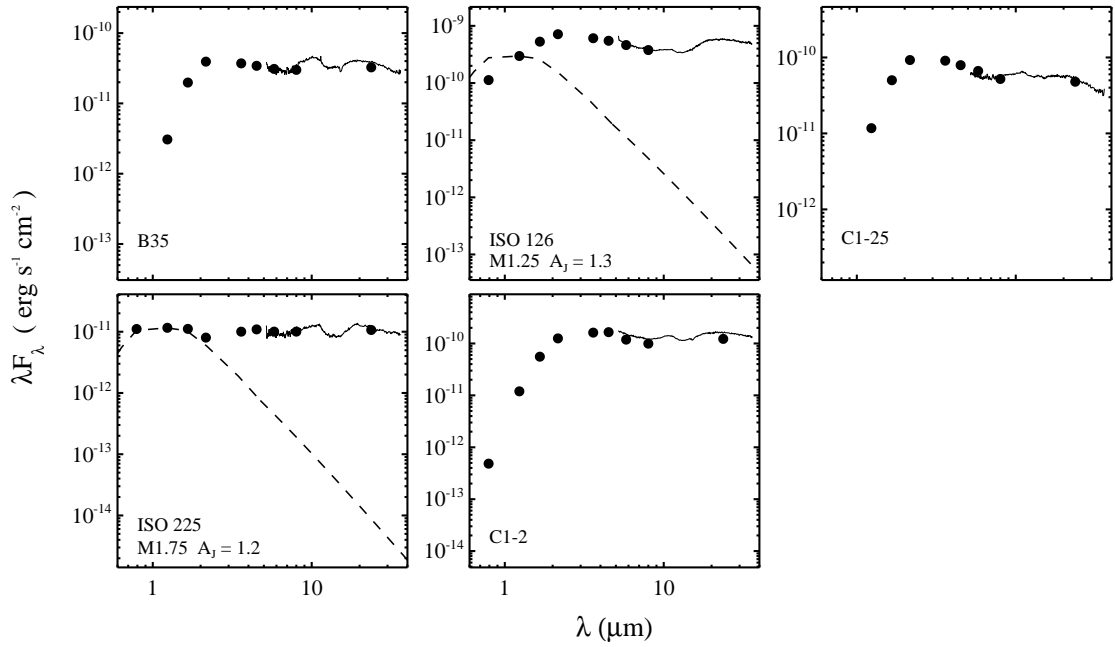}

\caption{SEDs of Flat spectrum sources in the Chamaeleon~I sample. Symbols and line types have the same meaning as in Fig.~\ref{class1_sed_fig}.}
\label{flat_sed_fig}
\end{figure}

\clearpage

\begin{figure}
\centering

\plotone{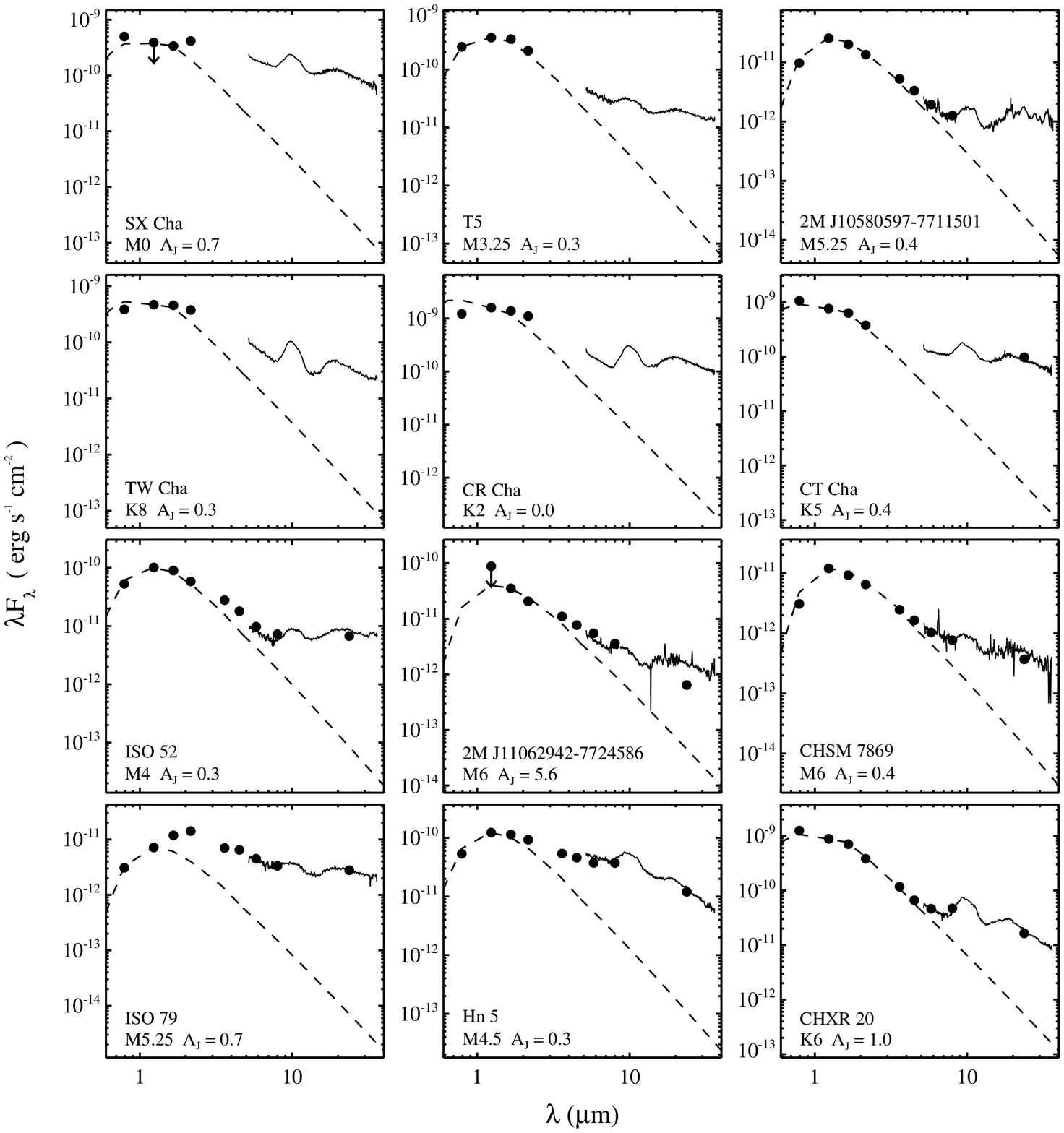}

\caption{SEDs of Class II objects in the Chamaeleon I sample. Symbols and line types have the same meaning as in Fig.~\ref{class1_sed_fig}.}
\label{class2_sed_fig}
\end{figure}

\clearpage

\begin{figure}
\centering
\figurenum{5}

\plotone{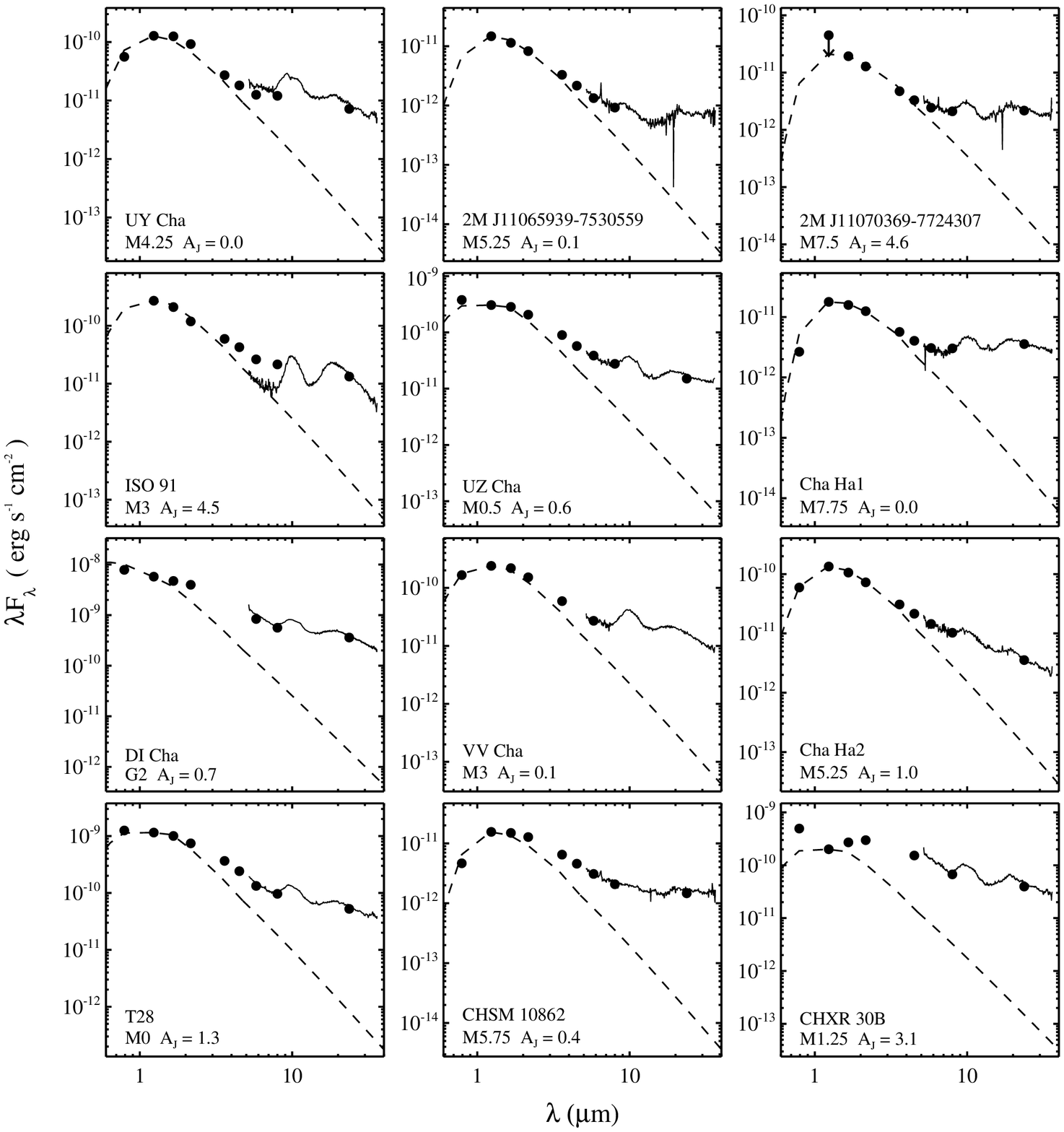}

\caption{Continued...}
\end{figure}

\clearpage

\begin{figure}
\centering
\figurenum{5}

\plotone{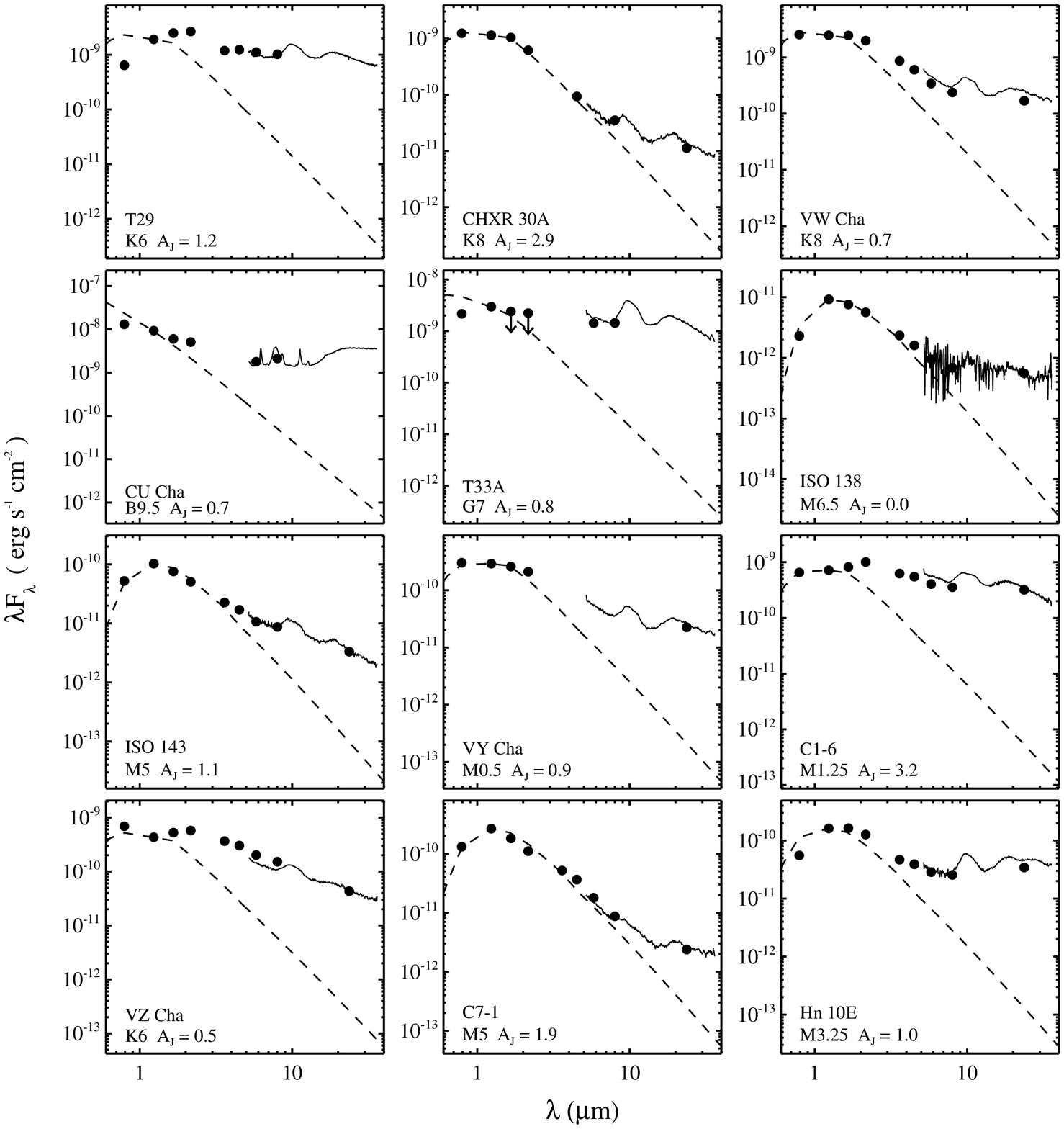}

\caption{Continued...}
\end{figure}

\clearpage

\begin{figure}
\centering
\figurenum{5}

\plotone{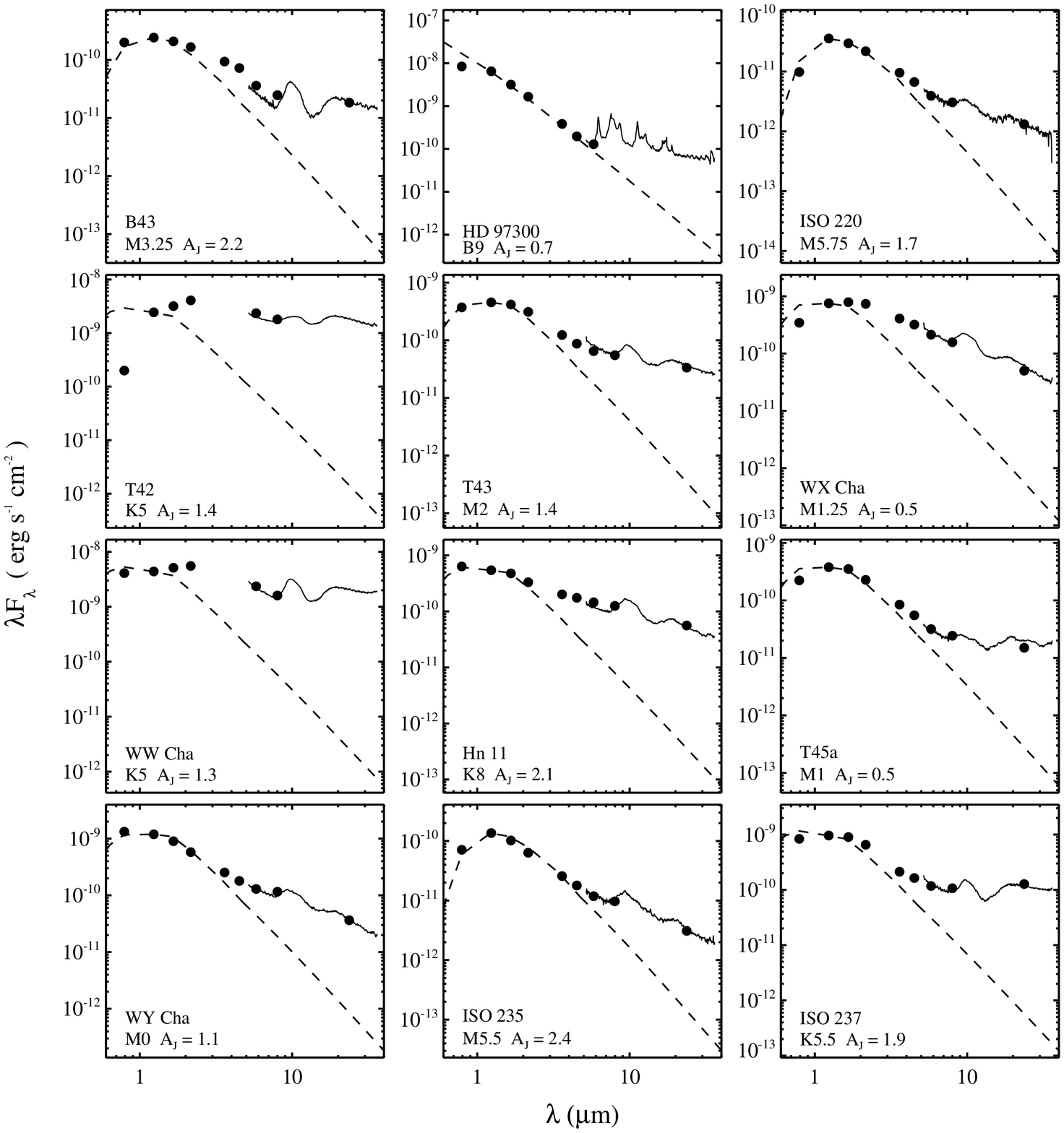}

\caption{Continued...}
\end{figure}

\begin{figure}
\centering
\figurenum{5}

\plotone{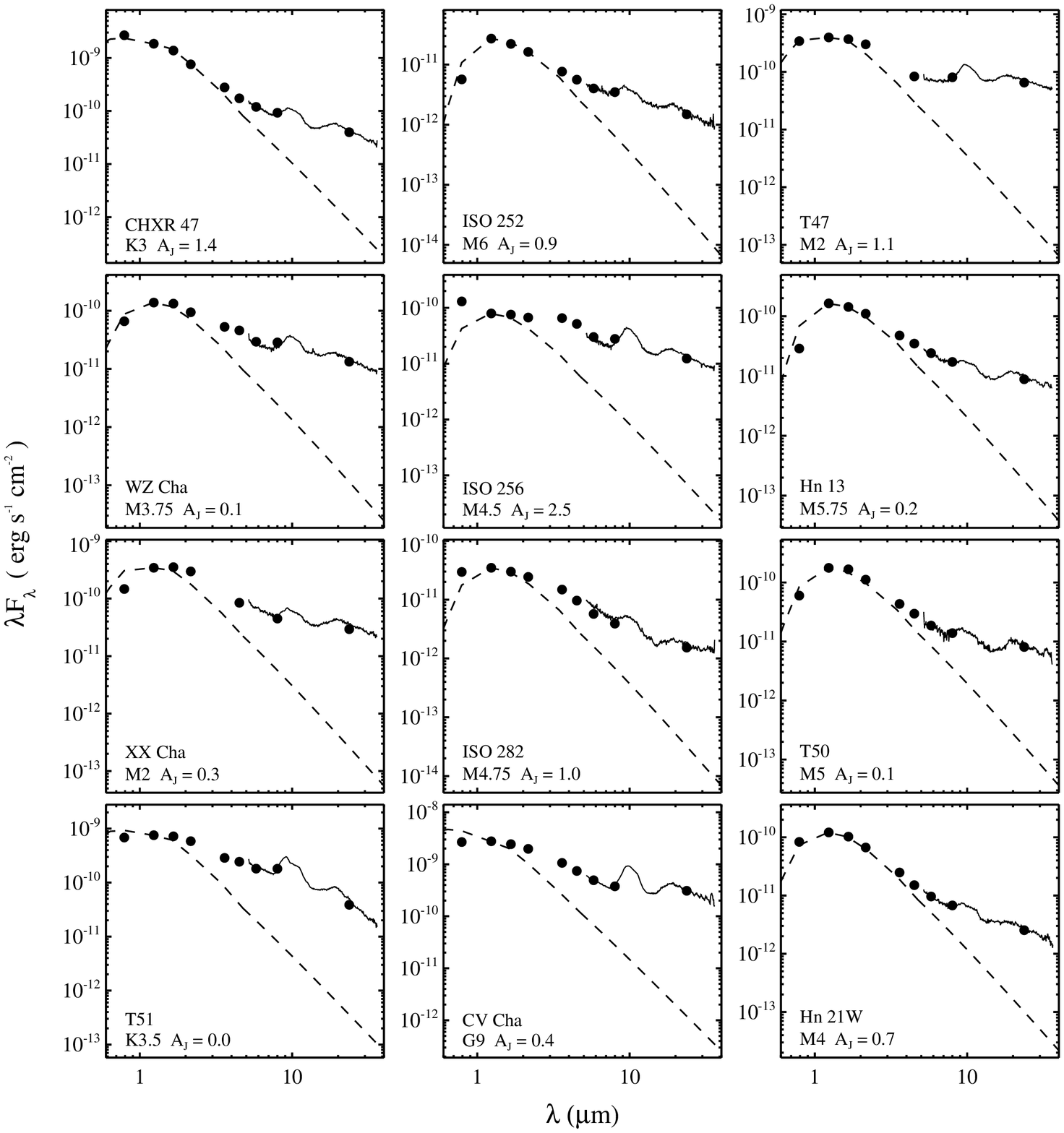}

\caption{Continued...}
\end{figure}

\clearpage

\begin{figure}
\centering

\plotone{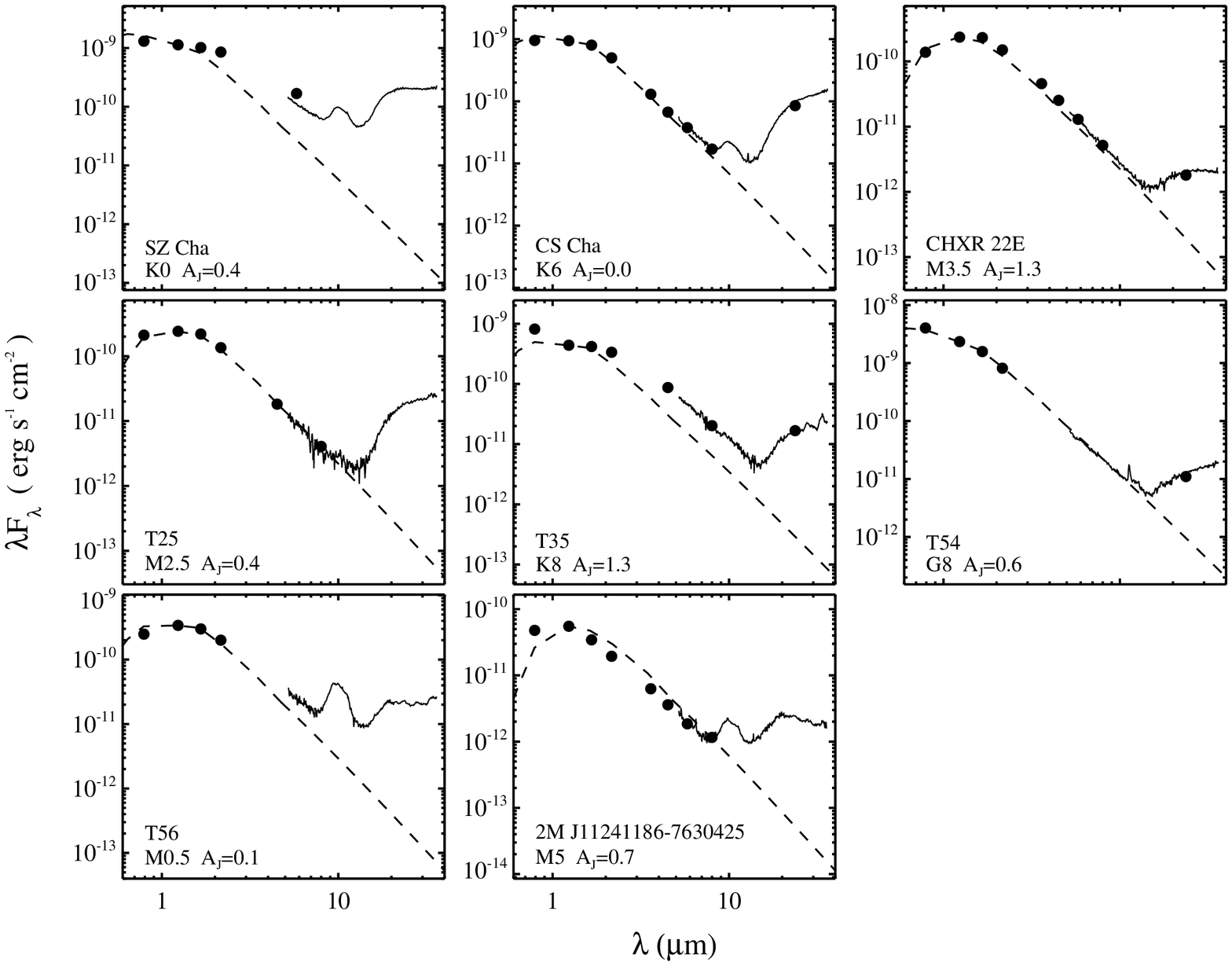}

\caption{SEDs of transitional disks in the Chamaeleon I sample. Symbols and line types have  the same meaning as in Fig.~\ref{flat_sed_fig}.}
\label{trans_sed_fig}
\end{figure}

\begin{figure}
\centering

\plotone{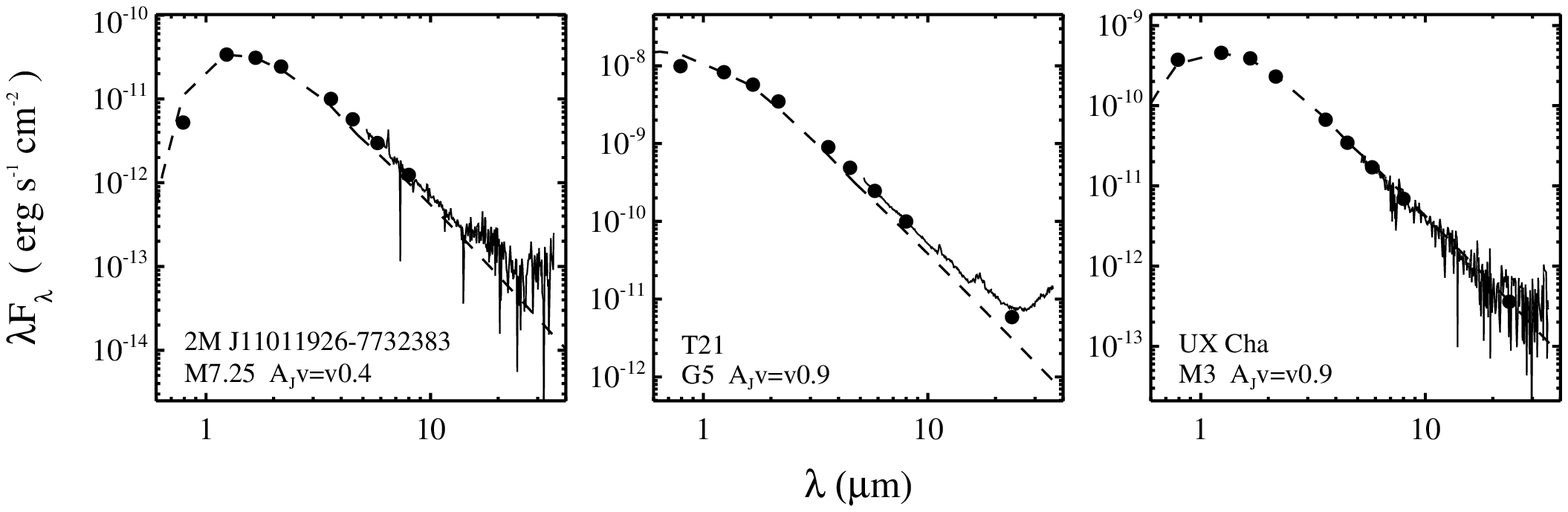}

\caption{SEDs of Class III objects in the Chamaeleon I sample. Symbols and line types have the same meaning as in Fig.~\ref{flat_sed_fig}.}
\label{class3_sed_fig}
\end{figure}

\begin{figure}
\centering

\plotone{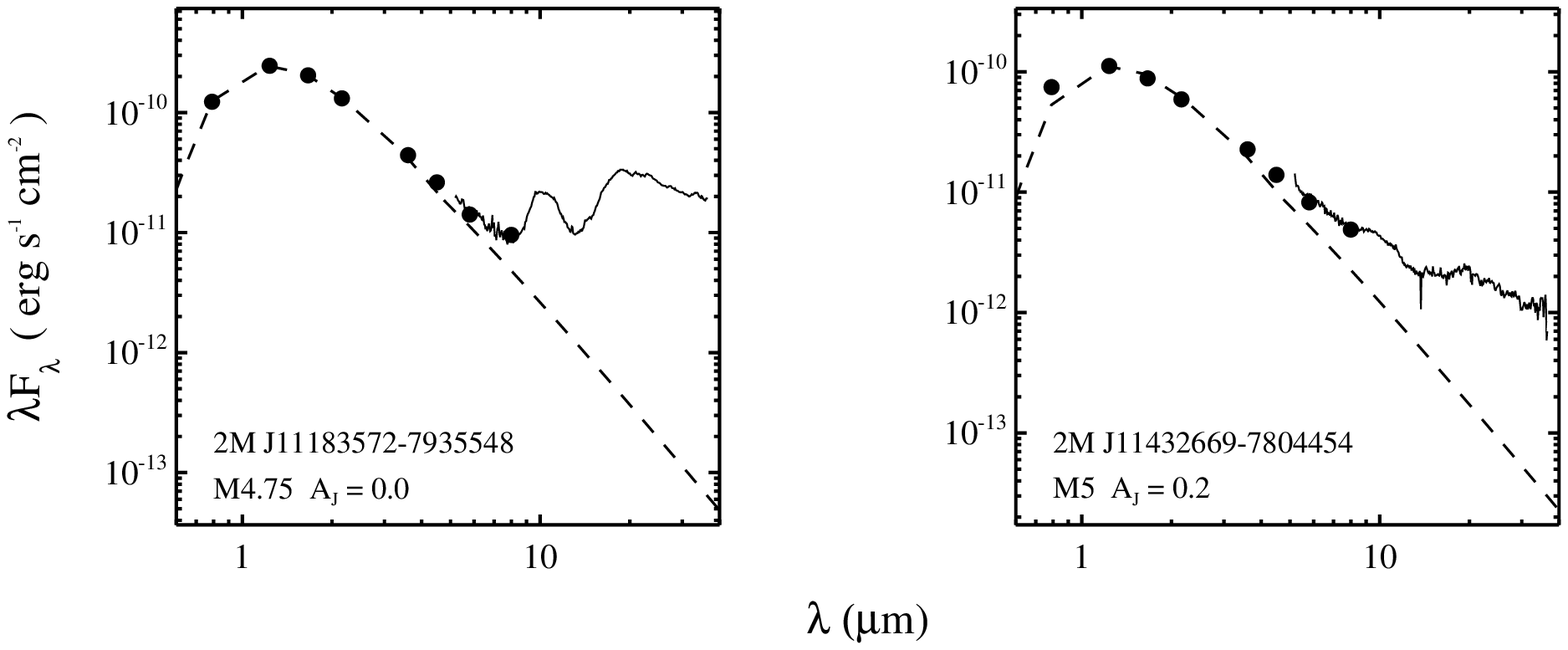}

\caption{SEDs of stars which are probable members of $\epsilon$ Cha group. Symbols and line types have the same meaning as in Fig.~\ref{flat_sed_fig}.}
\label{epscha_sed_fig}
\end{figure}

\begin{figure}
\centering

\plotone{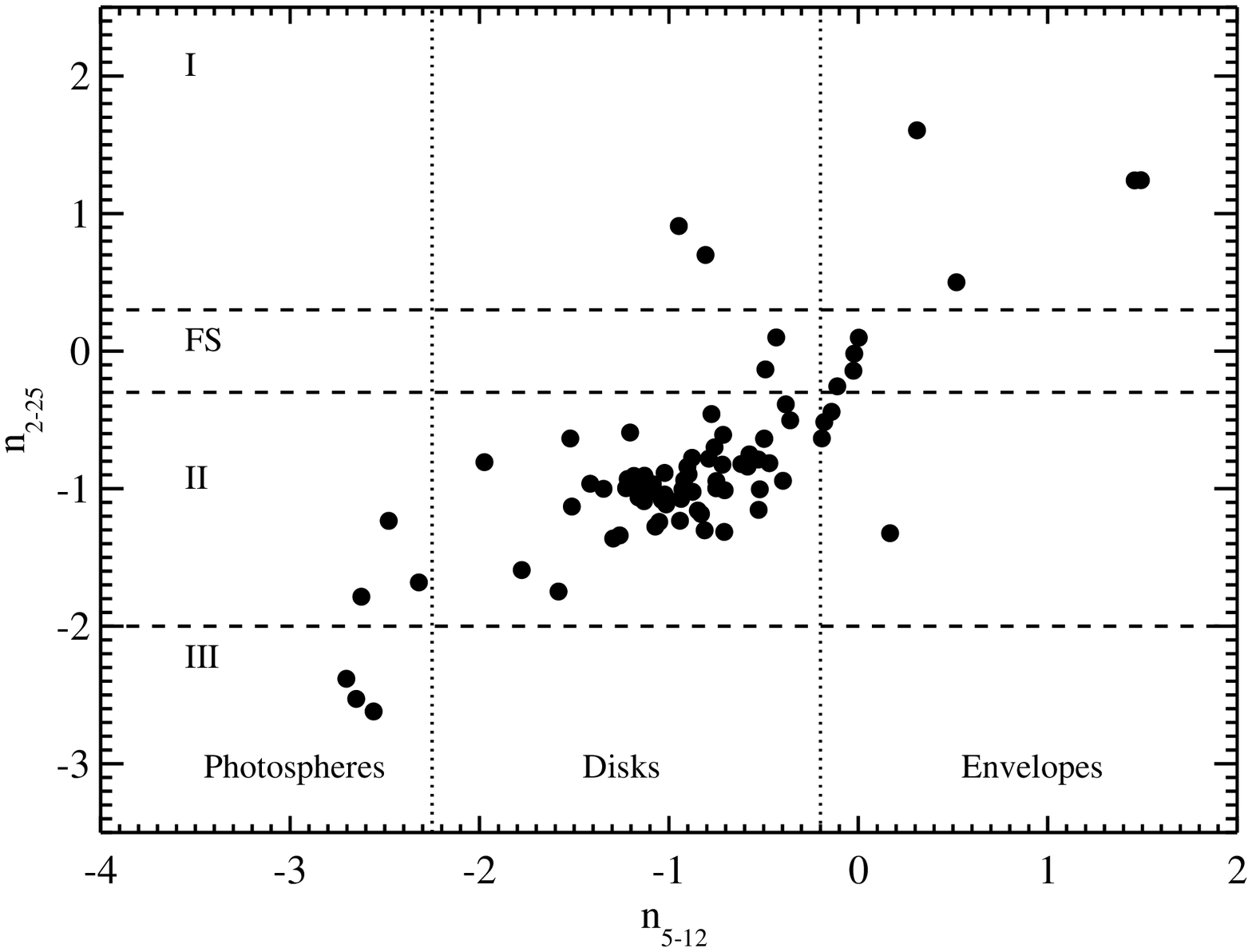}

\caption{Comparison of the observed spectral indices $n_{2-25}$ and
  $n_{5-12}$. Dashed horizontal lines indicate the regions occupied by
  Class I, Flat spectrum, Class II and Class III sources. Dotted
  vertical lines separate envelope, disk and photospheric sources.}
\label{extinct_free_fig}
\end{figure}

\clearpage

\begin{figure}
\epsscale{1.0}
\plotone{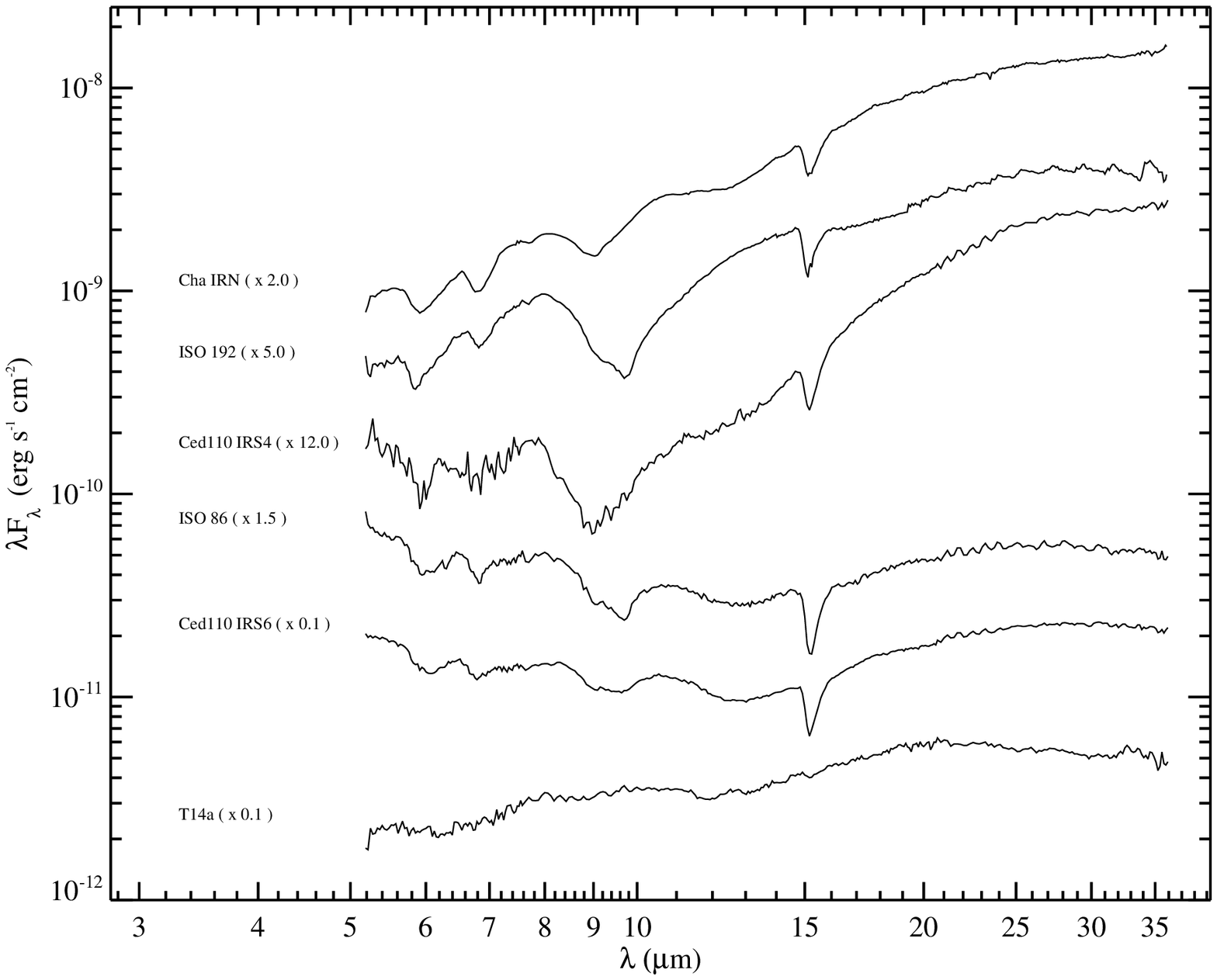}
\caption{IRS spectra of the Class I objects in our sample. For T14a,
   the dereddened spectrum is shown. Spectra have been scaled to fit
  the figure. \label{class1_irs}}
\end{figure}

\clearpage
\begin{figure}
\epsscale{1.0}
\plotone{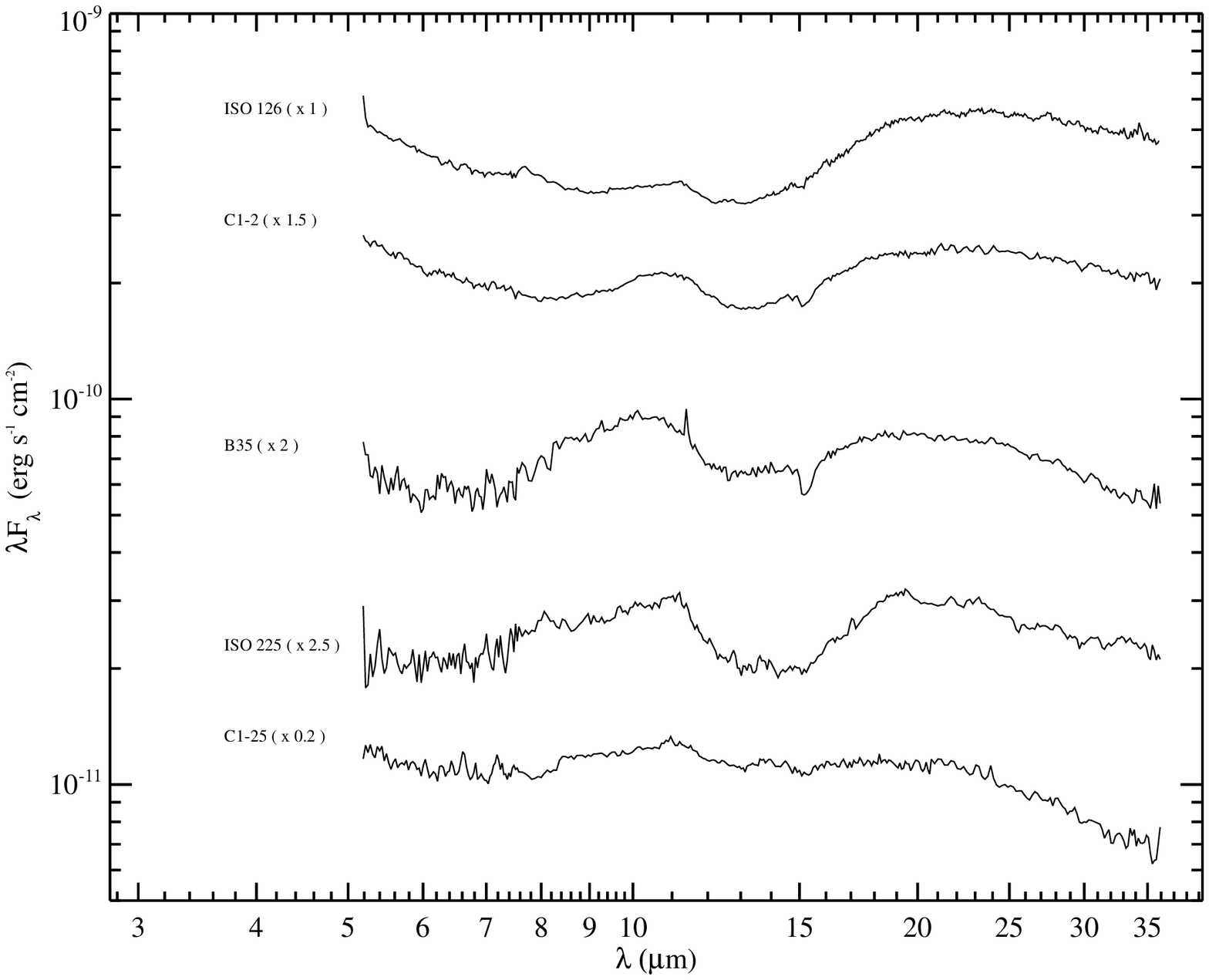}
\caption{IRS spectra of the Flat spectrum sources in our sample. Objects for which A$_J$ values are listed in Table 2, the dereddened spectra are shown. Spectra have been scaled to fit the figure. \label{flat_irs}}
\end{figure}

\clearpage
\begin{figure}
\epsscale{1.0}
\plotone{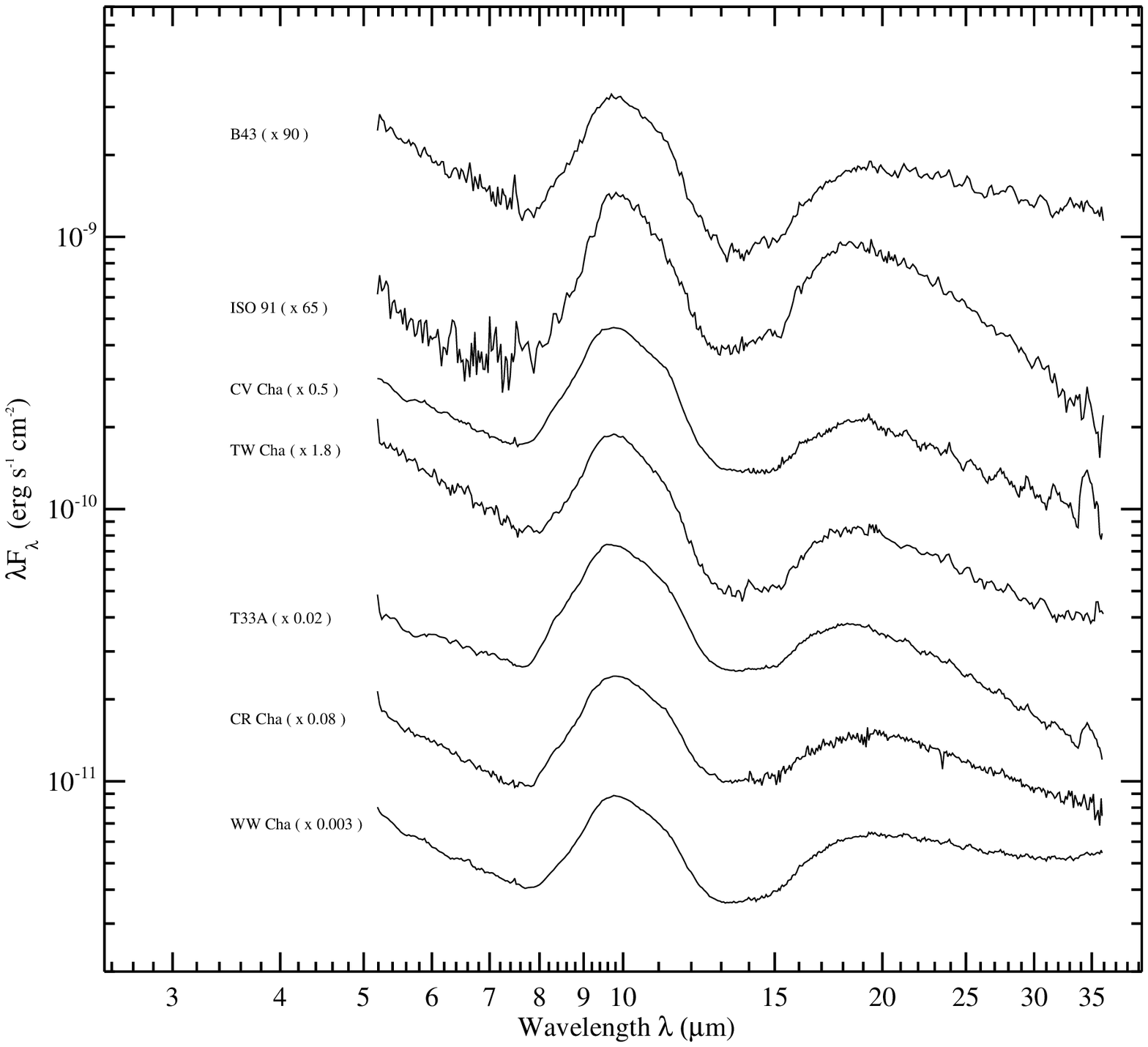}
\caption{Dereddened IRS spectra of Class II objects which show enhanced 10 $\micron$ silicate emission (W$_{10}$ $\ga$ W$_{10,upper\:octile}$). \label{grA}}
\end{figure}

\clearpage
\begin{figure}
\epsscale{1.0}
\plotone{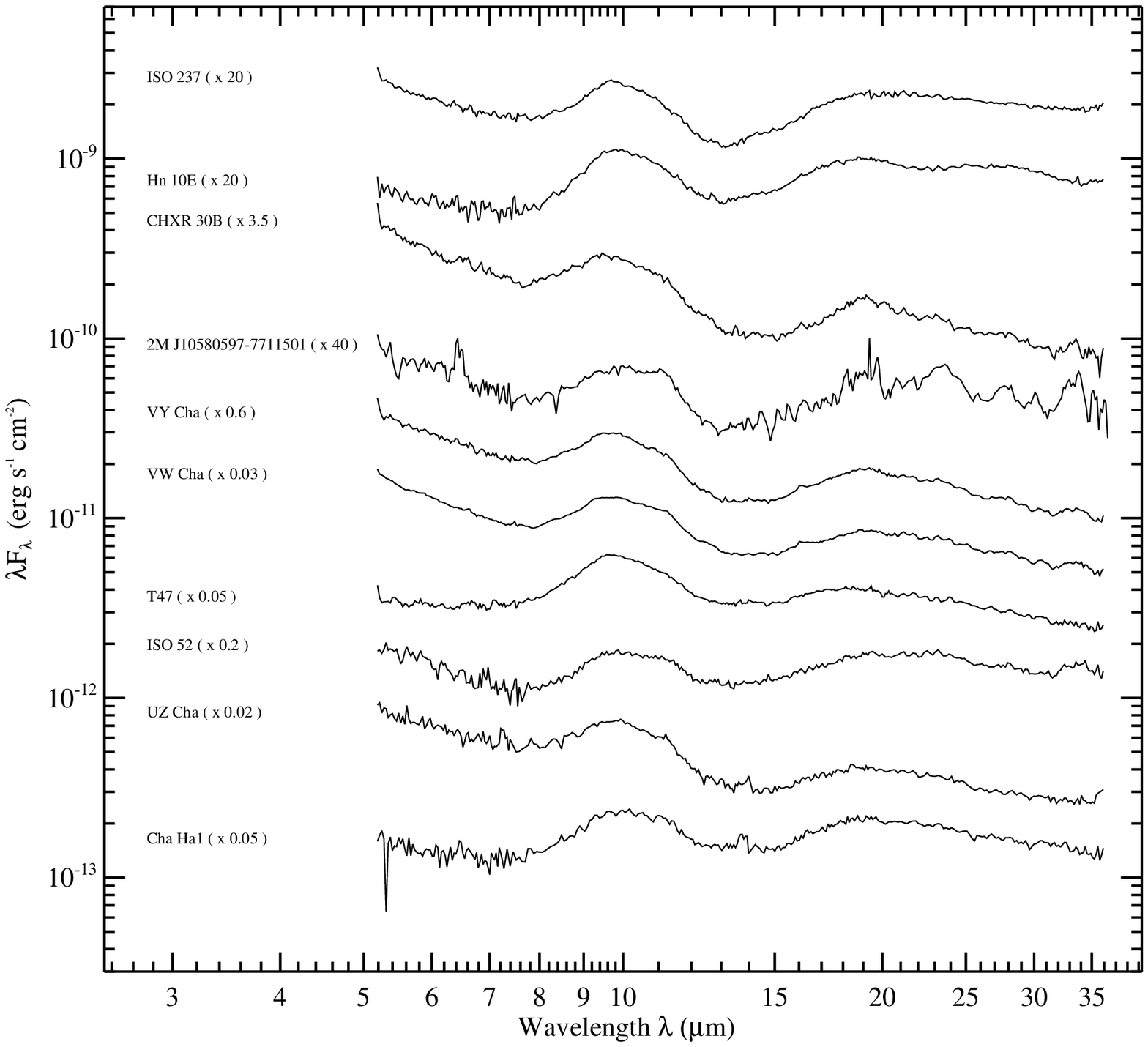}
\caption{Dereddened IRS spectra of Class II objects with moderate 10 $\micron$ feature strength and flatter continuum  \label{grB}}
\end{figure}

\clearpage
\begin{figure}
\epsscale{1.0}
\plotone{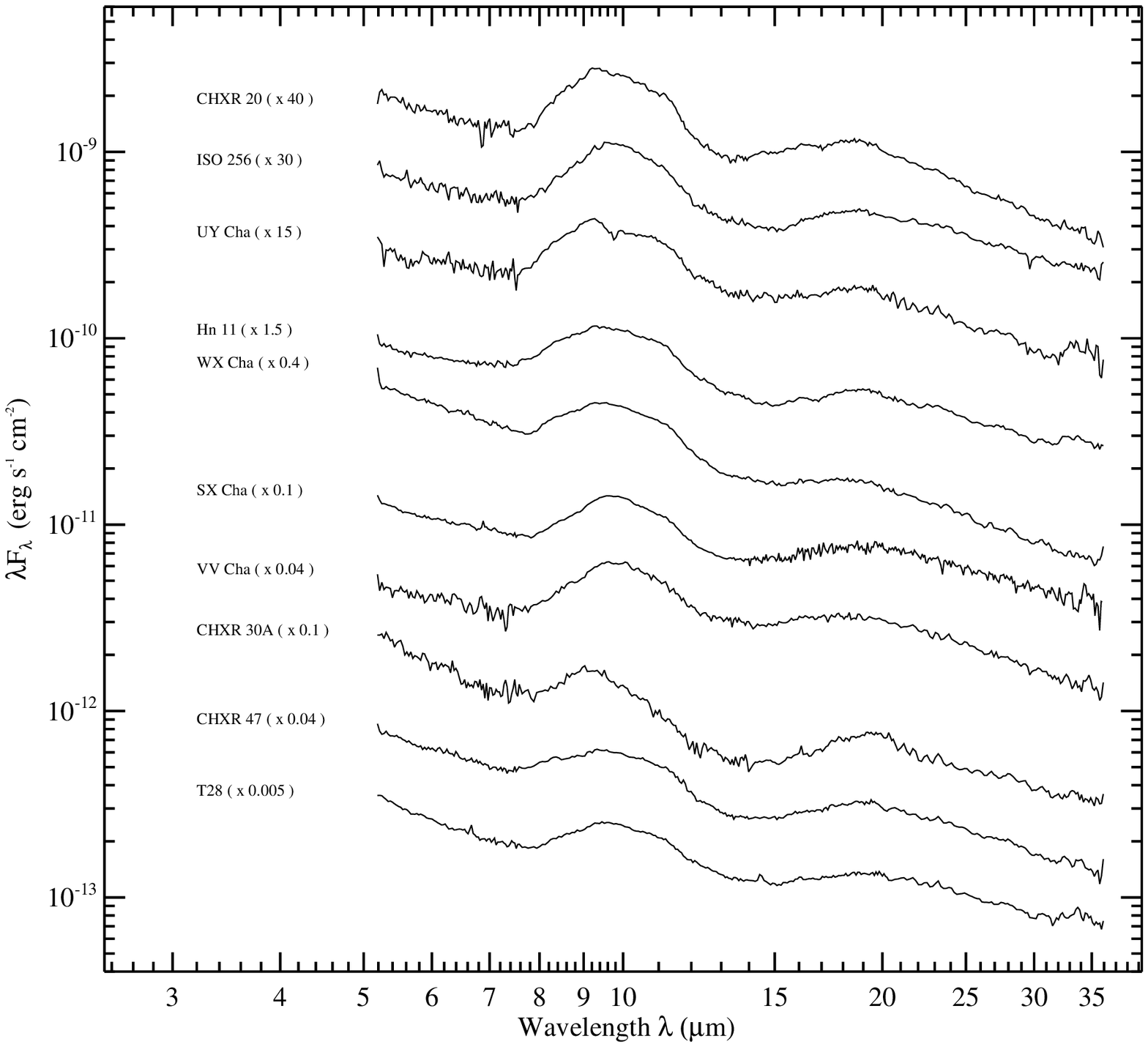}
\caption{Dereddened IRS spectra of Class II objects with moderate 10 $\micron$ feature strength and steeper (more negative) continuum slope \label{grC}}
\end{figure}

\clearpage
\begin{figure}
\epsscale{1.0}
\plotone{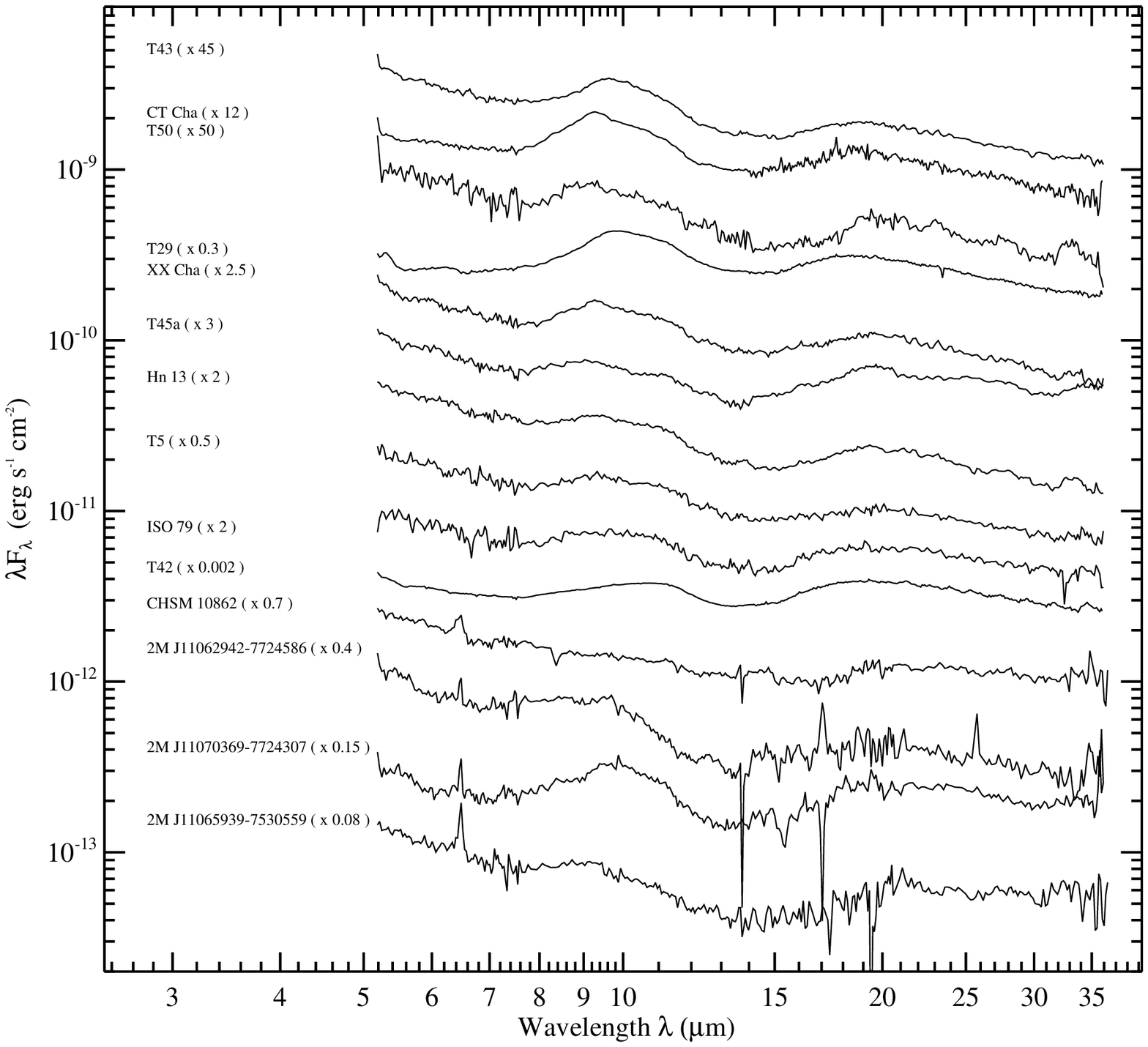}
\caption{Dereddened IRS spectra of Class II objects with weak 10 $\micron$ silicate emission  and flatter continuum \label{grD}}
\end{figure}

\clearpage
\begin{figure}
\epsscale{1.0}
\plotone{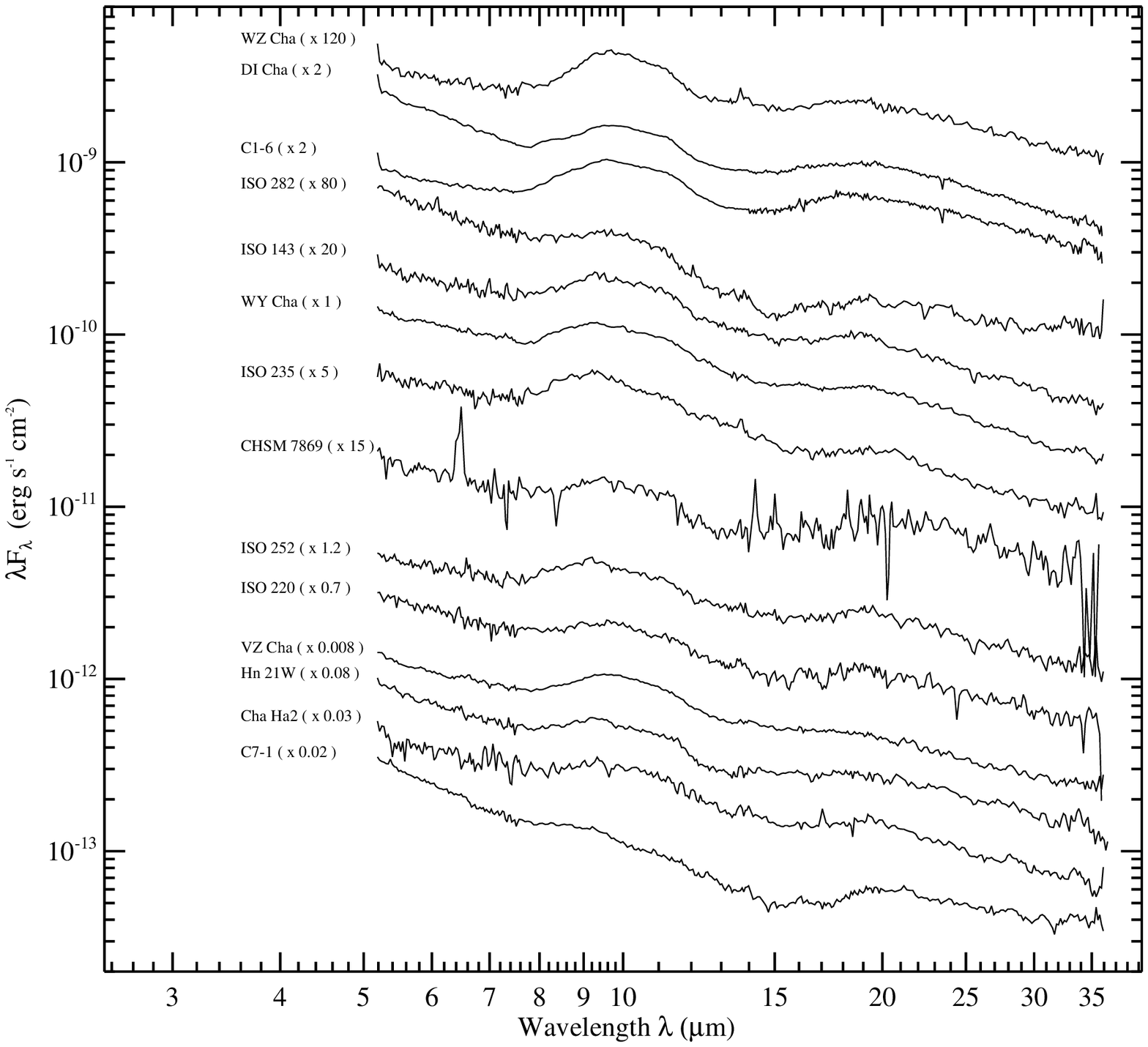}
\caption{ Dereddened IRS spectra of Class II objects with weak 10 $\micron$ silicate emission  and steeper (more negative)  continuum\label{grE}}
\end{figure}

\clearpage
\begin{figure}
\epsscale{1.0}
\plotone{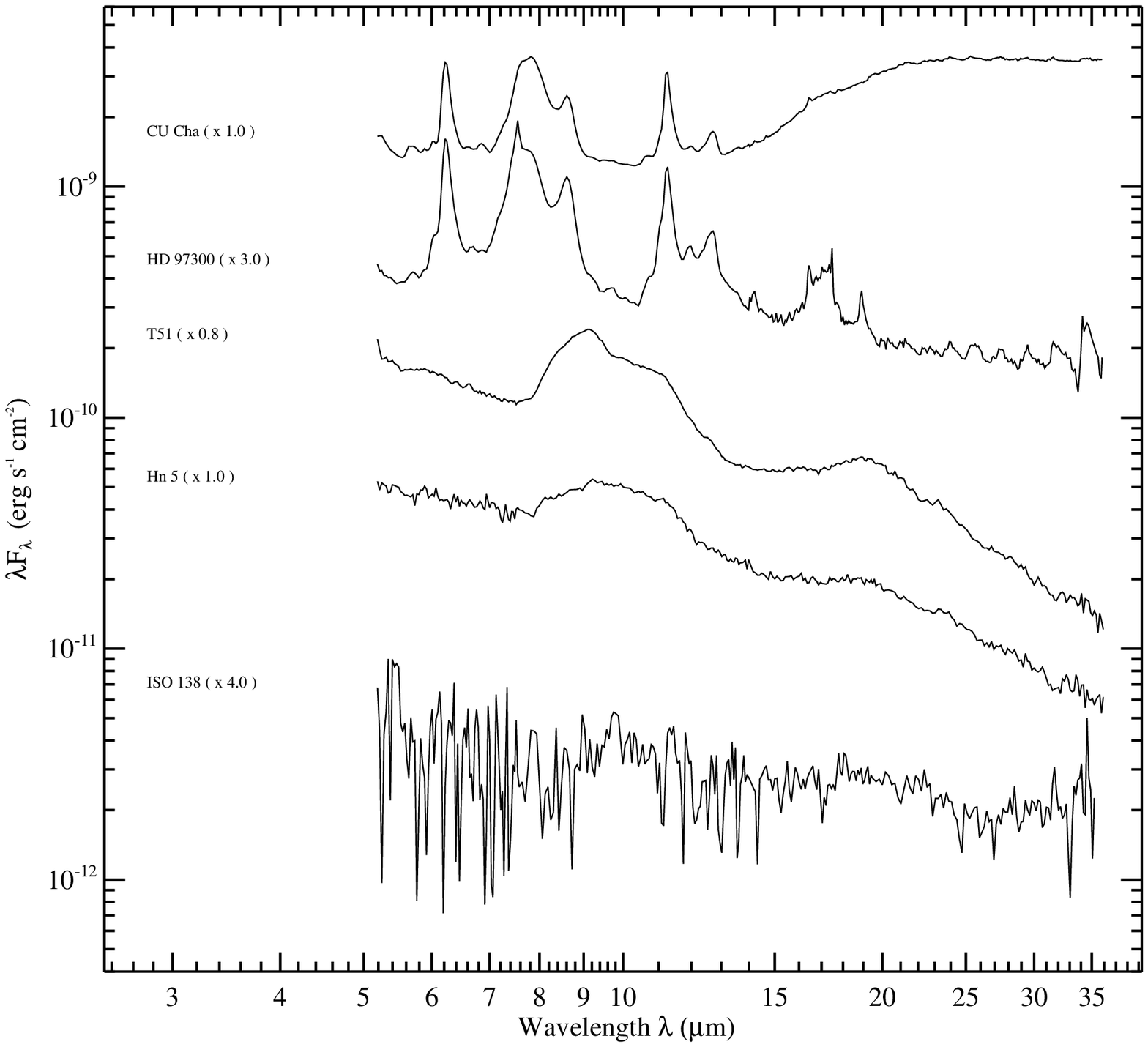}
\caption{Dereddened IRS spectra of Class II objects which could not fit in to the
  other groups. The first two are B type stars in the sample and the
  next two are outwardly truncated disks.  \label{out}}
\end{figure}

\begin{figure}
\epsscale{1.0}
\plotone{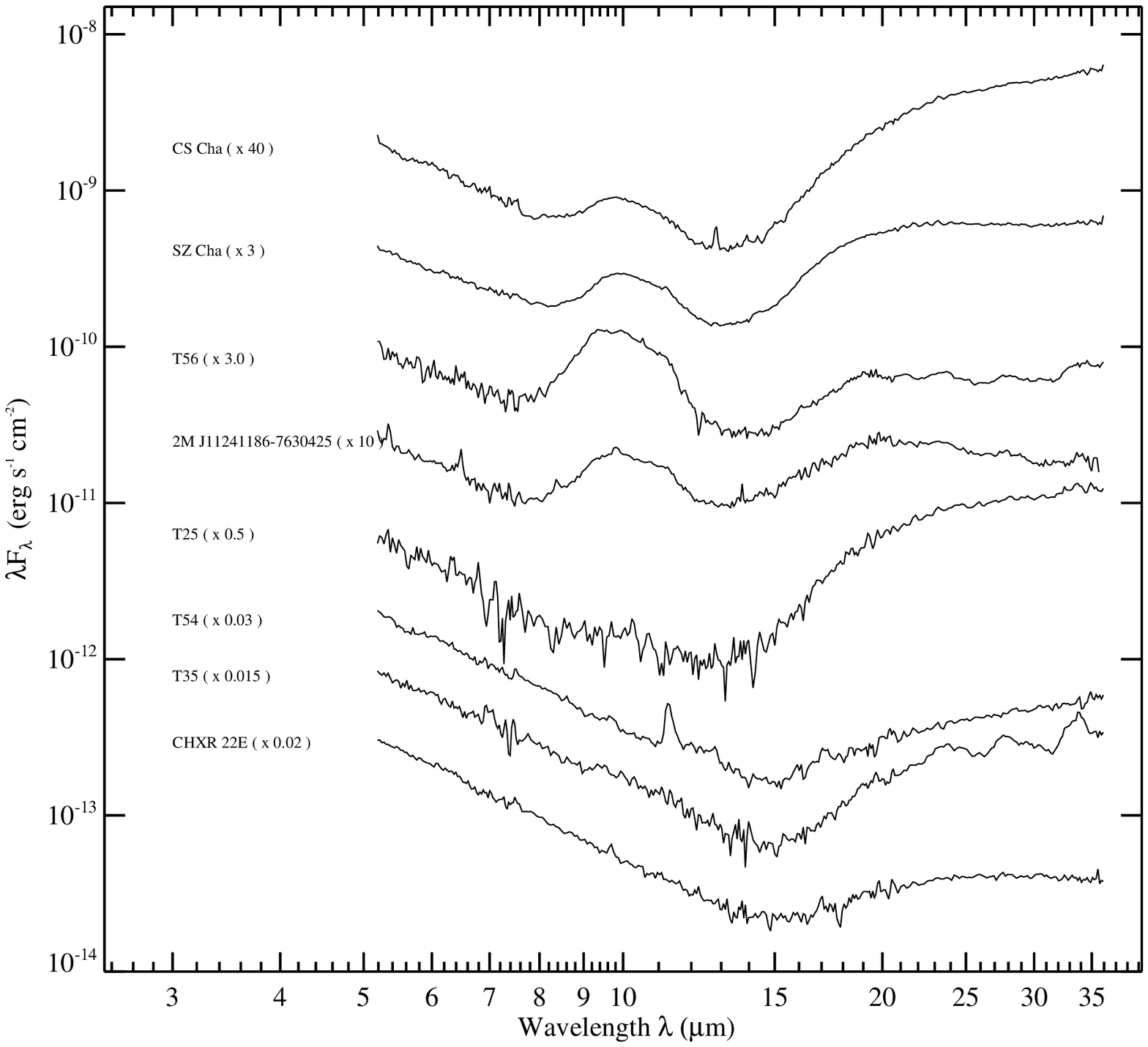}
\caption{IRS spectra of transitional disk candidates \label{trans_spec}}
\end{figure}

\clearpage 
\begin{figure}
\epsscale{1.0}
\plotone{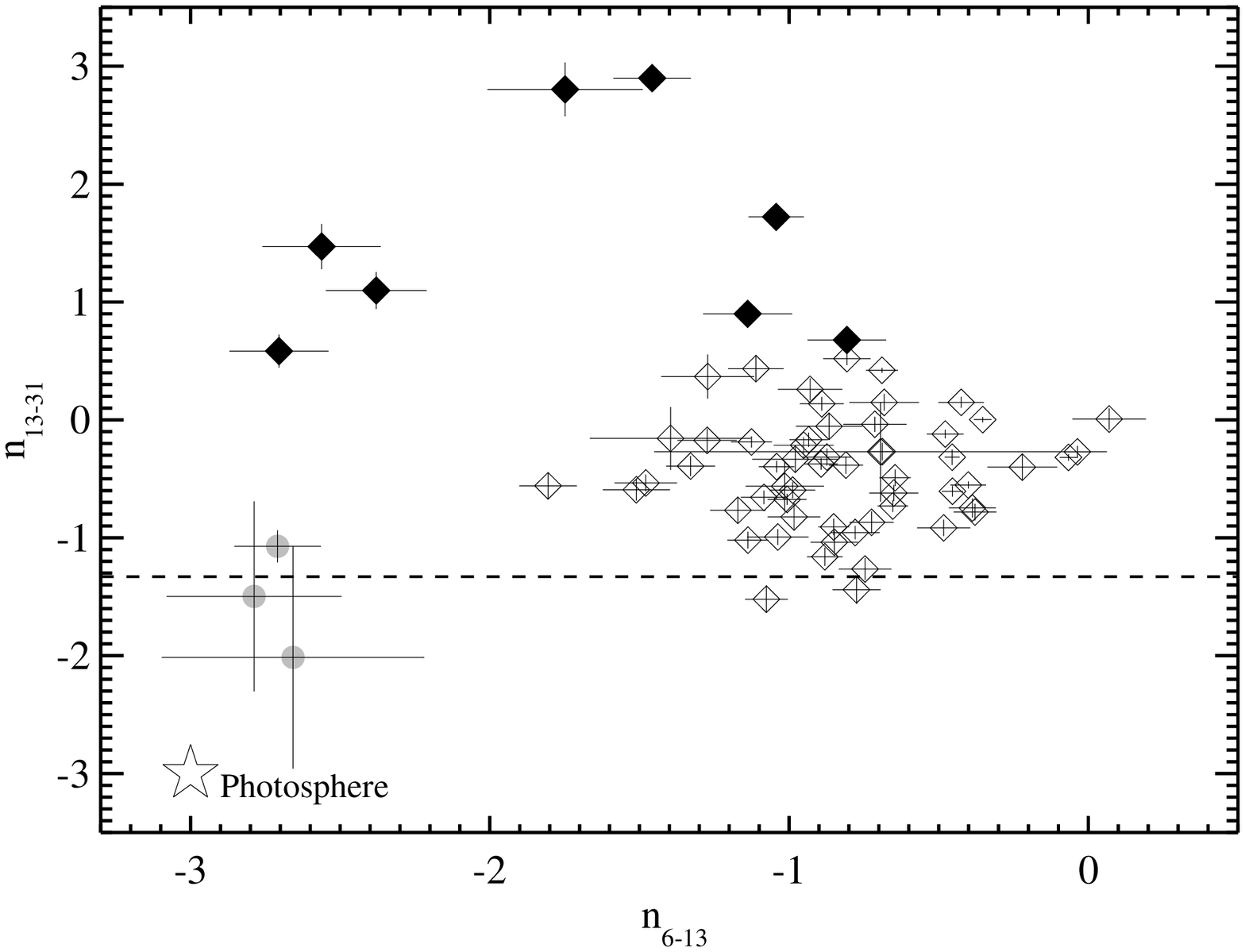}
\caption{Continuum spectral index $n_{13-31}$ plotted against the index $n_{6-13}$ (see text for definitions) for Class II {\it(open diamonds)} and Class III {\it(solid gray circles)} objects.  Transitional disk candidates are shown as {\it solid diamonds}. The star symbol represents location of the stellar photosphere ($n = -3$). The dashed line indicate $n_{13-31}$~=~$-4/3$ appropriate for a geometrically thin, optically thick, flat disk\label{fig_indices}. }
\end{figure}

\begin{figure}
\epsscale{1.0}
\plotone{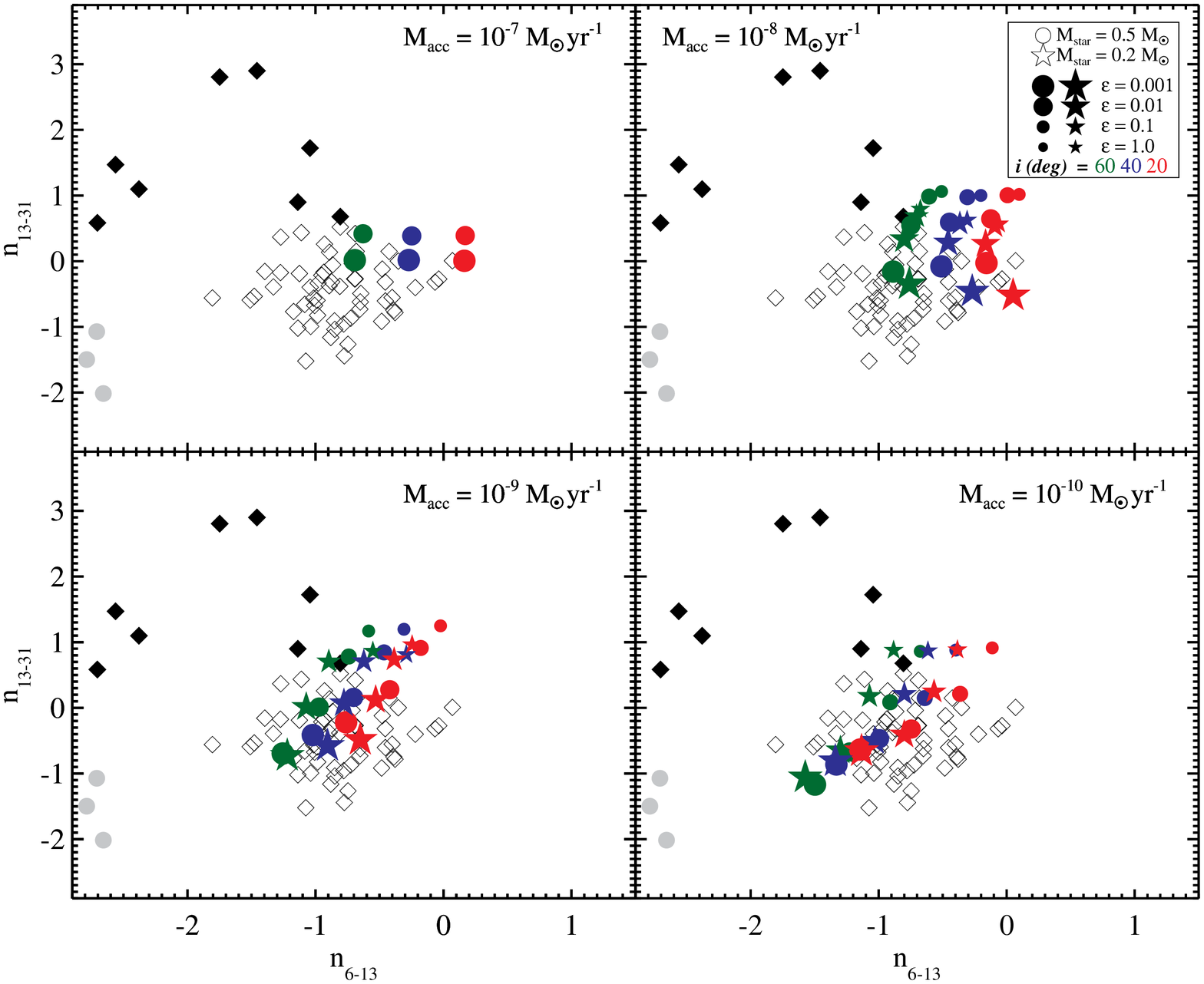}
\caption{ n$_{13-31}$ versus  n$_{6-13}$ for Cha I objects. The symbols have the same meaning as in Figure~\ref{fig_indices}. Overplotted are the  spectral indices computed from accretion disks models (see text for details) for 0.2 M$_{\odot}$ {\it(solid star symbol)} and 0.5 M$_{\odot}$ stars{\it(solid circles)}. The model indices computed for accretion rates of 10$^{-7}$ M$_{\odot}$yr$^{-1}$ {\it(top left panel)}, 10$^{-8}$ M$_{\odot}$yr$^{-1}$ {\it(top right panel)}, 10$^{-9}$ M$_{\odot}$yr$^{-1}$ {\it(bottom left panel)} and 10$^{-10}$ M$_{\odot}$yr$^{-1}$ {\it(bottom right panel)} and for inclination angles of 60$\degr$ {\it(green)}, 40$\degr$ {\it(blue)} and 20$\degr$ {\it(red)} are shown. Model indices for the settling parameter $\epsilon$ of 1, 0.1, 0.01 and 0.001 are represented by the sizes of the solid symbols from the smallest to the largest. \label{fig_index_model}}
\end{figure}

\clearpage
\begin{figure}
\epsscale{0.9}
\plotone{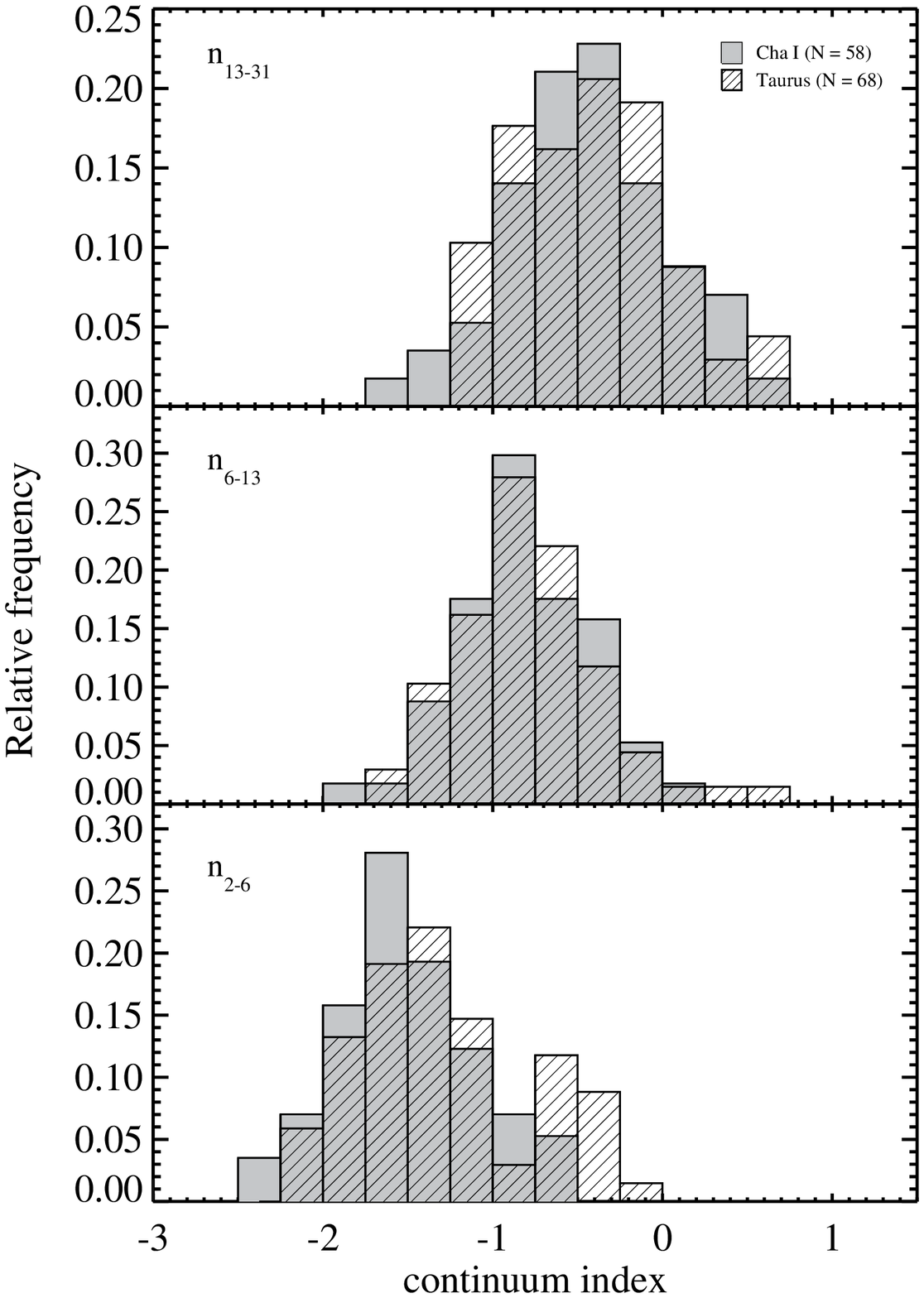}
\caption{Distribution of continuum indices for the Class II (full) disks in Cha I {\it (gray solid histogram)} and Taurus {\it (line histogram)}. \label{fig_cont_indices_hist}}
\end{figure}

\clearpage
\begin{figure}
\epsscale{0.9}
\plotone{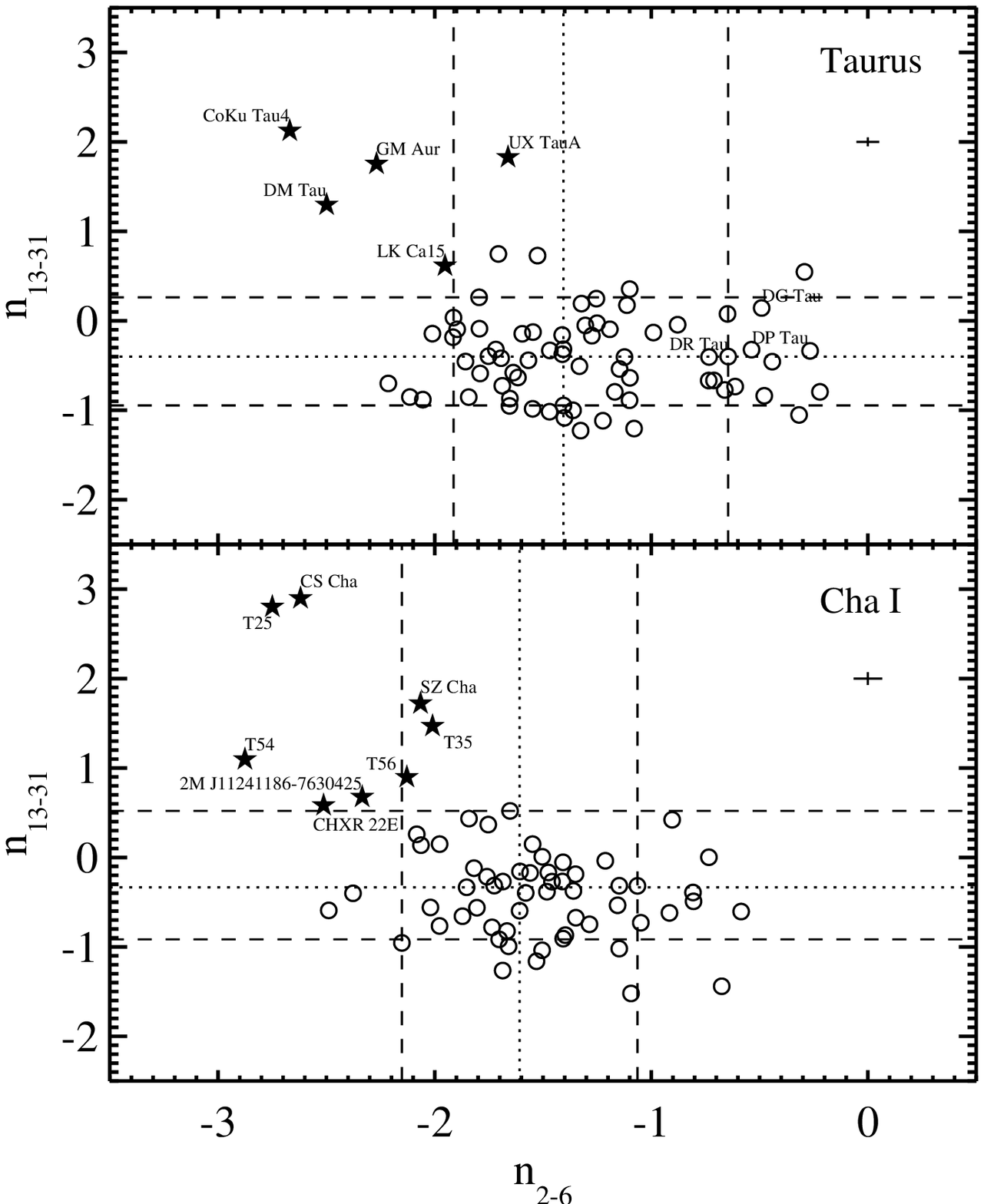}
\caption{Continuum spectral indices $n_{2-6}$ and $n_{13-31}$ plotted against each other for Class II objects in Taurus {\it(top)} and Cha I {\it(bottom)}. The dotted lines represent the median of the indices and the dashed lines represent lower and higher octiles for the two regions. Transitional disks in both regions are shown as  solid star symbols.\label{fig_tau_cha_cont_indices}}
\end{figure}

\clearpage
\begin{figure}
\epsscale{1.0}
\plotone{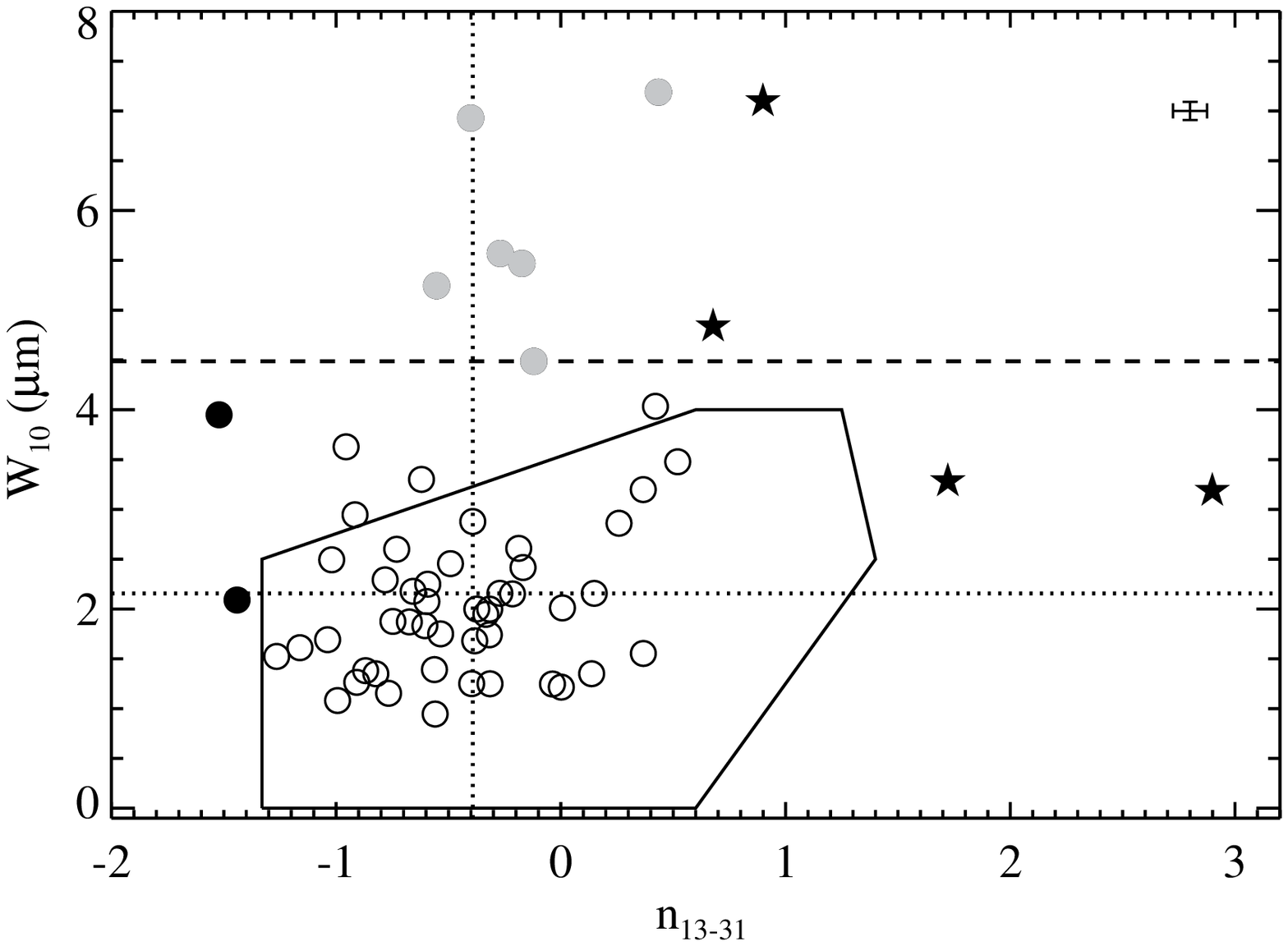}
\caption{Equivalent width of the 10 $\micron$ silicate feature,
  W$_{10}$, plotted against the continuum spectral index $n_{13-31}$
  for the Class II objects in Cha I. TDs and PTDs are shown as {\it
    solid black star symbols} and outwardly truncated disks are shown
  as {\it solid black circles}.  The dotted lines represent the median
  in W$_{10}$ and $n_{13-31}$ values and the dashed line the upper
  octile of W$_{10}$ distribution. The {\it solid gray circles}
  represents objects with enhanced 10 $\micron$ silicate emission
  (W$_{10}$ $\ga$ W$_{10,upper\:octile}$). The polygon defines the region in
  the W$_{10}$ - n$_{13-31}$ plane, allowed by the irradiated,
  accretion disk models for typical range of accretion rates, stellar
  masses, inclination angles and settling parameters $\epsilon$ (see
  text). Typical errors are shown on the top-right corner (see
  Table~\ref{tab5_index}).
\label{gr_ew10_index}}
\end{figure}

\clearpage
\begin{figure}
\epsscale{1.0}
\plotone{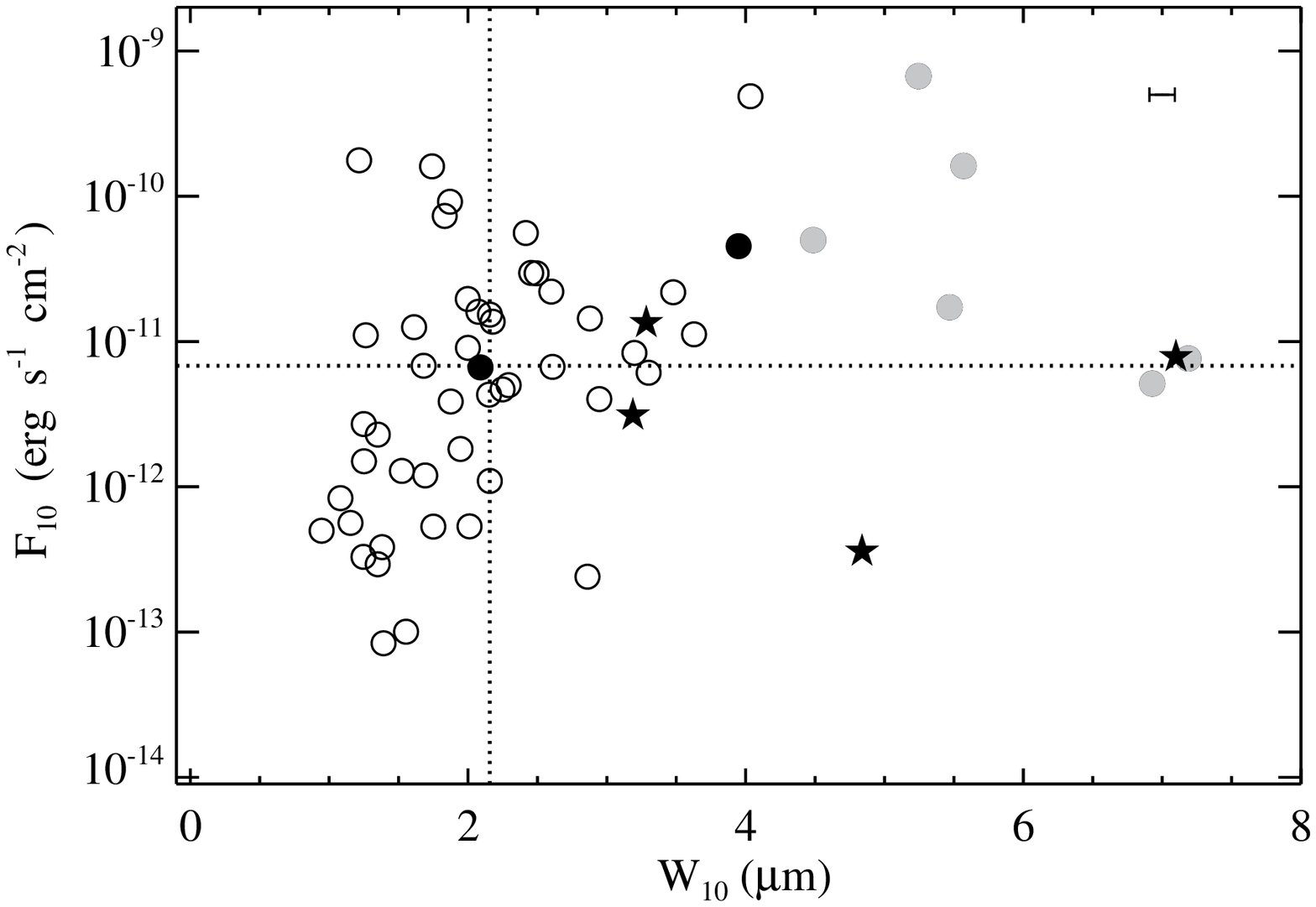}
\caption{Integrated flux of the 10 $\micron$ silicate feature F$_{10}$ plotted against the equivalent width W$_{10}$. The symbols have same meaning as in Figure~\ref{gr_ew10_index}.  Typical errors are shown on the top-right corner (see
  Table~\ref{tab5_index}). \label{flx10_ew10}}
\end{figure}

\clearpage
\begin{figure}
\epsscale{1.0}
\plotone{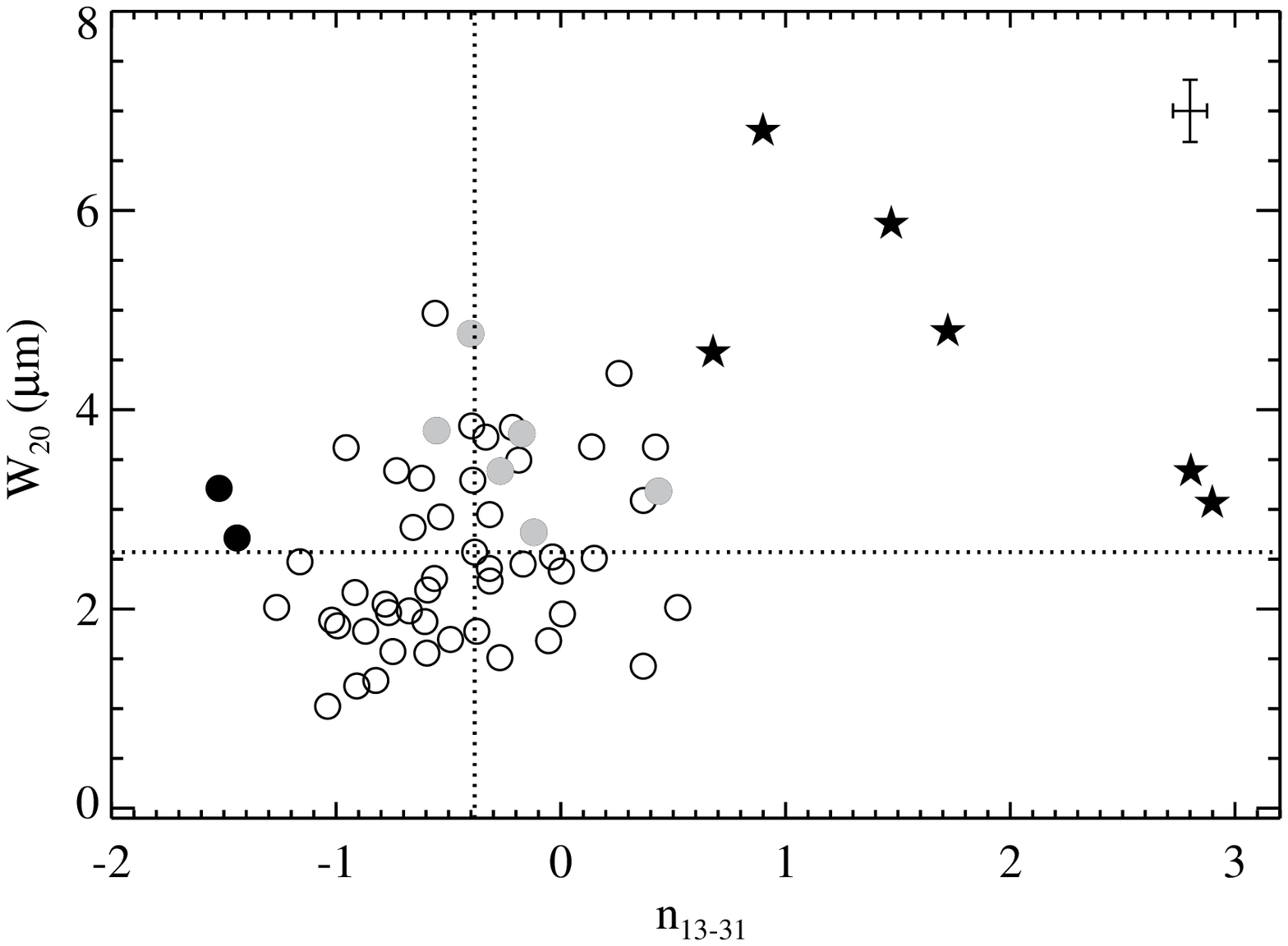}
\caption{Equivalent width of 20 $\micron$ silicate feature W$_{20}$ plotted against the continuum index $n_{13-31}$. The symbols have same meaning as in Figure~\ref{gr_ew10_index}.  Typical errors are shown on the top-right corner (see
  Table~\ref{tab5_index}). \label{gr_ew20_index}}
\end{figure}

\clearpage
\begin{figure}
\epsscale{1.0}
\plotone{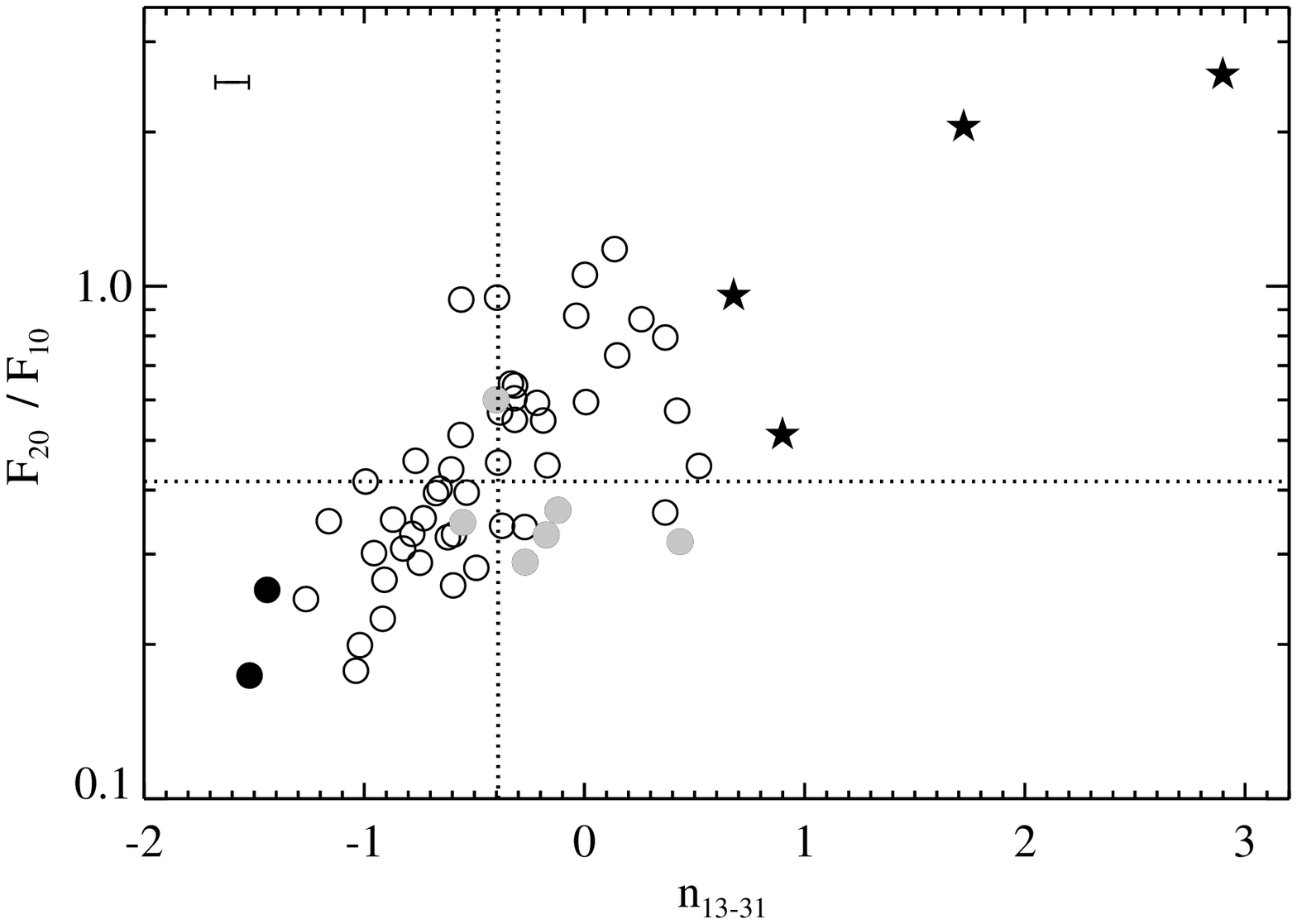}
\caption{Ratio of the integrated flux of 20 $\micron$ silicate feature to that of the 10 $\micron$ silicate feature F$_{20}$/F$_{10}$ versus  the continuum index $n_{13-31}$. The symbols have same meaning as in Figure~\ref{gr_ew10_index}.  Typical errors are shown on the top-left corner (see
  Table~\ref{tab5_index}).  \label{gr_f20f10_index}}
\end{figure}

\begin{figure}
\epsscale{1.0}
\plotone{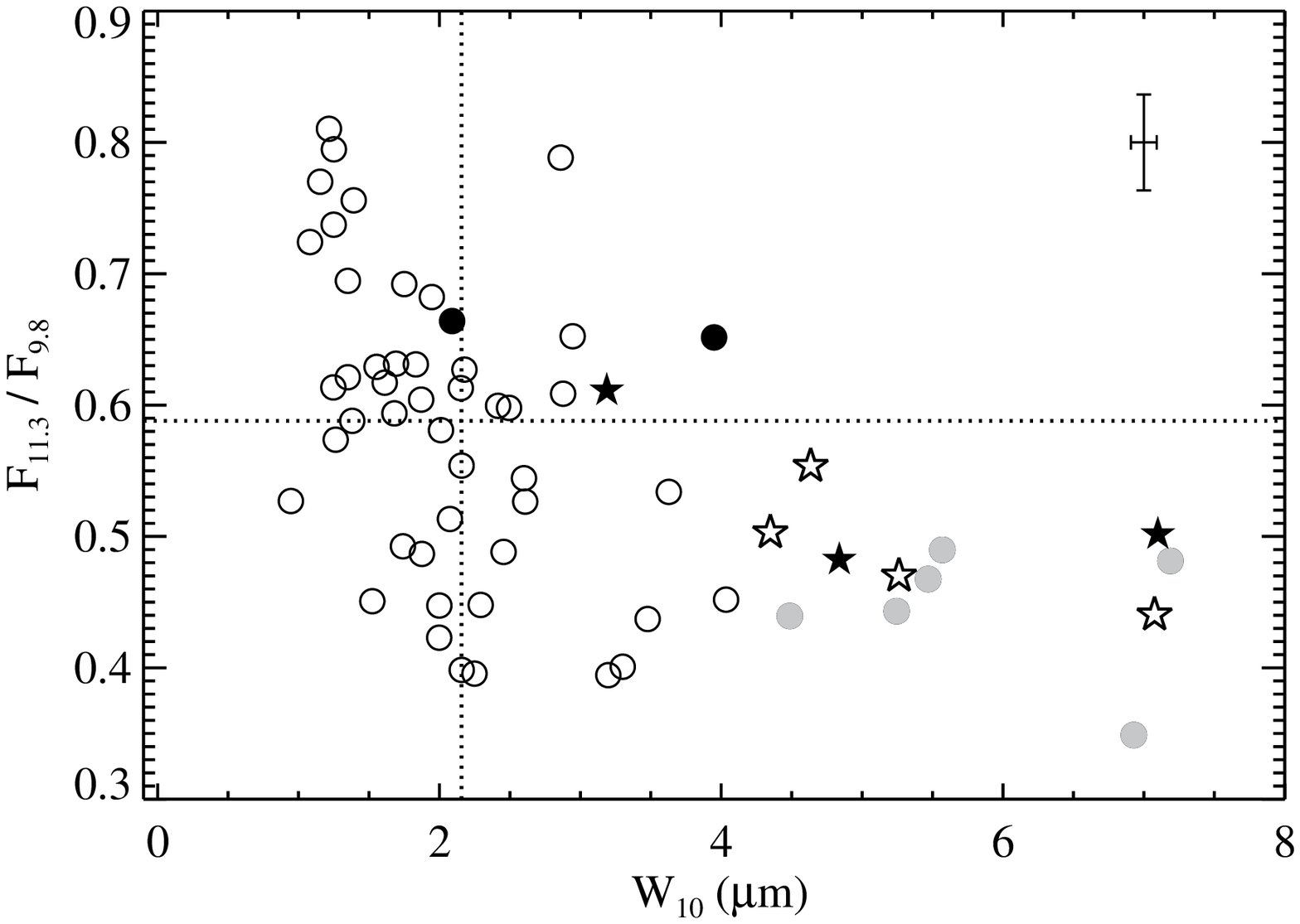}
\caption{$F_{11.3}/F_{9.8}$ plotted against the equivalent width of the 10 $\micron$ silicate feature W$_{10}$. The {\it open star symbols} represent TDs and PTDs in Tau-Aur region. The other symbols have same meaning as in Figure~\ref{gr_ew10_index}. SZ~Cha is not shown here as it has a strong PAH feature at 11.3 $\micron$ and a reliable estimate of $F_{11.3}/F_{9.8}$ could not be obtained.  Typical errors are shown on the top-right corner (see
  Table~\ref{tab5_index}). \label{gr_shape_ew10}}
\end{figure}

\begin{figure}
\epsscale{1.0}
\plotone{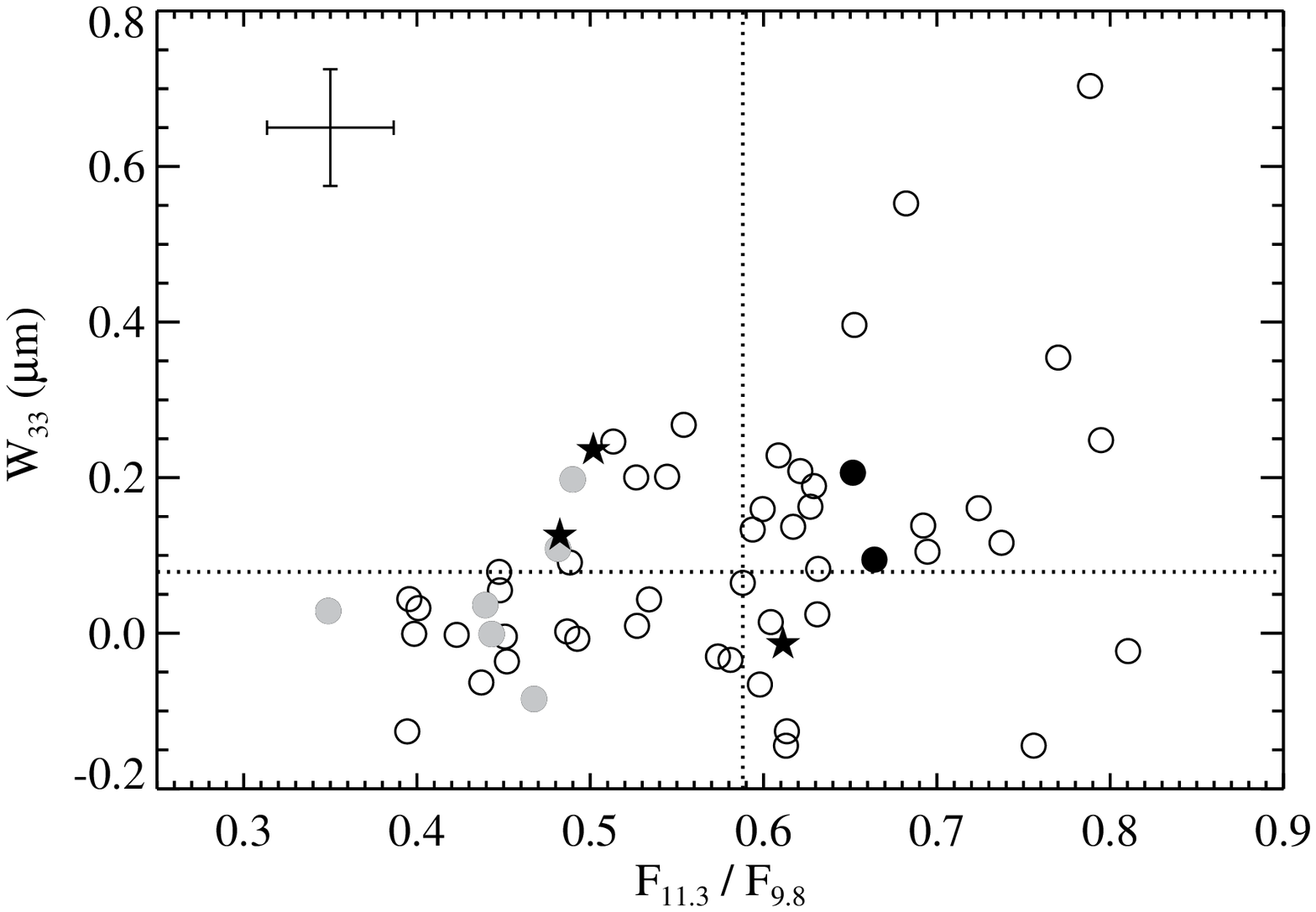}
\caption{ $F_{11.3}/F_{9.8}$ plotted against the equivalent width of the 33 $\micron$ silicate feature W$_{33}$ for Class II disks in Cha~I. The other symbols have the same meaning as in Figure~\ref{gr_ew10_index}. SZ~Cha is not shown here as it has a strong PAH feature at 11.3 $\micron$ and a reliable estimate of $F_{11.3}/F_{9.8}$ could not be obtained.  Typical errors are shown on the top-left corner (see
  Table~\ref{tab5_index}). \label{gr_shape_ew33}}
\end{figure}

\begin{figure}
\epsscale{1.0}
\plotone{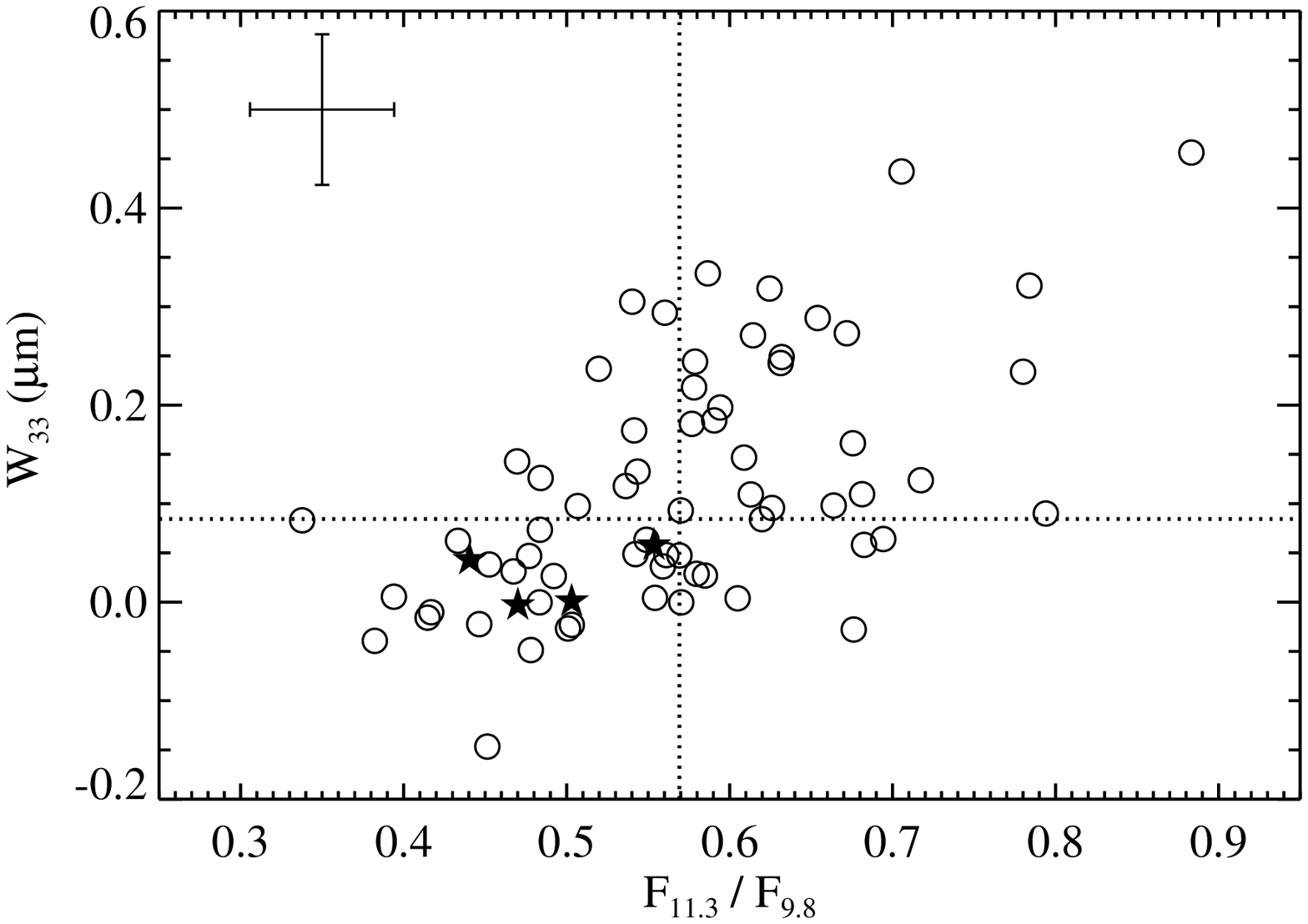}
\caption{$F_{11.3}/F_{9.8}$ plotted against W$_{33}$ for Class~II disks in Taurus. The {\it solid star symbols} represent TDs and PTDs.  Typical errors are shown on the top-left corner (see
  Table~\ref{tab5_index}). \label{tau_shape_ew33}}
\end{figure}

\begin{figure}
\epsscale{0.6}
\plotone{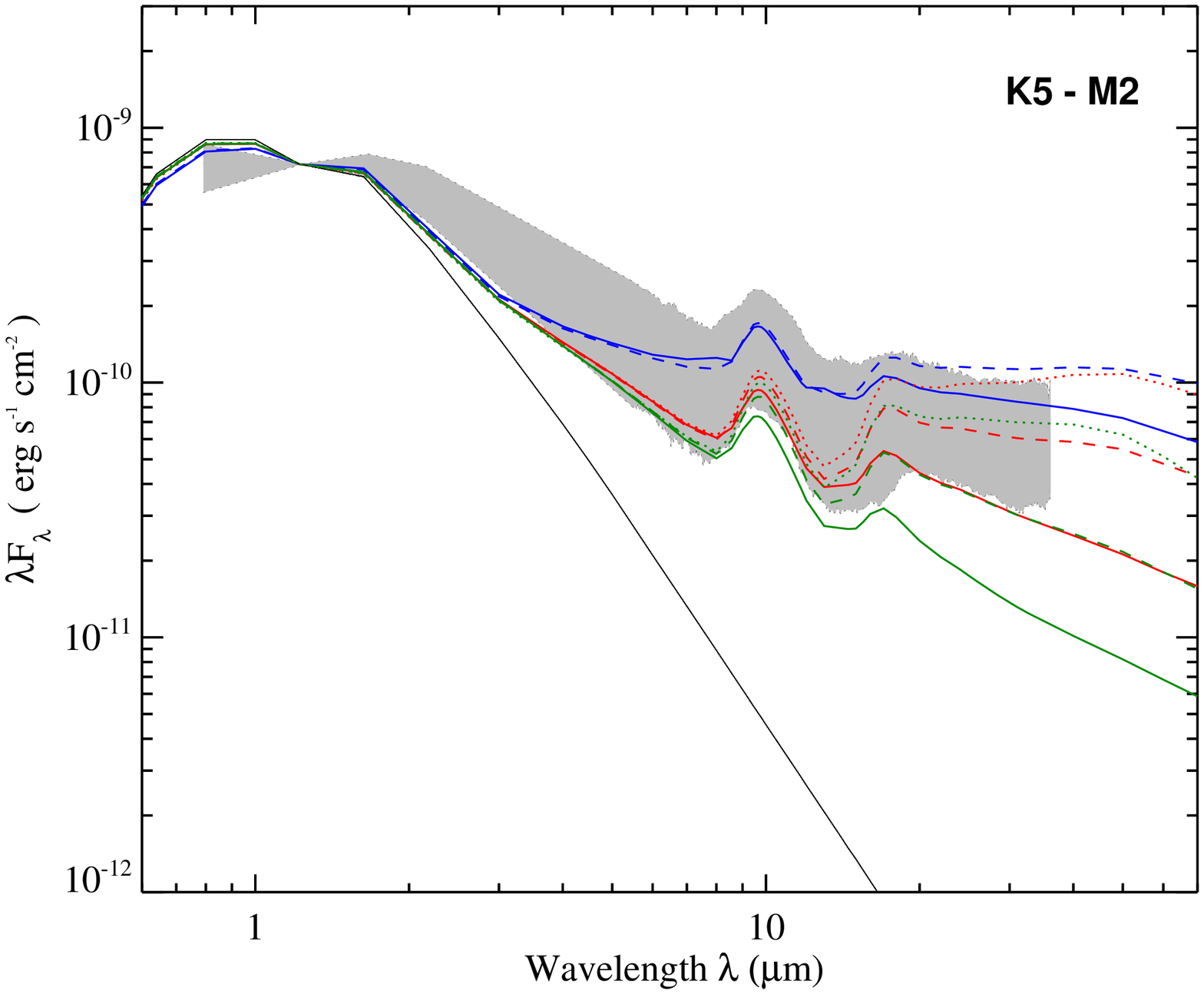}
\plotone{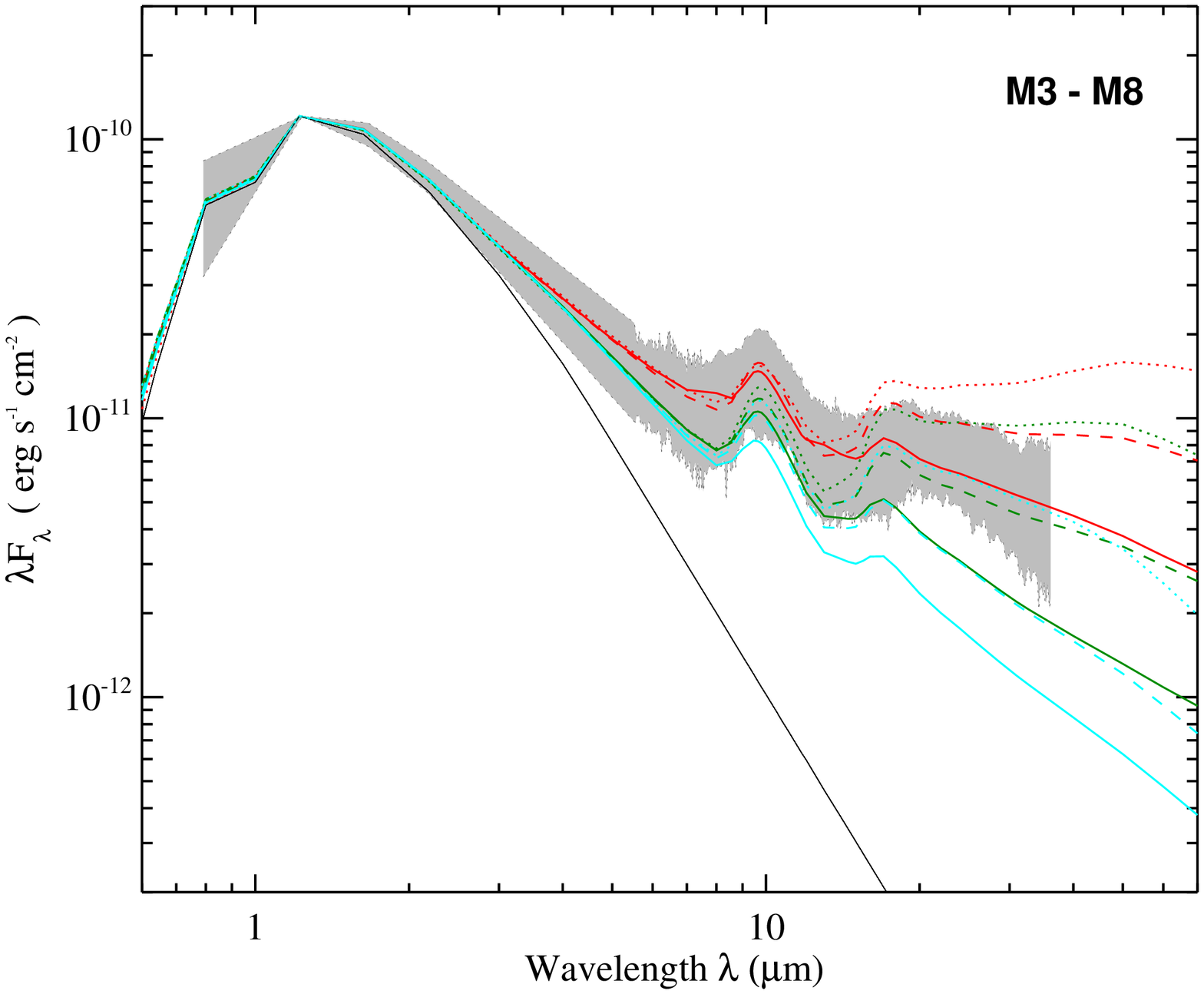}
\caption{Median SEDs of Class II objects in the spectral type range of K5-M2 {\it(top panel)} and M3-M8 {\it(bottom panel)}. The gray shaded regions are bound by the lower and upper quartiles of the median in both the panels. Synthetic SEDs generated from accretion disk models for a 0.5~M$_{\odot}$ star {\it(top panel)} for accretion rates of 10$^{-7}${\it(blue)},  10$^{-8}$ {\it(red)} and 10$^{-9}$ {\it(green)} ~M$_{\odot}$yr$^{-1}$ and  for depletion factors $\epsilon$ = 0.001 {\it(solid line)},  $\epsilon$ = 0.01 {\it(dashed  line)} and $\epsilon$ = 0.1 {\it(dotted line)} are overplotted on the K5-M2 median. SEDs generated from accretion disk models for a 0.2~M$_{\odot}$ star {\it(bottom panel)} for accretion rates of 10$^{-8}$ {\it(red)}, 10$^{-9}$ {\it(green)} and 10$^{-10}$ {\it(cyan)} M$_{\odot}$yr$^{-1}$  and for depletion factors $\epsilon$ = 0.001 {\it(solid line)},  $\epsilon$ = 0.01 {\it(dashed  line)} and $\epsilon$ = 0.1 {\it(dotted line)} are shown for the M3-M8 median. The models shown are for an inclination angle of $i$ = 60$\degr$. The photospheres plotted {\it(solid black line)} are for a K7 star (K5-M2; top panel) and for a M5 star (M3-M8; bottom panel). The photospheres are from \citet{kenhart95} and are normalized at J-band. \label{fig_medsed}}
\end{figure}

\begin{figure}
\centering
\epsscale{0.8}
\plotone{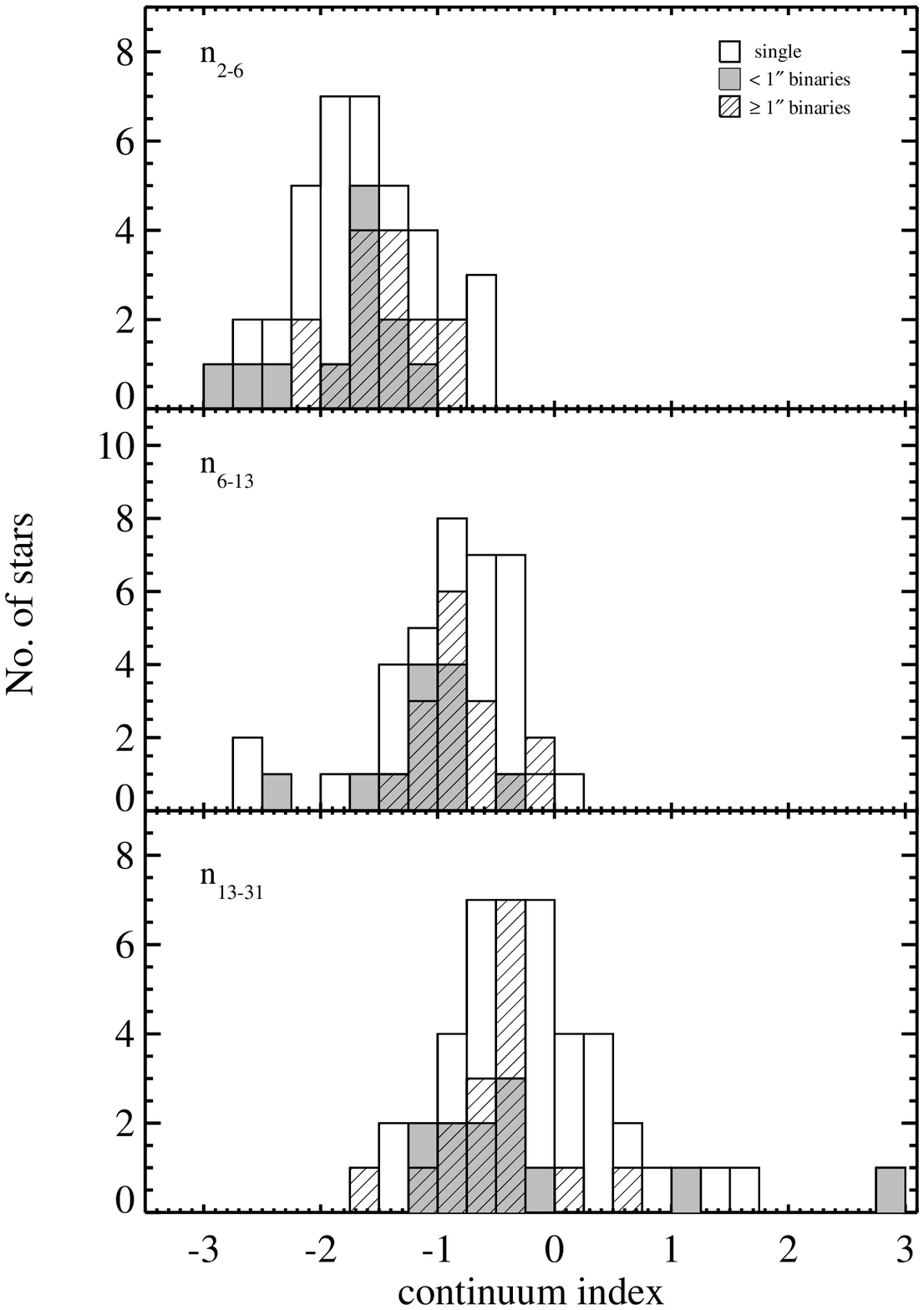}
\caption{Distribution of the continuum indices for close ($<$ 1$\arcsec$) and  wide ($\ge$ 1$\arcsec$) binaries and single stars among the Class~II objects in Cha~I} \label{bin_ind}
\end{figure}

\begin{landscape}
\begin{figure}
\centering
\epsscale{0.32}
\plotone{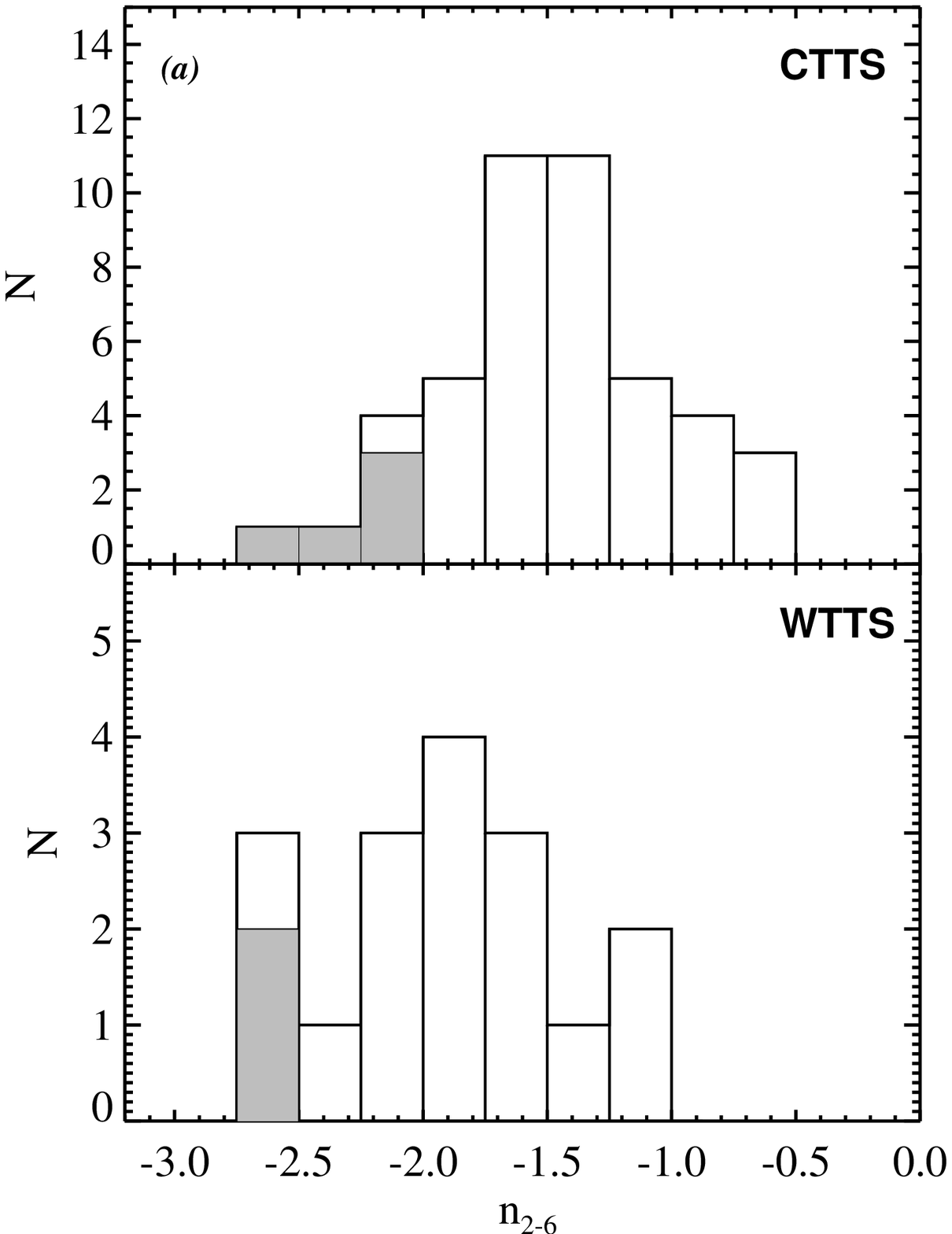}
\plotone{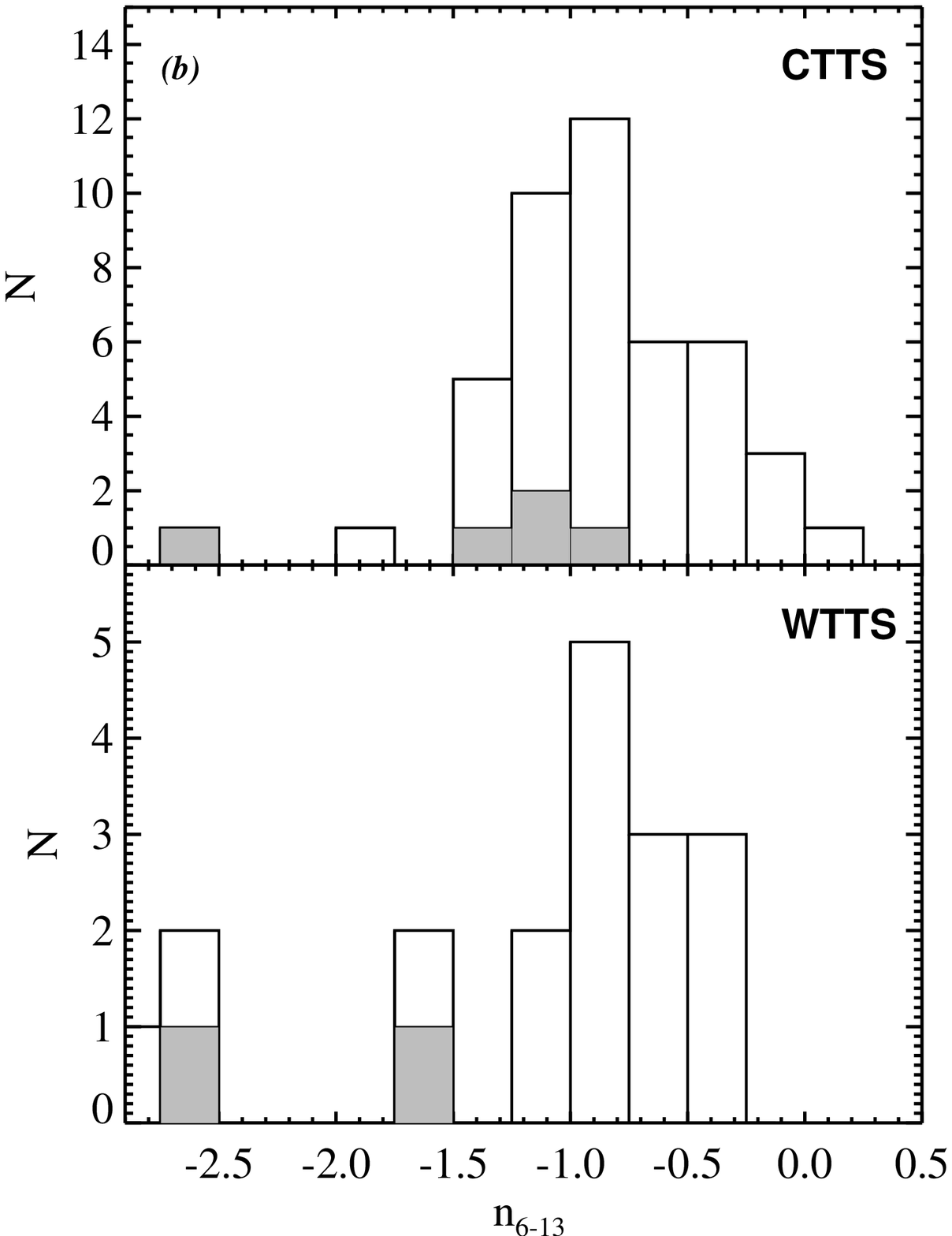}
\plotone{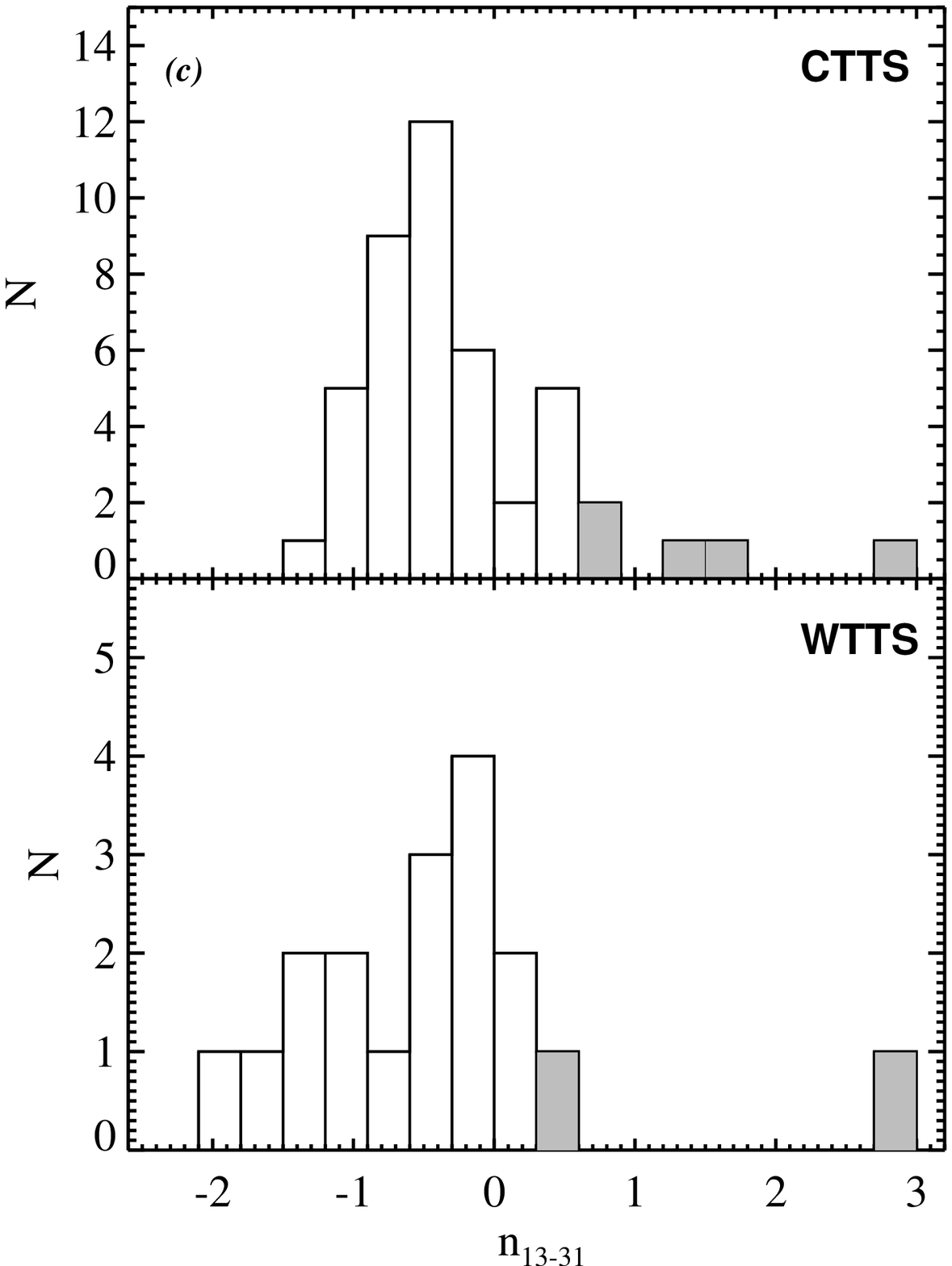}
\caption{Frequency distribution of the spectral indices $n_{2-6}$ {\it(a)}, $n_{6-13}$ {\it(b)} and $n_{13-31}$ {\it(c)} for the CTTS {\it(top panel)} and WTTS {\it(bottom panel)} in our sample. The {\it solid histogram } represents TDs and PTDs among CTTS and WTTS. \label{fig_cwtts}}
\end{figure}
\end{landscape}

\end{document}